\documentclass[twocolumn]{aastex63}

\usepackage{mathtools,hyperref}
\usepackage{amsmath}
\usepackage{graphicx}
\usepackage[figuresright]{rotating}
\usepackage{natbib}
\usepackage[all]{hypcap}								
\usepackage{fancyhdr}
\usepackage{bm}
\usepackage{url}
\usepackage{epsfig}
\usepackage{amsmath}
\usepackage{verbatim}
\usepackage{comment}

\pdfoutput=1

\newcommand{\cmfast}{{\tt 21cmFAST}}
\newcommand{\cmmc}{{\tt 21cmMC}}

\newcommand{\aveTs}{\overline{T}_{\rm S}}

\newcommand{\hone}{\mathrm{H}\textsc{i}}

\newcommand{\Rom}[1]{\uppercase\expandafter{\romannumeral #1}}
\newcommand{\rom}[1]{\lowercase\expandafter{\romannumeral #1}}

\newcommand{\avenf}{\overline{x}_{\hone}}

\newcommand{\lya}{Ly$\alpha$}

\newcommand{\Msun}{M_\odot}

\newcommand{\Tcmb}{T_\gamma}

\newcommand\lsim{\mathrel{\rlap{\lower4pt\hbox{\hskip1pt$\sim$}}
        \raise1pt\hbox{$<$}}}
\newcommand\gsim{\mathrel{\rlap{\lower4pt\hbox{\hskip1pt$\sim$}}
        \raise1pt\hbox{$>$}}}
\def\myputfigure#1#2#3#4#5%
{\vskip#5pt\makebox[0pt]{\hskip#2in
\includegraphics[width=#3\textwidth]{#1}}\vskip#4pt\hfill}

\newcommand{\referee}[1]{#1}

\def\bd{\mathbf{d}}
\def\bm{\mathbf{m}}
\def\br{\mathbf{r}}
\def\bu{\mathbf{u}}
\def\bSigma{\mathbf{\Sigma}}
\def\bGamma{\mathbf{\Gamma}}
\def\bW{\mathbf{W}}
\def\btheta{\mathbf{\theta}}
\def\Dsq{\Delta^2}

\defcitealias{HERA2021}{H21}

\shorttitle{HERA Phase I limits: theory interpretation}
\shortauthors{The HERA Collaboration}

\begin{document}
\title{HERA Phase I Limits on the Cosmic 21-cm Signal:\\Constraints on Astrophysics and Cosmology During the Epoch of Reionization} 

\collaboration{100}{The HERA Collaboration:}

\author{Zara  Abdurashidova}
\affiliation{Department of Astronomy, University of California, Berkeley, CA}

\author{James E. Aguirre}
\affiliation{Department of Physics and Astronomy, University of Pennsylvania, Philadelphia, PA}

\author{Paul  Alexander}
\affiliation{Cavendish Astrophysics, University of Cambridge, Cambridge, UK}

\author{Zaki S. Ali}
\affiliation{Department of Astronomy, University of California, Berkeley, CA}

\author{Yanga  Balfour}
\affiliation{South African Radio Observatory (SARAO), 2 Fir Street, Observatory, Cape Town, 7925, South Africa}

\author{{ Rennan Barkana}}
\affiliation{School of Physics and Astronomy, Tel-Aviv University, Tel-Aviv, 69978, Israel}

\author{Adam P. Beardsley}
\affiliation{Department of Physics, Winona State University, Winona, MN}
\affiliation{School of Earth and Space Exploration, Arizona State University, Tempe, AZ}

\author{Gianni  Bernardi}
\affiliation{Department of Physics and Electronics, Rhodes University, PO Box 94, Grahamstown, 6140, South Africa}
\affiliation{INAF-Istituto di Radioastronomia, via Gobetti 101, 40129 Bologna, Italy}
\affiliation{South African Radio Observatory (SARAO), 2 Fir Street, Observatory, Cape Town, 7925, South Africa}

\author{Tashalee S. Billings}
\affiliation{Department of Physics and Astronomy, University of Pennsylvania, Philadelphia, PA}

\author{Judd D. Bowman}
\affiliation{School of Earth and Space Exploration, Arizona State University, Tempe, AZ}

\author{Richard F. Bradley}
\affiliation{National Radio Astronomy Observatory, Charlottesville, VA}

\author{Philip Bull}
\affiliation{School of Physics \& Astronomy, Queen Mary University of London, London, UK}

\author{Jacob  Burba}
\affiliation{Department of Physics, Brown University, Providence, RI}

\author{Steve  Carey}
\affiliation{Cavendish Astrophysics, University of Cambridge, Cambridge, UK}

\author{Chris L. Carilli}
\affiliation{National Radio Astronomy Observatory, Socorro, NM}

\author{Carina  Cheng}
\affiliation{Department of Astronomy, University of California, Berkeley, CA}

\author{David R. DeBoer}
\affiliation{Department of Astronomy, University of California, Berkeley, CA}

\author{Matt  Dexter}
\affiliation{Department of Astronomy, University of California, Berkeley, CA}

\author{Eloy  de~Lera~Acedo}
\affiliation{Cavendish Astrophysics, University of Cambridge, Cambridge, UK}

\author{Joshua S. Dillon}
\altaffiliation{NSF Astronomy and Astrophysics Postdoctoral Fellow}
\affiliation{Department of Astronomy, University of California, Berkeley, CA}

\author{John  Ely}
\affiliation{Cavendish Astrophysics, University of Cambridge, Cambridge, UK}

\author{Aaron  Ewall-Wice}
\affiliation{Department of Astronomy, University of California, Berkeley, CA}

\author{Nicolas  Fagnoni}
\affiliation{Cavendish Astrophysics, University of Cambridge, Cambridge, UK}

\author{{Anastasia Fialkov}}
\affiliation{Institute of Astronomy, University of Cambridge, Madingley Road, Cambridge CB3 0HA, United Kingdom}
\affiliation{Kavli Institute for Cosmology, Madingley Road, Cambridge CB3 0HA, UK}

\author{Randall  Fritz}
\affiliation{South African Radio Observatory (SARAO), 2 Fir Street, Observatory, Cape Town, 7925, South Africa}

\author{Steven R. Furlanetto}
\affiliation{Department of Physics and Astronomy, University of California, Los Angeles, CA}

\author{Kingsley  Gale-Sides}
\affiliation{Cavendish Astrophysics, University of Cambridge, Cambridge, UK}

\author{Brian  Glendenning}
\affiliation{National Radio Astronomy Observatory, Socorro, NM}

\author{Deepthi  Gorthi}
\affiliation{Department of Astronomy, University of California, Berkeley, CA}

\author{Bradley  Greig}
\affiliation{School of Physics, University of Melbourne, Parkville, VIC 3010, Australia}
\affiliation{ARC Centre of Excellence for All-Sky Astrophysics in 3 Dimensions (ASTRO 3D)}

\author{Jasper  Grobbelaar}
\affiliation{South African Radio Observatory (SARAO), 2 Fir Street, Observatory, Cape Town, 7925, South Africa}

\author{Ziyaad  Halday}
\affiliation{South African Radio Observatory (SARAO), 2 Fir Street, Observatory, Cape Town, 7925, South Africa}

\author{Bryna J. Hazelton}
\affiliation{Department of Physics, University of Washington, Seattle, WA}
\affiliation{eScience Institute, University of Washington, Seattle, WA}

\author{{Stefan Heimersheim}}\thanks{heimersheim@ast.cam.ac.uk}
\affiliation{Institute of Astronomy, University of Cambridge, Madingley Road, Cambridge CB3 0HA, UK}

\author{Jacqueline N. Hewitt}
\affiliation{Department of Physics, Massachusetts Institute of Technology, Cambridge, MA}

\author{Jack  Hickish}
\affiliation{Department of Astronomy, University of California, Berkeley, CA}

\author{Daniel C. Jacobs}
\affiliation{School of Earth and Space Exploration, Arizona State University, Tempe, AZ}

\author{Austin  Julius}
\affiliation{South African Radio Observatory (SARAO), 2 Fir Street, Observatory, Cape Town, 7925, South Africa}

\author{Nicholas S. Kern}
\affiliation{Department of Physics, Massachusetts Institute of Technology, Cambridge, MA}

\author{Joshua  Kerrigan}
\affiliation{Department of Physics, Brown University, Providence, RI}

\author{Piyanat  Kittiwisit}
\affiliation{School of Chemistry and Physics, University of KwaZulu-Natal, Westville Campus, Durban, South Africa}

\author{Saul A. Kohn}
\affiliation{Department of Physics and Astronomy, University of Pennsylvania, Philadelphia, PA}

\author{Matthew  Kolopanis}
\affiliation{School of Earth and Space Exploration, Arizona State University, Tempe, AZ}

\author{Adam  Lanman}
\affiliation{Department of Physics, Brown University, Providence, RI}

\author{Paul  La~Plante}
\affiliation{Department of Physics and Astronomy, University of Pennsylvania, Philadelphia, PA}

\author{Telalo  Lekalake}
\affiliation{South African Radio Observatory (SARAO), 2 Fir Street, Observatory, Cape Town, 7925, South Africa}

\author{David  Lewis}
\affiliation{School of Earth and Space Exploration, Arizona State University, Tempe, AZ}

\author{Adrian  Liu}
\affiliation{Department of Astronomy, University of California, Berkeley, CA}
\affiliation{Department of Physics and McGill Space Institute, McGill University, 3600 University Street, Montreal, QC H3A 2T8, Canada}

\author{Yin-Zhe Ma}
\affiliation{School of Chemistry and Physics, University of KwaZulu-Natal, Westville Campus, Private Bag X54001, Durban 4000, South Africa}
\affiliation{NAOC-UKZN Computational Astrophysics Centre (NUCAC), University of KwaZulu-Natal, Durban 4000, South Africa}

\author{David  MacMahon}
\affiliation{Department of Astronomy, University of California, Berkeley, CA}

\author{Lourence  Malan}
\affiliation{South African Radio Observatory (SARAO), 2 Fir Street, Observatory, Cape Town, 7925, South Africa}

\author{Cresshim  Malgas}
\affiliation{South African Radio Observatory (SARAO), 2 Fir Street, Observatory, Cape Town, 7925, South Africa}

\author{Matthys  Maree}
\affiliation{South African Radio Observatory (SARAO), 2 Fir Street, Observatory, Cape Town, 7925, South Africa}

\author{Zachary E. Martinot}
\affiliation{Department of Physics and Astronomy, University of Pennsylvania, Philadelphia, PA}

\author{Eunice  Matsetela}
\affiliation{South African Radio Observatory (SARAO), 2 Fir Street, Observatory, Cape Town, 7925, South Africa}

\author{Andrei  Mesinger}
\affiliation{Scuola Normale Superiore, 56126 Pisa, PI, Italy}

\author{{Jordan Mirocha}}\thanks{jordan.mirocha@mcgill.ca}
\affiliation{Department of Physics and McGill Space Institute, McGill University, 3600 University Street, Montreal, QC H3A 2T8, Canada}

\author{Mathakane  Molewa}
\affiliation{South African Radio Observatory (SARAO), 2 Fir Street, Observatory, Cape Town, 7925, South Africa}

\author{Miguel F. Morales}
\affiliation{Department of Physics, University of Washington, Seattle, WA}

\author{Tshegofalang  Mosiane}
\affiliation{South African Radio Observatory (SARAO), 2 Fir Street, Observatory, Cape Town, 7925, South Africa}

\author{{Julian B.~Mu\~noz}}\thanks{julianmunoz@cfa.harvard.edu}
\affiliation{Center for Astrophysics | Harvard \& Smithsonian, Cambridge, MA}

\author{Steven G. Murray}
\affiliation{School of Earth and Space Exploration, Arizona State University, Tempe, AZ}

\author{Abraham R. Neben}
\affiliation{Department of Physics, Massachusetts Institute of Technology, Cambridge, MA}

\author{Bojan  Nikolic}
\affiliation{Cavendish Astrophysics, University of Cambridge, Cambridge, UK}

\author{Chuneeta D. Nunhokee}
\affiliation{Department of Astronomy, University of California, Berkeley, CA}

\author{Aaron R. Parsons}
\affiliation{Department of Astronomy, University of California, Berkeley, CA}

\author{Nipanjana  Patra}
\affiliation{Department of Astronomy, University of California, Berkeley, CA}

\author{Samantha  Pieterse}
\affiliation{South African Radio Observatory (SARAO), 2 Fir Street, Observatory, Cape Town, 7925, South Africa}

\author{Jonathan C. Pober}
\affiliation{Department of Physics, Brown University, Providence, RI}

\author{{Yuxiang Qin}}\thanks{yuxiang.l.qin@gmail.com}
\affiliation{Scuola Normale Superiore, 56126 Pisa, PI, Italy}
\affiliation{School of Physics, University of Melbourne, Parkville, VIC 3010, Australia}
\affiliation{ARC Centre of Excellence for All-Sky Astrophysics in 3 Dimensions (ASTRO 3D)}

\author{Nima  Razavi-Ghods}
\affiliation{Cavendish Astrophysics, University of Cambridge, Cambridge, UK}

\author{{Itamar Reis}}
\affiliation{School of Physics and Astronomy, Tel-Aviv University, Tel-Aviv, 69978, Israel}

\author{Jon  Ringuette}
\affiliation{Department of Physics, University of Washington, Seattle, WA}

\author{James  Robnett}
\affiliation{National Radio Astronomy Observatory, Socorro, NM}

\author{Kathryn  Rosie}
\affiliation{South African Radio Observatory (SARAO), 2 Fir Street, Observatory, Cape Town, 7925, South Africa}

\author{{Mario G. Santos}}
\affiliation{Department of Physics and Astronomy, University of Western Cape, Cape Town, 7535, South Africa}
\affiliation{South African Radio Observatory (SARAO), 2 Fir Street, Observatory, Cape Town, 7925, South Africa}

\author{{Sudipta Sikder}}
\affiliation{School of Physics and Astronomy, Tel-Aviv University, Tel-Aviv, 69978, Israel}

\author{Peter  Sims}
\affiliation{Department of Physics, Brown University, Providence, RI}

\author{Craig  Smith}
\affiliation{South African Radio Observatory (SARAO), 2 Fir Street, Observatory, Cape Town, 7925, South Africa}

\author{Angelo  Syce}
\affiliation{South African Radio Observatory (SARAO), 2 Fir Street, Observatory, Cape Town, 7925, South Africa}

\author{Nithyanandan  Thyagarajan}
\affiliation{CSIRO, Space \& Astronomy, P. O. Box 1130, Bentley, WA 6102, Australia}
\affiliation{National Radio Astronomy Observatory, Socorro, NM}
\affiliation{School of Earth and Space Exploration, Arizona State University, Tempe, AZ}

\author{Peter K.~G. Williams}
\affiliation{Center for Astrophysics | Harvard \& Smithsonian, Cambridge, MA}
\affiliation{American Astronomical Society, Washington, DC}

\author{Haoxuan  Zheng}
\affiliation{Department of Physics, Massachusetts Institute of Technology, Cambridge, MA}

\begin{abstract}

Recently, the Hydrogen Epoch of Reionization Array (HERA)  has produced the experiment's first upper limits on the power spectrum of 21-cm fluctuations at $z \sim 8$ and 10. Here, we use several independent theoretical models to infer constraints on the intergalactic medium (IGM) and galaxies during the epoch of reionization (EoR) from these limits. We find that the IGM must have been heated above the adiabatic cooling threshold by $z \sim 8$, independent of uncertainties about IGM ionization and the radio background. Combining HERA limits with complementary observations constrains the spin temperature of the $z \sim 8$ neutral IGM to 27 K $< \aveTs <$ 630 K (2.3 K $< \aveTs <$ 640 K) at 68\% (95\%) confidence. They therefore also place a lower bound on X-ray heating, a previously unconstrained aspects of early galaxies. For example, if the CMB dominates the $z\sim8$ radio background, the new HERA limits imply that the first galaxies produced X-rays more efficiently than local ones. The $z\sim 10$ limits require even earlier heating if dark-matter interactions cool the hydrogen gas.
If an extra radio background is produced by galaxies, we rule out (at 95\% confidence) the combination of high radio and low X-ray luminosities of $L_{r,\nu}$/{\rm SFR}$ > 4 \times 10^{24}$ W Hz$^{-1}$ $M_\odot^{-1}$ yr and $L_{\rm X}$/SFR$<7.6 \times 10^{39}$ erg s$^{-1}$ $M_\odot^{-1}$ yr.
The new HERA upper limits neither support nor disfavor a cosmological interpretation of the recent EDGES measurement.
The framework described here provides a foundation for the interpretation of future HERA results.  

\end{abstract}


\section{Introduction}
\label{sec:intro}	

One of the final frontiers of observational cosmology is the Cosmic Dawn, during which the first luminous sources formed and grew into galaxies. This era ended with the reionization of the intergalactic medium (IGM), when ultraviolet photons from these sources ionized virtually all the neutral hydrogen -- and hence when stars and black holes affected every baryon in the Universe. This constitutes the last baryonic phase transition in the Universe's history and has important implications for later generations of galaxies.

Observations are now beginning to probe this era. Measurements of the large-scale polarization of the cosmic microwave background (CMB) imply that reionization reached its midpoint at $z \sim 7$--$8$ \citep{Planck18,debelsunce21,heinrich21}. Models of Lyman-$\alpha$ emission lines of galaxies \citep{Stark10, Schenker12, Jensen13, Caruana14, Pentericci14, Mesinger15, mason18, Mason19} and quasars \citep{mesinger04,Bolton11,Greig17b,Davies18, Yang20, Wang20, GMB21} also suggest a relatively large neutral fraction at $z \sim 7$. While the conventional wisdom has long held that the reionization process ends at $z \sim 6$ (e.g. \citealt{McGreer2015}; though see \citealt{LOF06, Mesinger10}), recent measurements of the Lyman-$\alpha$ forest suggest that it may continue to somewhat later times \citep{becker15, bosman18, kulkarni19, keating20, nasir20, choudhury21, Qin21}. 

However, our understanding of this era is still incomplete: models and empirical extrapolations suggest that even the deepest \emph{Hubble Space Telescope} (and upcoming {\it JWST}) observations probe only a fraction of the total star formation in the early Universe \citep{Robertson2015, behroozi15, Mason15, Furlanetto17, GMP20}. This could mean that the galaxies providing most of the reionizing photons will remain {\it unseen}.  Moreover, while reionization is the most dramatic effect of the first galaxies, their X-ray and ultraviolet radiation fields can affect the IGM even while it remains neutral -- a phase that cannot be observed directly by many cosmological probes. 

A complete understanding of the Cosmic Dawn therefore requires complementary measurements of the IGM gas. The most powerful potential probe is the 21-cm spin-flip line of neutral hydrogen \citep{Field59, Madau97}. The 21-cm line is particularly sensitive to \citep{Furlanetto06, Morales12, Pritchard12}: (1) structure formation in the Universe, which can be observed through density fluctuations; (2) the reionization process, which eliminates the 21-cm signal inside the large ionized bubbles that grow throughout that era; (3) the X-ray background (or other exotic heating or cooling mechanisms), which likely sets the IGM temperature before reionization and hence determines whether the 21-cm line is seen in absorption or emission; (4) the non-ionizing ultraviolet background, as photons that redshift into the hydrogen Lyman-$\alpha$ transition mix the hyperfine level populations; and (5) the radio background at high redshifts, including the CMB but also potential contributions from astrophysical sources or exotic processes. 

Because the spin-flip cosmological signal is very weak compared to other astrophysical radio backgrounds, mapping these IGM fluctuations is extremely challenging, and early efforts to observe it have focused on two complementary directions. One is the ``global" all-sky signal, measuring the sky averaged spectral signature of the line, covering Gpc$^3$-sized comoving volumes~\citep{Shaver:1999gb,Munoz:2020itp}. 
Several such experiments are underway~\citep{LEDA,Singh:2017syr,PRIZM,Voytek:2013nua,DiLullo:2020owx}. Of these, only the EDGES collaboration has made a tentative detection \citep{Bowman18}, although the cosmological interpretation of the measurement is subject to significant instrumental and systematic uncertainties (e.g. \citealt{Hills18, SP19, Bradley19, Singh:19, Tauscher20}). Interestingly, the claimed signal is much stronger than expected, requiring either that the IGM temperature is smaller than allowed by adiabatic cooling (from, e.g.,  energy exchange with dark matter; \citealt{Munoz:2018pzp,Barkana18,Berlin:2018sjs,Slatyer:2018aqg,Kovetz:2018zan}), or that an additional radio background (beyond the CMB) is present in the early Universe (e.g., \citealt{Feng:2018,Pospelov:2018,EW18,FB19,Mebane2020}).

A number of other experiments hope to use interferometers to measure statistical fluctuations in the 21-cm background, most often quantified through the power spectrum, which measures the variance in the field as a function of smoothing scale. Several experiments have now published upper limits from $z \sim 6$--10, though these limits so far probe only a small fraction of the parameter space spanned by  ``standard" models of early galaxies  \citep{Mondal:2020,Ghara2020,Greig21LOFAR,Greig21MWA, Ghara21}. 

Recently in \citet{HERA2021} (hereafter \citetalias{HERA2021}), we presented the first upper limits on the 21-cm power spectrum from a new experiment, the Hydrogen Epoch of Reionization Array (HERA). HERA is now under construction in the Karoo Desert of South Africa \citep{DeBoer17}. Its phased construction allowed an initial observing campaign in 2017-18, whose results are considered here. Note that we base these results on data from just 39 antennas; HERA is now expanding to $\sim 350$ antennas, so the interpretation here provides a framework for improved analyses in the future.

This paper is organized as follows. We introduce the physics of the 21-cm signal in section \ref{sec:21cm-overview}. Then, in section \ref{sec:limits}, we describe HERA's limits and our inference tools.  In section \ref{sec:bias}, we use a very simple model to motivate the most important implications of HERA's upper limit. In the following four sections, we present several complementary interpretations to elucidate these results: we use the {\tt 21cmMC} code to infer constraints on early-galaxy populations and the IGM (section \ref{sec:21cmmc}) and a phenomenological model that directly parametrizes IGM properties to better understand the IGM constraints (section \ref{sec:phenom_models}). Then, we examine the implications of the HERA limits for exotic dark-matter models (section \ref{sec:exotic_DM}), and finally we consider constraints derived from models with an enhanced radio background (section \ref{sec:radio}). In section \ref{sec:disc}, we summarize these results and their implications for the epoch of reionization.

Throughout this work, we assume a standard flat $\Lambda$CDM cosmology, consistent with the latest CMB measurements \citep{Planck18}. The separate analyses use slightly different cosmological parameters, but these have little effect on our constraints. We denote comoving Mpc with cMpc. 


\section{The 21-cm signal}
\label{sec:21cm-overview}

HERA and other low-frequency instruments aim to observe emission or absorption of the neutral hydrogen hyperfine transition at an observed wavelength of $\lambda_{\rm obs} = 21(1+z)$~cm. The intensity of this line is conventionally expressed as the differential brightness temperature, $\delta T_{21}$, relative to the low-frequency radio background, which we assume has a brightness temperature $T_{\rm rad}$ at the relevant frequency. Then the brightness of a patch of the IGM can be expressed approximately as \citep{Madau97,Furlanetto06}
\begin{equation}
    \begin{aligned}
        \delta T_{21}(\nu) &= \frac{T_{\mathrm{S}}-T_{\rm rad}}{1+z}\left(1-e^{-\tau_{\nu_{0}}}\right) \\
        &\approx T_0 x_{\mathrm{HI}}\left(1+\delta \right)\left(\frac{H}{\mathrm{d} v_{r} / \mathrm{d} r+H}\right)\left(1-\frac{T_{\rm rad}}{T_{\mathrm{S}}}\right),
        \end{aligned}
    \label{eq:21cm_signal}
\end{equation}
where $T_0 = 27 [(1+z) / 10]^{1/2}$~mK is the overall normalization, $H$ is the Hubble parameter at the appropriate redshift, and we have assumed $\Omega_m h^2=0.15$ and $\Omega_b h^2 = 0.023$ (with $H_0 = 100 h$~km/s/Mpc). Here, $x_{\rm HI}$ is the neutral fraction of the patch, $\delta = (\rho - \bar{\rho})/\bar{\rho}$ is its fractional overdensity, $T_S$ is the spin temperature (or the excitation temperature of the 21-cm transition), and $dv_r/dr$ is the gradient of the proper velocity along the line of sight. 

The spin temperature is determined by (e.g., \citealt{Madau97,Furlanetto06,Pritchard12,Venumadhav18})
\begin{equation}
T_S^{-1} = \frac{x_{\rm rad} T_{\rm rad}^{-1} + x_c T_K^{-1} + x_\alpha T_K^{-1}}{x_{\rm rad} + x_c + x_\alpha},
\label{eq:Tspin}
\end{equation}
where $x_{\rm rad}$, $x_\alpha$, and $x_c$ are coupling constants describing the strength of the relevant interactions.  This equation reflects the competition between several processes: (1) interactions with radio photons tend to drive $T_S$ to $T_{\rm rad}$, with a coupling constant $x_{\rm rad}$; (2) collisions drive $T_S$ toward the kinetic temperature of the gas, $T_K$ with a coupling constant $x_c$; and (3) absorption and re-emission of Lyman-$\alpha$ photons mixes the hyperfine states and also drives $T_S$ toward $T_K$ with a coupling $x_\alpha$, through a process known as the Wouthuysen-Field effect \citep{Wouthuysen52,Field58,Field59, Hirata06}. Meanwhile, the kinetic temperature is affected by the expansion cooling of the IGM and interactions with several radiation backgrounds -- most importantly, prior to reionization, any X-ray background generated by early sources. A proper accounting of the temperature requires tracking both the IGM properties and the radiation backgrounds generated by galaxy formation or exotic processes in the early Universe. 

It is important to note that, in the standard picture, reionization by UV photons is an inhomogeneous process -- (nearly) completely ionized regions around the first galaxies expand into (nearly) completely neutral IGM patches as the source population grows.  The values of  $\avenf$ we quote below can therefore be considered as approximately corresponding to the volume filling factor of the remaining neutral IGM patches during the EoR. However, X-rays have much longer mean free paths than UV photons and can deposit their energy in the neutral IGM, partially ionizing and heating that phase, so the relation between the true neutral fraction and the filling factor of the ionized bubbles is not exact.

Given the sensitivity of current experiments, the focus of interferometric observations to date has been on measuring the spatial power spectrum  of the 21-cm signal,
\begin{equation}
    \langle \tilde{\delta T}_{21}({\bf k}_1) \tilde{\delta T}_{21}({\bf k}_2) \rangle = (2 \pi)^3 \delta^D({\bf k}_1 + {\bf k}_2) P_{21}({\bf k}_1),
\end{equation}
where tildes denote Fourier transforms, angular brackets denote ensemble averages, and $\delta^D$ is the Dirac delta function. We will typically plot $\Delta_{21}^2({\bf k}) \equiv k^3 P_{21}({\bf k})/(2 \pi^2)$, with units of mK$^2$. 

The velocity term in equation~(\ref{eq:21cm_signal}) accounts for the mapping between redshift and real space, which is complicated by redshift-space distortions (RSDs; \citealt{Kaiser87,BA04,BL05}). Crudely, overdense regions expand more slowly than the average Universe, so they appear compressed along the radial direction, while underdense regions appear larger in that direction. Because these distortions occur only along the line of sight, they make the power spectrum anisotropic. The modes used in the HERA analysis are mostly aligned along the line of sight and care must be taken when comparing to the theoretical models, as we discuss further below.


\begin{figure}
		\includegraphics[width=0.48\textwidth]{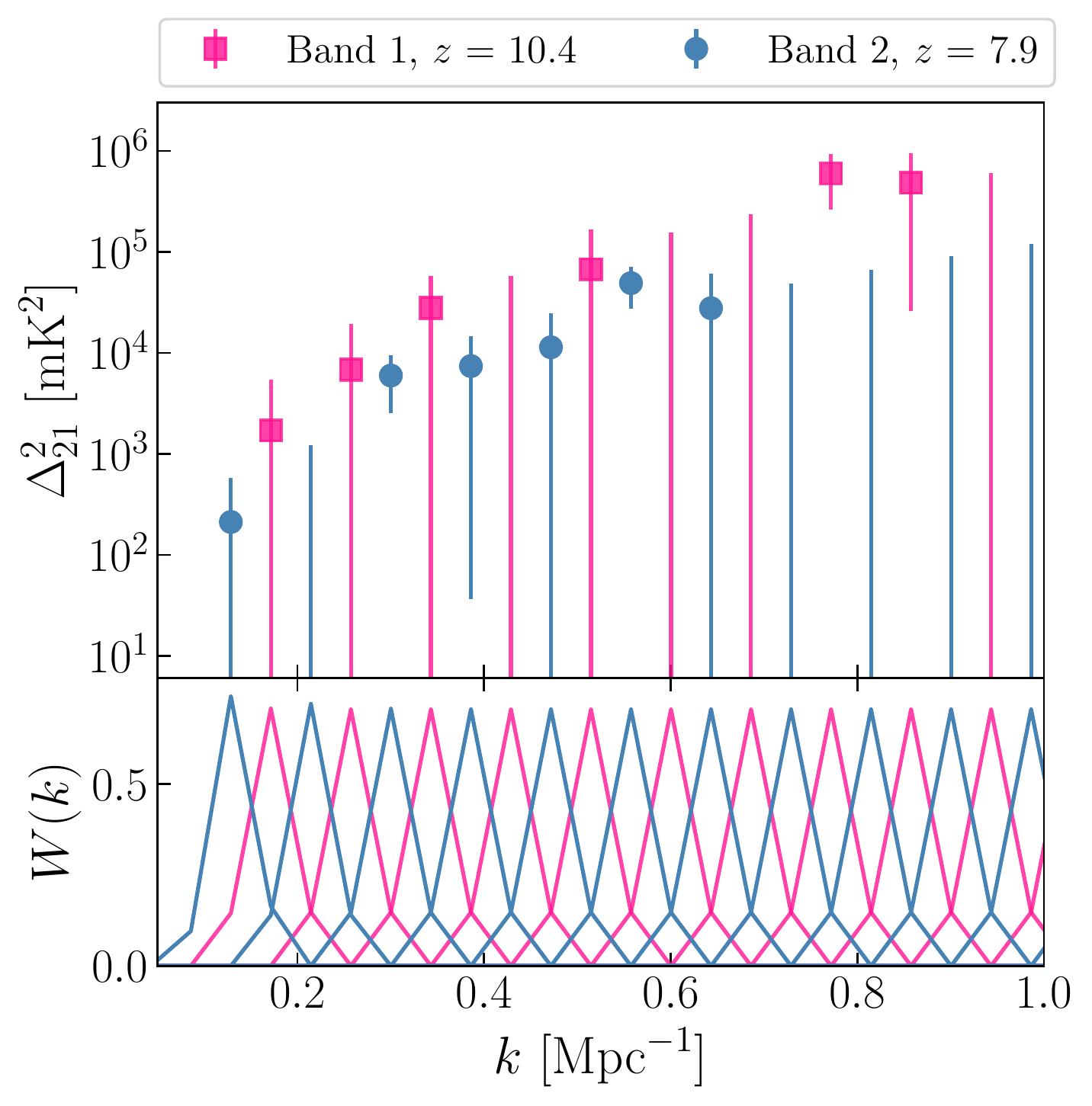}
    \caption{\label{fig:data-limits} Reported limits on the 21\,cm power spectrum from \citetalias{HERA2021} used to place constraints on the various models explored in this work (top panel). The bottom panel shows the derived window functions of the limits, showing a peaked sensitivity with compact support around each $k$ mode. Note that in the present analysis we only include every other $k$ bin from the limits quoted in \citetalias{HERA2021} in order to mitigate the effect of non-zero covariance between neighboring $k$ modes. We start this decimation at $k=0.179$/cMpc and $k=0.134$/cMpc for Bands 1 and 2, respectively. Filled points represent positive measurements, and errorbars without points represent negative measurements. The errorbars show $\pm1\sigma$ uncertainty. 
	}
\end{figure}

\section{HERA Phase I Power Spectrum Limits}
\label{sec:limits}	

Next we establish the formalism that will be used to interpret our observables. Section~\ref{sec:ObsCampaign} describes HERA's data products that are used in this paper, and section~\ref{sec:likelihood} defines the likelihood that links these observable quantities to theoretical models. The goal is therefore to provide the necessary machinery to interpret our measurements in a model-agnostic way before introducing our theoretical models in subsequent sections.

\subsection{Observational Campaign}
\label{sec:ObsCampaign}
The power spectrum upper limits analyzed in this paper have been published in \citetalias{HERA2021}. Here, we describe some of the essential features of the data for convenience, but we refer the reader to \citetalias{HERA2021} for more details.

The upper limits relevant to the paper are reproduced in Figure~\ref{fig:data-limits}. These were based on 18 nights of data (Julian Dates 2458098 to 2458116) taken as part of an observing campaign from October 2017 to April 2018 when HERA was in its Phase I observing configuration. In Phase I, HERA observed with ``hybrid'' antenna elements which consisted of HERA's 14-m parabolic antennae with modified cross-dipole feeds and a front-end from the PAPER experiment \citep{Parsons10, DeBoer17}. HERA Phase I also inherited PAPER's back-end system, which processed 100 MHz of bandwidth from 100--200 MHz. For these observations, HERA consisted of 52 operating antennas, 39 of which were deemed science-ready after passing our data quality metrics \citepalias{HERA2021}. 
Note that these 52 antennas make up a small fraction of the experiment at full capacity of $\sim$350 antennas, which will observe from 50 -- 225 MHz \citep{DillonParsons2016, DeBoer17}.

The analysis and reduction of these data are discussed in \citetalias{HERA2021} and in several supporting papers in more detail \citep{Kern2020a, Kern2020b, Dillon2020, Tan2021, Aguirre2021}.
For the purposes of this work, the important takeaway is that, while nearly the full band is processed in the data reduction pipeline, only two portions of the band are largely free of radio frequency interference (RFI), which sets the redshift ranges studied in this work (Band 2, centered at $z=7.9$, and Band 1, centered at $z=10.4$).
Additionally, the power spectra studied in this work come from only one of the fields reported in \citetalias{HERA2021}. Because HERA observes in a drift scan mode, it surveys a $\sim 10^\circ$-wide stripe centered on declination $-30.7^\circ$. However, to avoid the brightest portions of the sky (including foregrounds from our Galaxy as well as bright sources such as Fornax A), \citetalias{HERA2021} made further cuts to the data in local sidereal time (LST). This yields three fields (with LST ranges from 1.25 to 2.7 hours, 4.5 to 6.5 hours, and 8.5 to 10.75 hours) worth of data that were propagated through to the power spectrum pipeline. The parameter inference discussed in this work comes solely from the limits presented from the first cut (Field 1; see Fig.~1 in \citetalias{HERA2021}), as these showed the least amount of foreground contamination and therefore produced the most stringent limits.

For the $z\sim8$ band, the data presented in \citetalias{HERA2021} provide the most sensitive upper limits on the 21\,cm power spectrum to date, improving upon previous limits at that redshift by roughly one order of magnitude. Another important feature of the \citetalias{HERA2021} analysis is that they report measurements consistent with the thermal noise floor at intermediate and high Fourier $k$ wavevectors.  The dynamic range between that noise floor and the peak measured foreground signal is $\sim 10^9$ in power, in spite of the fact that they perform no explicit foreground subtraction in their analysis.
Upper limits on the 21\,cm power spectrum are also currently best constrained by the MWA at lower redshifts \citep{Trott20} and by LOFAR at higher redshifts \citep{Gehlot19, Mertens20}.

\subsection{Data Likelihood}
\label{sec:likelihood}

To relate our power spectrum measurements to theoretical models, we first group our data at all $k$ bins and redshifts into a column vector, i.e.,
\begin{align}
\label{eq:data_vec}
\bd = \left(\begin{array}{c}\Dsq_{21}(k_1, z_1) \\ \Dsq_{21}(k_1, z_2) \\ \Dsq_{21}(k_2, z_1) \\ \Dsq_{21}(k_2, z_2) \\ \vdots \end{array}\right),  
\end{align}
which has a length of $N_d = N_k \times N_z$.
In this work, we use the power spectrum data tabulated in \citetalias{HERA2021} Tables 3 and 4 for Field 1 only, spanning a $k$ range of 0.13--0.64 ${\rm cMpc}^{-1}$ and the two redshift bins $z=10.4$ and $z=7.9$.
Furthermore, we also make use of the associated window function and covariance matrices, which are included with the data and will be publicly accessible.
In this work, we assume the thermal noise on the data to be Gaussian distributed and thus adopt a Gaussian likelihood.
This is a fair approximation as the large amounts of averaging performed in the analysis Gaussianizes the data due to the central limit theorem.
Having adopted a model $\mathcal{M}$ for the cosmic 21\,cm signal (e.g. one of the simulations described in later sections), ${\mathbf m}$, and a model for any extant systematics, $\bu$, we can write the probability distribution for the data given the parameters (i.e., the likelihood function ) as
\begin{align}
\label{eq:likelihood_func}
\mathcal{L}(\bd | \btheta, \mathcal{M},\mathbf{u}) \propto \exp\left(-\frac{1}{2}\br(\btheta,\bu)^T\bGamma\br(\btheta,
\bu)\right),
\end{align}
where $\br(\btheta,\bu) = \bd- \bu - \bW\bm(\btheta)$, $\btheta$ are the parameters of $\mathcal{M}$, $\bm$ is the simulation's deterministic prediction of the data vector mean given $\btheta$, $\bW$ is the $N_d\times N_d$ window function matrix of the data\footnote{In general, the window function matrix can be of shape $N_d \times N_m$, with $N_m$ the number of $k$-bins predicted by the model. In our case, we estimate the window function along with the data power spectrum and discretize into the same space.}, and $\bGamma = \bSigma^{-1}$ is the $N_d\times N_d$ precision matrix, which is the inverse of the covariance matrix of the data.
The window functions account for the corrections to the predicted mean vector due to the telescope measurement and data reduction process \citep[c.f.][]{Tegmark1997, Liu2011, Dillon13, LiuShaw2020, Kern2021}; in other words, it is the point spread function of the power spectrum measurement in Fourier $k$ space.
The covariance matrix accounts for the variance of the measured power spectrum and the correlation of that uncertainty between band powers, irrespective of non-thermal systematics. This covariance is assumed to be diagonal given the analysis methods in \citetalias{HERA2021} (see Section~\ref{sec:marginal_likelihood} for details).  The on-diagonal elements (i.e., the variances) are estimated using antenna auto-correlation data to model the instrument noise. Because the power spectrum is a quadratic statistic, the sky signal enters in various signal-noise cross terms even if our variance model is due entirely to instrumental noise. For this contribution it is the total sky signal (including foregrounds) that matters, and we model this using the empirically measured power spectrum, as detailed in \citet{Tan2021}. Note that while we write $\bSigma$ as model independent, there are some terms that can be model dependent, and thus it can take on an explicit dependence on $\btheta$ (cosmic variance, for example, is dependent on the amplitude of the predicted mean signal). For the current limits, we do not expect cosmic variance to be important \citepalias{HERA2021}.

Ultimately, one is interested in the probability distribution of the parameters $\btheta$ \emph{given} the data $\bd$, i.e. the posterior probability distribution $p(\btheta,\bu | \bd, \mathcal{M})$. This is related to the likelihood via Bayes' theorem:
\begin{equation}
\label{eq:Bayes}
p(\btheta, \bu | \bd, \mathcal{M}) \propto \mathcal{L}(\bd | \btheta, \mathcal{M},\bu) p(\btheta | \mathcal{M}) p(\bu),
\end{equation}
where $p(\btheta | \mathcal{M})$ is our prior distribution on the parameters and $p(\bu)$ is our prior on the systematics (assumed to be independent of the physical parameter prior).

\subsubsection{Marginalizing over systematics}
\label{sec:marginal_likelihood}

The likelihood as expressed in equation~(\ref{eq:likelihood_func}) (and thus the posterior in eq.~\ref{eq:Bayes}) has a dependence on the systematics, $\bu$. 
In this paper, we have no explicit way of modelling $\bu$, so we desire a likelihood which is dependent only on the astrophysical parameters. 
This suggests \textit{marginalizing} over the prior range of the unknown systematics. 
In principle, we would express $\bu = \bu(\phi)$, i.e. we would have some physically motivated set of parameters $\phi$ that produce a set of systematics in $\Delta_{21}^2(k,z)$, and we would marginalize the posterior over these parameters. 
In the absence of such a physically motivated model, we marginalize directly over the binned values $\bu$:

In particular, taking a multivariate uniform prior on $\bu$ gives
\begin{align}
  p(\btheta|\bd,\mathcal{M})  \propto p(\btheta | \mathcal{M}) \int_{\bu_{\rm min}}^{\bu_{\rm max}} \exp\left(-\frac{1}{2}\br(\btheta,\bu)^T\bGamma \br(\btheta,\bu) \right) d\bu.
    \label{eq:marg_like_integral}
\end{align}
Note the assumptions that have been made in obtaining this expression. Here, our multivariate uniform prior on $\bu$ allows each $k$ and $z$ bin to vary independently, thus allowing random fluctuations of arbitrary form.
Although this is the form we employ for this paper, future analyses would be considerably improved with detailed physical models for systematics that might be present, for example by imposing smoothness priors (in $k$ and/or $z$) when appropriate.

If we also assume that $\bGamma$ is diagonal, and writing $\mathbf{t} = \bd - \bW\bm(\btheta)$, then this equation reduces to 
\begin{align}
     p(\btheta|\bd, \mathcal{M}) &\propto  p(\btheta | \mathcal{M}) \int_{\bu_{\rm min}}^{\bu_{\rm max}} \exp\left(-\frac{1}{2} \sum_i^{N_d} (t_i - u_i)^2\Gamma_{ii}\right) d\bu \nonumber \\
    &= p(\btheta | \mathcal{M}) \prod_i^{N_d} \int_{u_{i, {\rm min}}}^{u_{i, {\rm max}}} \exp\left(-\frac{\left[t_i - u_i\right]^2}{2\sigma_i^2}\right) du_i,
    \label{eq:product_integral}
\end{align}
where the second line follows due to the separability of the factors in $u_i$, and $\sigma_i = (\bGamma^{-1/2})_{ii}$ is the standard deviation of $d_i$. In this paper, we utilize bandpowers that are widely separated in wavenumber (see below) so that a diagonal covariance matrix is a good approximation. 

In order to provide a systematics-marginalized likelihood, we must choose prior ranges for the systematics on each $(k,z)$-bin. 
Allowing for unbounded (possibly negative) systematics would not allow us to constrain the cosmological signal as the systematic would be
completely degenerate with the model. Thus, we should look to our understanding of the data analysis process to set this prior.
Calibration errors causing residual phase differences or chromatic effects can lead to biases that are positive or negative. Though negative biases or systematics in the power spectrum have been observed in previous experiments (see, e.g., \citealt{Kolopanis2019}), for the present H21 dataset and analysis pipeline the most likely causes of such issues have been mitigated by the application of absolute calibration and other improvements; large negative detections are not observed in null tests or validation simulations. The most likely remaining source of systematic bias is unmodeled signal chain chromaticity common to all elements, and this would couple \emph{positive} foreground power beyond the wedge. Given this expectation of positive-only systematics, we set the prior constraint that $\mathbf{u} \ge 0$, yielding
\begin{align}
\label{eq:final_marg_likelihood}
  p(\btheta|\bd,\mathcal{M}) \propto p(\theta|\mathcal{M}) \prod_i^{N_d}\frac{1}{2}\left(1+{\rm erf}\left[\frac{t_i}{\sqrt{2}\sigma_i}\right]\right),
\end{align}
where ${\rm erf}$ is the error function. It is worth making clear that this form of the posterior is relatively flat once $t_i \gtrsim \sigma_i$. Since $t_i$ represents the data minus the theory model, in effect this means that our posterior produces close to equal probability for any scenario in which the model is less than or equal to measured values (within error bars). Our treatment of systematics therefore leads to a well-defined posterior that naturally treats data points as ``upper limits". 
This result is the same form as that derived in Appendix B of \citet{Ghara2020} in the interpretation of LOFAR data.
A similar derivation of the marginal upper-limit likelihood can also be found in Appendix A of \citet{Li2019}.

\label{sec:datapoint_decimation}
If the off-diagonal components of $\bGamma$ are not zero, the integral of \autoref{eq:marg_like_integral} is not tractable in closed form.
For the HERA data used in this work, we specifically use band powers that are widely separated in $k$ such that their error correlations are negligibly small;
concretely we use only every second $k$-bin (``decimation'').
\referee{Quantitatively, Figure 20 of \citet{HERA2021} shows an example of the normalized covariance between $k$ bins, demonstrating that after decimation the remaining modes have negligible covariance, on the order of 1-2\%.}
In decimating, we could in principle choose either the even or odd $k$-bins from each band,
and as each of the four choices would be a slight underestimation of the constraints,
we choose the combination providing the strongest limits. This includes
$k=0.17\,\mathrm{cMpc}^{-1}$ in Band 1 and $k=0.13\,\mathrm{cMpc}^{-1}$ in Band 2;
as a matter of convention we refer to the former as even and the latter as odd $k$-bins.
As we will show later (see Fig.~\ref{fig:example_slices} and \ref{fig:extraradiobackground_appendix_emulator_error}), the constraints on realistic models are primarily driven by the
two most stringent limits,  so we can expect the decimation to have a negligible effect.

\subsubsection{``Inverse" Likelihood}
\label{sec:inverse_likelihood}

In practice, given that the upper limits presented in \citetalias{HERA2021} are still roughly two orders of magnitude above fiducial 21\,cm models, the majority of the parameter space for standard models is left unconstrained. One way to illustrate how the new limits help is by combining them with existing constraints. Alternatively, to provide a clearer picture of the model parameter choices that \emph{exceed} the HERA limits we also consider an ``inverse likelihood'' defined as 
\begin{align}
  \label{eq:comp_likelihood}
  \mathcal{L}^{\rm inv}(\bd|\btheta) \equiv 1 - \frac{\mathcal{L}_m (\bd|\btheta)}{L_0}
\end{align}
where $L_0$ is the maximum of $\mathcal{L}_m$.\footnote{This typically is the likelihood of a model power spectrum equal to zero. Including $L_0$ here makes sure $\mathcal{L}^{\rm inv}$ is independent of the normalization of $\mathcal{L}_m$, e.g. of the number of data points used.}  With the inverse likelihood, the resulting marginalized distributions identify the parameter combinations that can be ruled out by the HERA limits alone (see Fig.~\ref{fig:reverse_posterior} for an example). However, these distributions must be treated with caution: models that lie inside of the projections of the full distribution are not necessarily excluded. The inverse likelihood should only be used to gain intuition about the utility of the HERA limits
and parameters that are necessary (but not sufficient) to drive a power spectrum beyond the HERA limits.


\section{Building physical intuition: A density-driven bias approach}
\label{sec:bias}

Before studying galaxy-driven models, we begin with a simple bias analysis, which will allow us to build intuition about the implications of the HERA measurements. As these limits are well above predictions of ``vanilla" models of the reionization era, the most important parameter that can be constrained is the IGM temperature, as the temperature ratio term in equation~(\ref{eq:21cm_signal}) can become arbitrarily large for gas that is very cold. In the spirit of simplicity, throughout this section we will assume efficient Wouthuysen-Field coupling ($T_K=T_S$), sourced by a non-ionizing ultraviolet background from early star formation. We then infer constraints from the most stringent HERA $k$-bin at each redshift, which we will interpret in terms of changes to the gas kinetic temperature (i.e., we will set $T_{\rm rad}=T_{\rm CMB}$).

\begin{figure}
		\includegraphics[width=0.48\textwidth]{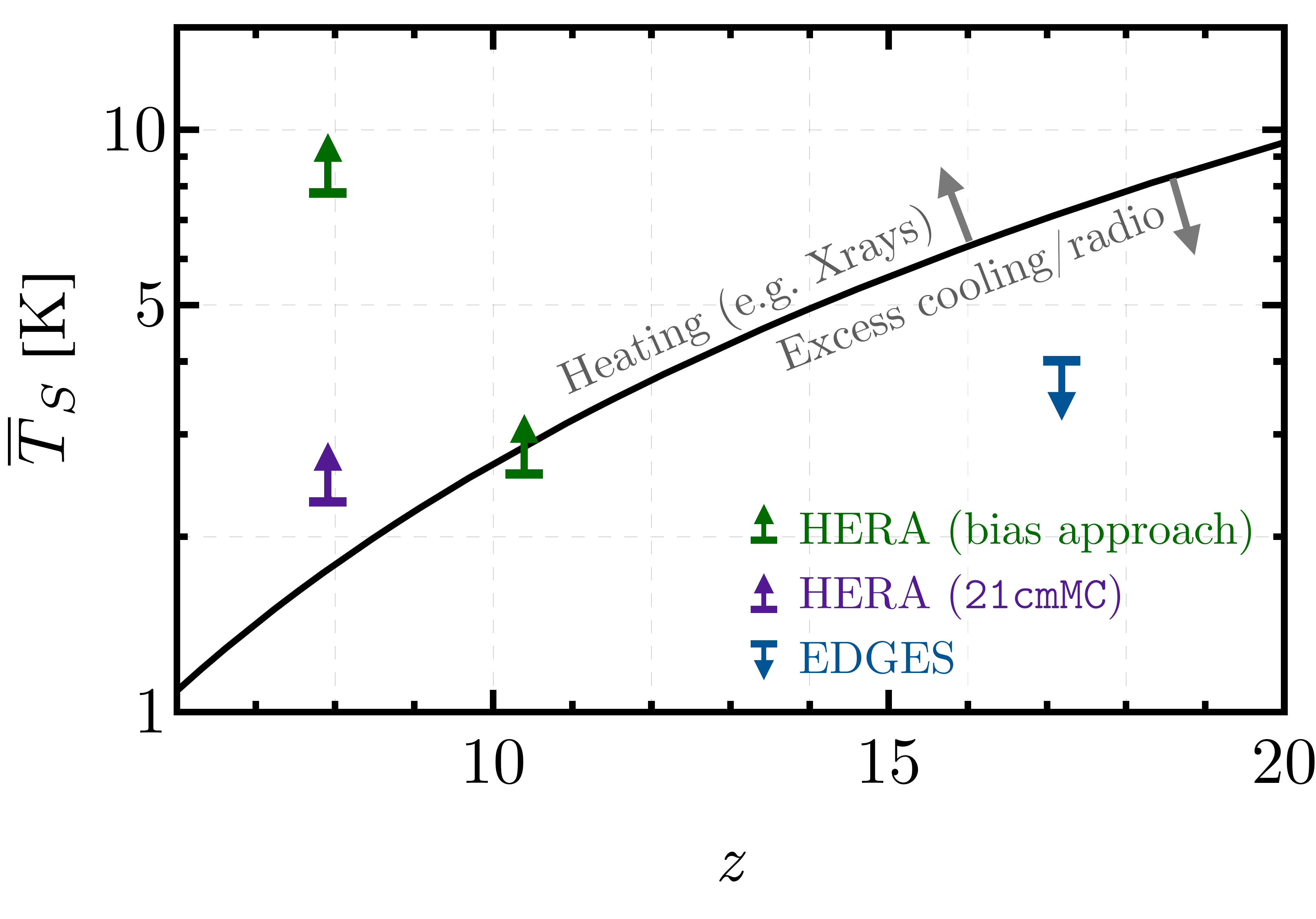}
    \caption{\label{fig:FirstLimits}	Lower
	limits on $\overline{T}_S$ from HERA data (95\% C.L., green and purple arrows) compared to the upper limit from EDGES (blue arrow).	
	The green HERA limits have been obtained by assuming that the IGM is fully neutral and at a constant temperature (aside from small fluctuations due to adiabatic expansion).
	The purple HERA limit is from full galaxy models at $z=7.9$ with {\tt 21cmMC} (see section~\ref{sec:21cmmc}); note that by construction the {\tt 21cmMC} models cannot cool below the adiabatic prediction, so we do not show a 21cmMC limit at $z=10.4$.
	Both assume spherically symmetric RSDs.
	The black line shows the standard-model prediction assuming full Wouthuysen-Field coupling ($T_S=T_K$) without any heating. The HERA Band 2 data can rule out adiabatic cooling in both approaches, requiring some heating to take place before $z=7.9$. 
	}
\end{figure}

Let us begin by supposing that the 21-cm power spectrum traces the matter power spectrum,
\begin{equation} \label{eq:d21-analytic}
    \Delta^2_{21}(k,z) = b_m^2(z) \Delta^2_m(k,z),
\end{equation}
which is appropriate when the fluctuations are sourced by the matter fluctuations. The key assumption here is that the bias parameter is scale-independent -- which is exact for ionization and temperature that vary linearly with density, and can be extended beyond this approximation~\citep{McQuinn05}. We then use the HERA measurements to constrain the bias parameter $b_m$. 
We compute the linear matter power spectrum from {\tt CAMB}\footnote{\url{https://camb.info/}}~\citep{Lewis:2002ah} at each $z$, and we find the 95\% CL limits of 
$b_m < \{156$, $529\}$ mK for $z=\{7.9,\ 10.4\}$ (an analogous analysis for the relative-velocity power spectrum can be found in Appendix~\ref{sec:vao}).
These can be translated into lower limits on the ratio of the spin-to-radio temperatures through the relation
\begin{equation}
    b_m  = T_0 x_{\rm HI} \left\{(1+\mu) - \dfrac{T_{\rm rad}}{\aveTs}[(1+\mu)-C_T]\right\}
    \label{eq:biasm}
\end{equation}
derived from equation~\eqref{eq:21cm_signal} (see e.g.,~\citealt{Pritchard:2008da}),
where $C_T$ is the adiabatic index, which accounts for the preferential cooling of underdense regions; and $\mu$ is the line-of-sight cosine of the wavenumbers observed, which accounts for RSDs.
Here, and throughout this text, an overbar represents average over volume.
We obtain the adiabatic index $C_T= \delta T_K/(T_K \delta) \approx 0.6$ as a function of $z$ following~\cite{Munoz:2015eqa}, which we correct for kinetic temperatures above the adiabatic threshold ($T_K > T_K^{\rm ad}$) by writing $C_T (T_K) \to C_T \times {\rm min} (1, T_K^{\rm ad}/T_K)$, assuming homogeneous heating.
We further assume negligible ionizations, setting $x_{\rm HI}=1$, and for RSDs we take spherically averaged modes ($\mu=0.6$) through this section to match the common procedure done in simulations.
We will show how the constraints shift when altering these two assumptions later in section~\ref{sec:exotic_DM}.

Under our assumptions, the $b_m$ upper limits translate into lower limits for the spin temperature of
\begin{equation}
 \aveTs \geq \{7.8,1.9\}\,{\rm K   \quad for} \quad z=\{7.9,10.4\}
 \label{eq:templims}
\end{equation}
at 95\% confidence, where we re-emphasize we have assumed $x_{\rm HI}=1$.
We show these limits in Fig.~\ref{fig:FirstLimits}, along with the adiabatic-cooling prediction in the standard CDM model.
These $T_{S}$ values have interesting implications for the thermal state of the IGM at high redshifts.
As is clear from Figure~\ref{fig:FirstLimits}, the HERA Band 2 ($z=7.9$) 95\% confidence limit is above the adiabatic-cooling prediction, which demands that some heating must have occurred before $z=7.9$.
Moreover, the HERA limits for Band 1 ($z=10.4$), while below the adiabatic limit at that $z$, can be used to clarify the state of the IGM in comparison with the claimed EDGES detection (also shown in Fig.~\ref{fig:FirstLimits}), which we will explore in section~\ref{sec:exotic_DM}.

We emphasize that these limits rest on three strong assumptions, which we will highlight here and will, in the following sections, explore with more physics-rich models. 
First, the limits assume full Wouthuysen-Field coupling ($T_S=T_K$), which is all but guaranteed by the redshifts we consider ($z\leq 10.4$, see section~\ref{sec:21cmmc}). 
Second, they assume a value of $x_{\rm HI}=1$, which can be varied at each $z$ in equation~\eqref{eq:biasm}, though only homogeneously. 
Lastly, in this analysis we have performed a spherical average of RSDs, whereas HERA data
mostly contains modes along the line of sight ($\mu\approx1$, see section~\ref{sec:21cm-overview}). Properly accounting for RSDs can result in stronger limits, as we will show in section~\ref{sec:exotic_DM}.

In summary, the bias approach here outlined is useful for building intuition, although reionization models and observations suggest that the spatial fluctuations in the ionization field (rather than the matter field) should drive the 21-cm signal at $z\sim8$.
We will explore such models in detail in the following sections, but for now we show the limit from {\tt 21cmMC} in Fig.~\ref{fig:FirstLimits}.  We describe below how this limit was obtained, but we see already that there is general agreement ($\sim$ factor of few) with the density-driven bias limit at $z \approx8$ (indeed our density-driven bias limit is  very close to the analogous density-driven \cmmc\ limit, denoted by the red contours in Figure~\ref{fig:reverse_xT}).
The reason these two approaches yield similar results -- despite their vastly different assumptions about the EoR -- is that the density and ionization power spectra are of the same magnitude at $z\sim 8$ and $k\sim0.1$ Mpc$^{-1}$~(e.g. \citealt{Furlanetto:2004}). Astrophysical models can only modify the peak power during the EoR by a factor of a few (e.g.~\citealt{GM15}). The only way to reach the power spectrum amplitudes probed by the HERA limits is by having a large $\sim (1-T_{\rm rad}/T_S)^2$ pre-factor, i.e., requiring $\aveTs \ll T_{\rm rad}$.  In this regime, model differences can be easily compensated by relatively small changes in $\aveTs$.
Therefore, in the regime of current HERA limits, constraints on $\aveTs$ are of the same magnitude whether the 21-cm power spectrum tracks density or ionization fluctuations.


\section{Galaxy and IGM Properties Inferred From HERA Observations}
\label{sec:21cmmc}	

 We next consider the HERA limits in light of ``standard" galaxy formation models using data-constrained \cmfast\ semi-numerical simulations.

\subsection{Galaxy-driven models of the cosmic 21-cm signal}
\label{sec:gal}

Here we briefly summarize how the 21-cm signal is computed using the galaxy-driven models of \cmfast\footnote{\url{https://github.com/21cmfast/21cmFAST}} \citep{MF07, MFC11, Murray20}.  The main ansatz of these 21-cm models is that {\it cosmic radiation fields are sourced by galaxies, hosted by dark matter halos} (whose relation to the large-scale matter field is comparably well-understood). We generate Eulerian density and velocity fields with second-order Lagrangian perturbation theory (2LPT; e.g. \citealt{Scoccimarro97}). Galaxy properties are then assigned to dark-matter halos via scaling relations with halo mass.  In section \ref{sec:phenom_models}, we explore toy models in which radiation fields are not directly associated with galaxies in order to study the robustness of our inferences and the ``value-added" by explicit models of structure formation.

Specifically, we use the empirical galaxy relations of \citet{Park19}, capable of reproducing the observed UV luminosity functions of galaxies during the EoR ($z=$ 6 -- 10), as well as the spatial distribution of IGM opacities seen in \lya\ forest spectra at $z=$ 5 -- 6 \citep{Qin21}.  Consistent with semi-analytic models and hydrodynamic simulations of high-$z$ galaxies (e.g. \citealt{Moster13, Xu16, SF16, Mutch16, Tacchella18, Behroozi19, Yung19, Ma20}), we describe the mean stellar to halo mass relation, $M_\ast / M_h$,  with a power law:
\begin{equation}
\label{eq:star_halo}
\frac{M_\ast}{M_h} = f_{\ast,10} \left( \frac{M_{\rm h}} {10^{10}{\rm M}_{\sun}} \right)^{\alpha_{\ast}} \left( \frac{\Omega_{\rm b}}{\Omega_{\rm m}}\right) ,
\end{equation}
where $(\Omega_{\rm b}/\Omega_{\rm m})$ is the mean baryon fraction, and the stellar fraction, $f_{\ast} = f_{\ast,10} (M_{\rm h}/ 10^{10}{\rm M}_{\sun})^{\alpha_\ast}$, is restricted to be between 0 and 1.
The corresponding star-formation rate assumes a characteristic star-formation time-scale that scales with the Hubble time, $H^{-1}$ (which during matter domination is equivalent to scaling with the halo free-fall time): $\dot{M}_\ast = M_\ast / (t_\ast H^{-1})$.  Furthermore, we assume only a fraction $f_{\rm duty} = \exp[- M_{\rm turn}/M_h]$ of halos host galaxies; the free parameter $M_{\rm turn}$ encodes the mass scale below which inefficient cooling and/or feedback suppresses efficient star formation (e.g. \citealt{HG97, SH03, OGT08, SM13a, Xu16, Ocvirk18, Ma20}).

We then compute the galactic emissivities (soft UV, ionizing UV and X-ray), assuming they scale with the star formation rates.    
We identify ionized regions with an excursion set approach \citep{FZH04}, comparing the cumulative (local) numbers of emitted photons and recombinations.  We slightly adjust the number of emitted photons to correct for the non-conservation of ionizing photons in excursion set algorithms (e.g. \citealt{Zahn07, PC14}; for details see Park et al. in prep). 
Sub-grid IGM recombinations are tracked according to  \citet{SM14}.  
We assume PopII stellar SEDs for the ionizing and soft UV emission, corresponding to $\sim$5000 ionizing photons produced per stellar baryon (e.g. \citealt{Leitherer99, BL05_WF})\footnote{\referee{Specifically, the PopII SEDs were generated with the Starburst99 code \citep{Leitherer99}, assuming a \citet{Scalo98} IMF and 0.05 solar metallicity.  The spectra in the Lyman bands were interpolated using broken power laws between each Lyman transition according to \citet{BL05_WF}.}}.  A fraction $1 - f_{\rm esc}$ of these photons is absorbed within the galaxy itself, and does not reach the IGM.  We allow the ionizing escape fraction to also scale with the halo mass:
\begin{equation}
\label{eq:f_esc}
f_{\rm esc}(M_{\rm h}) = f_{{\rm esc, 10}}\left( \frac{M_{\rm h}}{10^{10}{\rm M}_{\sun}}\right)^{\alpha_{\rm esc}},
\end{equation}
where $f_{{\rm esc, 10}}$ is the normalization and $\alpha_{\rm esc}$ is a power-law index.  The ionizing escape fraction is also restricted to values between 0 and 1.  Although there is currently no consensus on the ionizing escape fraction or its dependence on galaxy properties, simulations suggest that such a generic power law is an acceptable characterization of the population-averaged values (e.g. \citealt{PKD15, Kimm17, Lewis20}).

In contrast to ionizing UV photons, the soft UV and X-ray photons responsible for coupling the gas and spin temperatures and heating the gas, can have long mean free paths through even the neutral IGM.  We follow the corresponding ionization and heating rates for each simulation cell, by integrating the specific emissivities back along the lightcone, attenuated by the corresponding opacities. \referee{Our simulations track the spatial fluctuations in the X-ray and Lyman series backgrounds, with the IGM opacity computed assuming a standard “picket-fence” absorption for Lyman series photons and absorption from partially-ionized hydrogen and helium in a two-phased IGM for X-ray photons (e.g. \citealt{MFC11, MFS13, Qin20}).}
   The X-ray SED emerging from galaxies is approximated as a power-law whose luminosity scales with the SFR.  This is consistent with theoretical models and observations of local star forming galaxies, for which X-ray emission is dominated by high-mass X-ray binaries (HMXBs) and/or the hot ISM (e.g. \citealt{Fragos12, MGS12_HMXB, Pacucci14, BKP14, Lehmer16}).  Specifically, we parametrize the typical emerging X-ray SED of high-$z$ galaxies via their integrated soft-band ($< 2$~keV) luminosity per SFR (in units of ${\rm erg\,s^{-1}\,M_{\sun}^{-1}\,yr}$),
\begin{equation}\label{eq:soft_X-ray}
L_{\rm X<2\,keV}/{\rm SFR} = \int_{E_0}^{2\,{\rm keV}}{\rm d}E_{\rm e}\, L_{\rm X}/{\rm SFR},
\end{equation}
where $L_{\rm X}/{\rm SFR}$ is the specific X-ray luminosity per unit star formation escaping the host galaxies in units of ${\rm erg\,s^{-1}\,keV^{-1}\,M_{\sun}^{-1}\,yr}$, taken here to be a power law with energy index $\alpha_X$ and $E_0$ is the minimum energy for X-rays to be able to emerge from the galaxy and not be absorbed locally in the ISM.  For reference, the typical value of $E_0\sim0.5$~keV found in the simulations of \citet{Das17} corresponds to an HI column density of $\sim 10^{21.4}$ cm$^{-2}$, assuming zero metallicity.

In summary, our \cmfast\ galaxy models have nine free parameters:
\begin{enumerate}
\item $f_{\ast,10}$, the normalization of the stellar mass--halo mass relation, evaluated at $M_h = 10^{10} \Msun$
\item $\alpha_\star$, the power law index of the stellar mass--halo mass relation
\item $f_{\rm esc, 10}$, the normalization of the ionizing escape fraction--halo mass relation, evaluated at $M_h = 10^{10} \Msun$
\item $\alpha_{\rm esc}$,  the power law index of the ionizing escape fraction -- halo mass relation
\item $M_{\rm turn}$, the characteristic halo mass scale below which the abundance of active galaxies is exponentially suppressed
\item $t_\ast$, the characteristic star formation time scale, expressed in units of the Hubble time
\item $L_{\rm X<2\,keV}/{\rm SFR}$, the soft-band X-ray luminosity per unit SFR
\item $E_0$,  the minimum X-ray energy of photons capable of escaping their host galaxies
\item $\alpha_{\rm X}$, the energy power law index of the X-ray SED
\end{enumerate}

We emphasize that this flexible galaxy parametrization used in \cmfast\ enables us to set physically meaningful priors over the free parameters and use high-$z$ galaxy observations in our inference.  For instance, the common simplification of a constant stellar to halo mass relation is inconsistent with galaxy SFR and LF observations, and can thus bias parameter inference (c.f. \citealt{MFS17}; Fig. 1 in \citealt{Park19}).  Our galaxy model therefore allows us to use existing high-$z$ observations, in addition to HERA, when computing the model likelihood (see section \ref{sec:normal}).
This quantifies the ``added value" of HERA, given that existing observations already exclude a significant prior volume (e.g. \citealt{Park19}). Without them our posterior would strongly depend on our priors.

\subsection{Inference}
\label{sec:inference}

To perform Bayesian inference, we use \cmmc\footnote{\url{https://github.com/21cmfast/21CMMC}} \citep{GM15, GM17_21CMMC, GM18} with the recently implemented Multinest-based \citep{Feroz2009} sampler (\citealt{Qin21_mini2}; see also \citealt{BP19}). For a given sample of astrophysical parameters, we compute 4D realizations of the 21-cm signal in a cubic volume with a periodic boundary condition and a length of 250 cMpc.
The initial conditions and 2LPT are calculated on a $512^3$ grid, while the final radiation fields are computed on a $128^3$ grid.
Choosing a line-of-sight axis, we account for nonlinear RSDs via the real-to-redshift space sub-grid transformation described in \citet{GM18}, and first introduced in \citet{Mao12_RSD, Jensen13RSD}.\footnote{We note, however, that we spherically average the model power spectra before comparing with the data, which does not match the line-of-sight selection performed by the HERA analysis. In regimes dominated by density fluctuations, the HERA mode selection can substantially enhance the power \citep{LaPlante14, Pober15, Jensen16}.  However, as we will see in section 5.4, these regimes are excluded by current observations requiring reionization to be underway at $z\sim8$.  We therefore do not expect our main conclusions in section 5.4 to be impacted by these selection effects; nevertheless, in future analysis, we will compare the forward-modeled power spectra to the data in a like-to-like fashion, using the same mode sampling.}

When evaluating the likelihood according to equation (\ref{eq:likelihood_func}), we add in quadrature a conservative
20\% modeling error (c.f. \citealt{Zahn11}) as well as the sample variance from our simulation. 
In contrast to other simulation-based inference codes,
\cmmc\ forward models 4D realizations of the 21cm signal.  We  compute the power spectra (PS) on-the-fly from these 4D realizations {\it without} emulators, over our 9-dimensional parameter space; we therefore do not include emulator error/bias in our likelihood.

\begin{figure*}
		\includegraphics[width=\textwidth]{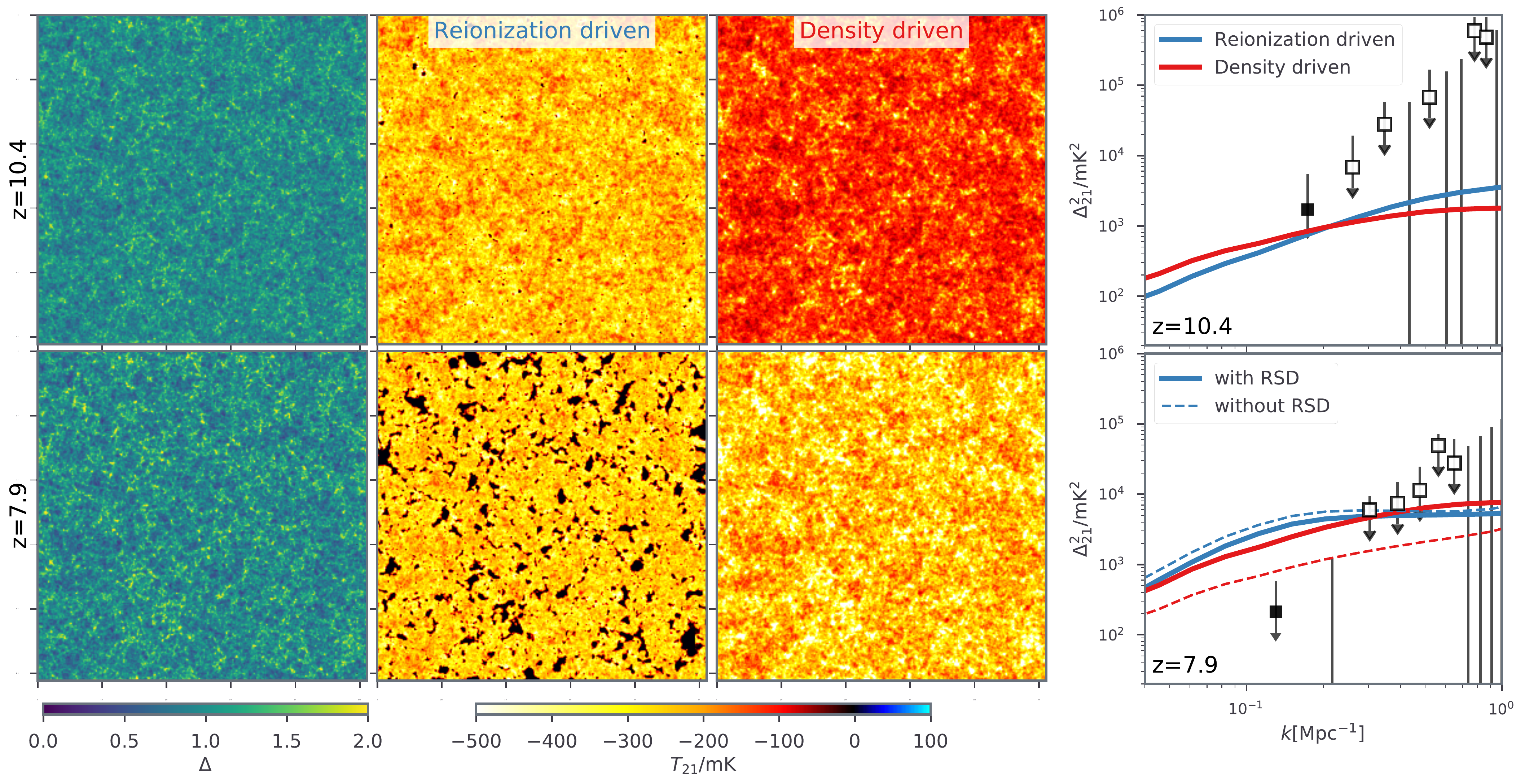}
		\vspace*{-6pt}
		\caption{\label{fig:example_slices}
Two examples of galaxy-driven models that are ruled out by HERA 2021 limits.  The rows correspond to Band 1 ($z=10.4$; {\it top}) and Band 2 ($z=7.9$; {\it bottom}).  The columns correspond to (from left to right): (i) the IGM density; (ii) the brightness temperature of a ``reionization driven" model ($\avenf {=} 0.73$, $\aveTs{=}1.42$ at $z{=}7.9$; black patches are cosmic HII regions); (iii) brightness temperature of a ``density driven" model ($\avenf {=} 0.98$, $\aveTs {=}1.99$ at $z{=}7.9$); and (iv) the corresponding power spectra together with the \citetalias{HERA2021} limits.  Note that the power spectra of the models are much flatter (with $k$) than the observational limits; thus the constraining power is entirely provided by the two filled squares at low $k$.  In the bottom right panel, we also show power spectra ignoring  redshift space distortions; RSDs are important for the density driven models but much less so for the reionization driven models.
The slices are 1 cGpc on a side and 2 cMpc thick and were generated with \cmfast\ v3.
}
\end{figure*}

When we include other observational constraints in our inference procedure (see Section \ref{sec:normal}), we calculate the total likelihood with $\mathcal{L}_{\rm total} = \mathcal{L}_m \times \mathcal{L}_{\rm LFs} \times \mathcal{L}_{\rm DF} \times \mathcal{L}_{\tau_e}$, where the last three terms reflect the comparison between the modeled results against (i) the observed faint galaxy ($M_{\rm UV}>-20$) UV luminosity functions at $z=6$--10 from \citet{Bouwens15, Bouwens16LFs} and \citet{Oesch18}; (ii) the upper limit on the neutral hydrogen fraction at $z{\sim}5.9$ measured by the dark fraction on high-reshift quasar spectra \citep{MMO15}, $x_{\hone}<0.06+0.05(1\sigma)$ where we consider a one-sided Gaussian likelihood function\footnote{\referee{A revision of these dark fraction limits from a larger QSO sample (Campo et al. in prep) as well as inference from the large-scale Lyman forest opacity fluctuations \citep{Qin21}, seem to favor a slightly later end to reionization (a delay of $\Delta z \sim$ 0.5).  Since reionization-driven PS amplitudes are maximized around the midpoint of the EoR, which for current observations occurs right around HERA's Band 2 at $z\sim8$, we expect that shifting the EoR towards later times could slightly weaken the HERA constraints we derive below.}}; and (iii) the Thomson scattering optical depth of CMB photons, using \citet{Planck18} data analysed by \citet{Qin20CMB}, $\tau_e=0.0569^{+0.0081}_{-0.0066}$.

\subsection{Models that exceed the HERA limits}
\label{sec:inverted}

\begin{figure*}
		\includegraphics[width=\textwidth]{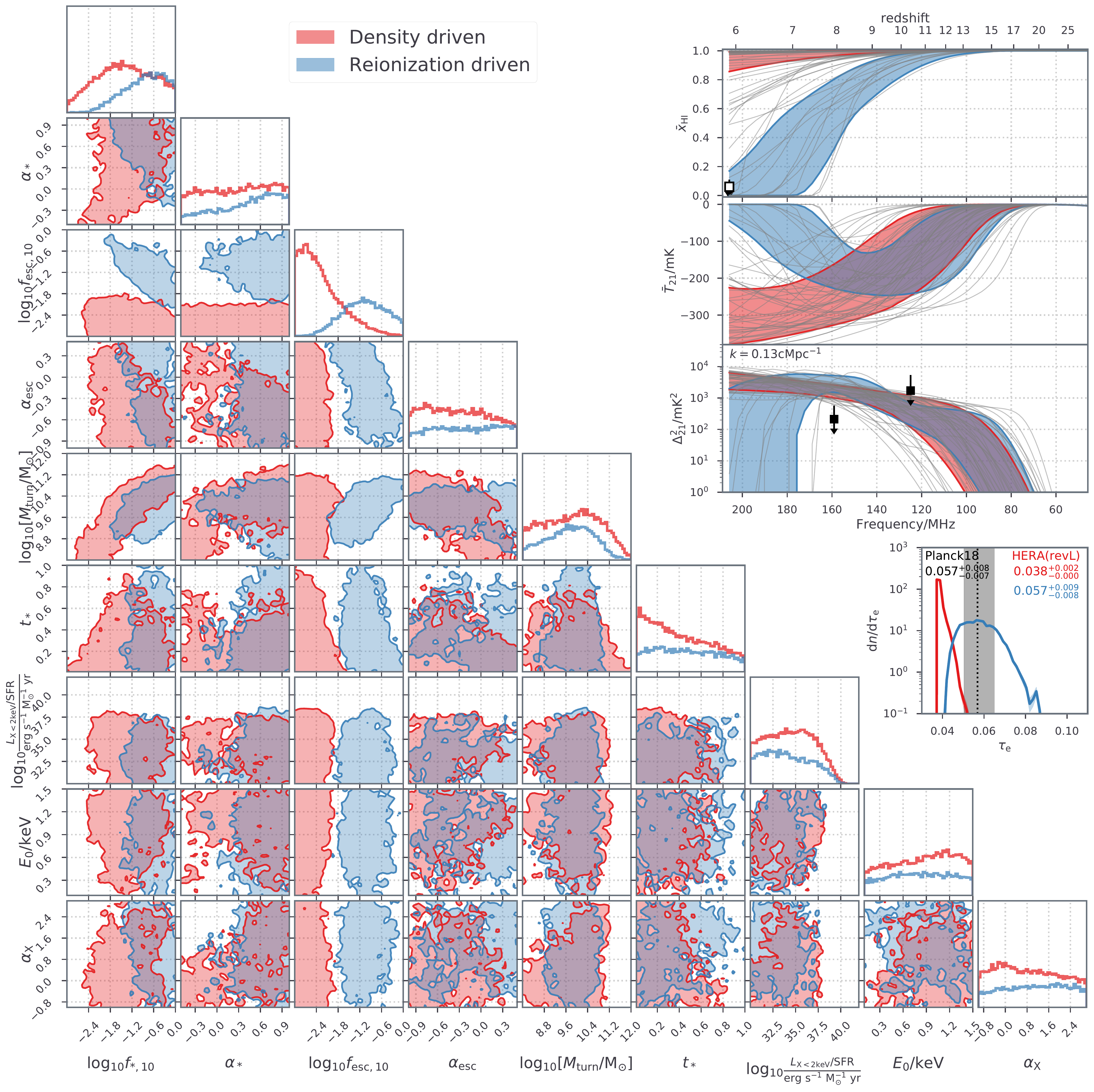}\\
		\vspace*{-6pt}
		\caption{\label{fig:reverse_posterior}
Distribution of models disfavored by  \citetalias{HERA2021}, calculated using the inverse likelihood (eq.~\ref{eq:comp_likelihood}) and using only the two most constraining data points in Bands 1 and 2 (c.f. filled squares in Fig. \ref{fig:example_slices}).  The 1D and 2D marginalized distributions were generated by assuming flat priors over the ranges shown by the figure; we caution that these marginalized inverse likelihood results should not be interpreted strictly as a ``posterior", but instead serve to illustrate where the models disfavored by \citetalias{HERA2021} sit in astrophysical parameter space.  In the bottom left we show the 2D and 1D distributions, while in the top right we show the  EoR history, global signal evolution and power spectrum evolution at $k=0.13$ cMpc$^{-1}$.  Red / blue curves denote ``density driven" / ``reionization driven" models, classified according to the value of the neutral fraction at $z=6$; shaded regions enclose 68\% of the disfavored models for each mode. In the power spectrum evolution plot we also show the two \citetalias{HERA2021} data points used to compute these distributions (note that the Band 1 data point is at a slightly higher wavenumber of $k=0.17$ cMpc$^{-1}$); this highlights that the Band 2 ($z=8$) data point has all of the constraining power.   In the EoR history panel, we also include the QSO dark fraction upper limits from \citet{MMO15} (empty square). In the bottom right, we also include the corresponding PDFs of the CMB optical depth, $\tau_e$, from both modes; the gray region spans 68\% C.L. of the observed value, implied by the galaxy-model recovery of \citet{Planck18} EE power spectra described in \citet{Qin20CMB}.  These two EoR observations were not used in the inverted likelihood; unlike the "density driven" modes, the ``reionization driven" modes are largely consistent with these limits.
}
\end{figure*}

In this section we highlight the astrophysical models  {\it disfavored} by the current HERA limits.  To do this, we use the inverse likelihood from equation~\eqref{eq:comp_likelihood}. Because the inverse likelihood is only illustrative, we also confine the analysis to the two most stringent limits at $z=7.9$ and $z=10.3$. In any case, these two data points provide all the constraining power because the observed limits rise much more steeply with $k$ than the model predictions.
This allows us to compare to similar analysis of recent LOFAR and MWA data, which also used an inverse likelihood and the same galaxy models \citep{Greig21MWA,Greig21LOFAR}. 

Before showing the full distribution of models, in Fig. \ref{fig:example_slices} we show examples of two classes of models capable of exceeding the HERA upper limits.  The top row corresponds to Band 1 ($z=10.4$) and the bottom to Band 2 ($z=7.9$).  Slices through the density field and 21-cm brightness temperature fields are shown on the left, with the 21-cm power spectra shown together with the data in the rightmost panels.
For visualization purposes, the maps are generated from larger boxes than used in the inference, corresponding to 1 cGpc on a side, but with the same 2 cMpc resolution.\footnote{\referee{We confirm that the power spectra in the 1 cGpc and 250 cMpc runs are converged to the percent level or better for the relevant wave-numbers, $k>$ 0.1 cMpc$^{-1}$.  This level of convergence is consistent with the results of \citet{KGM20}, who quantified the bias and scatter in the 21-cm signal resulting from missing large-scale modes (see also \citealt{Iliev14}), and is orders of magnitude smaller than the observational uncertainties.}}

The 21-cm power spectra in the two classes of models exceeding these upper limits are driven by spatial fluctuations in either: (i) the IGM ionized fraction, which we will refer to as ``reionization driven" (also referred to as ``cold reionization" in the literature; e.g. \citealt{ME-WH14}) ; or (ii) the gas density, which we will refer to as ``density driven" (see section~\ref{sec:bias}, and \citealt{Greig21MWA} for the same qualitative result using recent MWA limits).\footnote{\citet{Ghara2020} also consider a model in which highly biased AGN with luminous, soft X-ray SEDs but negligible UV emission dominate the radiation background.  Such extreme scenarios might also produce very strong temperature fluctuations, capable of exceeding the HERA limits; however, such models are not inside our prior volume.} Both scenarios {\it require} a cold IGM, which sets a lower limit on the heating rate (and hence on the X-ray emissivity within these models).

In the right panels, we confirm that the 21-cm power spectra of both scenarios are flatter than the observational limits.  Thus when the observational limits are consistent with thermal noise, the constraining power comes entirely from the deepest limits (c.f. \citealt{Mertens20, Trott20, Ghara21}) (in our case, primarily the deepest data point at $z=7.9$).

In the bottom right panel, we also show how the power spectra depend on RSDs.  For the ``reionization driven" model, RSDs are not important since the first HII regions are highly biased, zeroing out the signal from the densest regions with the strongest RSDs (\citealt{MFC11, Jensen13, GCD15, Ross20}).  However, the ``density driven" models have a negligible contribution from ionization and heating, with the 21-cm power spectrum driven entirely by the non-linear matter field.  By comparing the solid and dashed red curves, we see that non-linear RSDs can boost the spherically averaged
power by factors of $\sim$ 2--3, in excess of the linear prediction of $1.87$ (e.g. \citealt{BA04, BL05}).  Indeed without RSDs, this density driven model is consistent with the data at $\sim$ 1 $\sigma$. We explore the effect of different RSD assumptions for density-driven models in section~\ref{sec:adiabatic}.

In Figure \ref{fig:reverse_posterior} we show a corner plot corresponding to the inverted likelihood from equation~(\ref{eq:comp_likelihood}).  We caution that our parameter ranges in this figure/subsection do not correspond to a ``prior" belief of the distribution of disfavored models, and marginalizing over an inverse likelihood is different from an inversion of the 2D marginalized Bayesian posteriors.  Therefore Figure \ref{fig:reverse_posterior} should not be interpreted as a Bayesian posterior of disfavored models, and it is difficult to formally relate it to the normal likelihood results in the next section. However, the figure illustrates where the models that exceed HERA reside in our parameter space.  In the top right, we draw from these distributions the redshift evolution of the mean neutral fraction, the mean 21-cm signal, and 21-cm power spectrum at $k=0.13$ cMpc$^{-1}$.

Here we highlight the two modes discussed above: red and blue curves denote the ``density driven" and ``reionization driven" models,  classified on the basis of whether the Universe is mostly neutral or mostly ionized at $z=6$.\footnote{There is a clear bimodality in the $z\sim6$ neutral fraction of disfavored models (c.f. top right panel of Fig. \ref{fig:reverse_posterior}), allowing us to easily distinguish the ``density driven" and ``reionization driven" modes.  Models with intermediate values  $\avenf (z=6)\sim 0.5$, would generally have $\avenf(z=8) \sim$ 0.1 - 0.2.  In this early EoR regime, the negative contribution of the ionization-density cross-correlation can result in a decrease of large-scale 21-cm power (e.g. \citealt{Lidz08, Zahn11}), making it difficult for those models to exceed the HERA limits.  Thus the highest power is achieved when only one variable is dominating the fluctuations and the cross terms can be ignored.} The shaded regions enclose 68\% of the distributions.  Astrophysically, the two modes are most easily distinguished by the ionizing escape fraction parameter, $f_{\rm esc, 10}$, and to a lesser degree by their star formation efficiencies, here parametrized by the ratio $f_{\ast, 10}$/$t_\ast$.   All of the models require that the IGM was not heated significantly, as seen by the upper limits on the X-ray luminosity per SFR, $L_{\rm X, <2 kev}$/SFR.

The upper right panels show that the ``density driven" models are already ruled out by other observations, since they fail to reionize the Universe early enough.  In particular, we show the observed upper limit on the neutral fraction from the dark pixels in the Lyman forests \citep{MMO15}, as well as the Compton scattering optical depth from Planck 2018 (\citealt{Qin20CMB}).  Note that these observations were not used in computing the inverse likelihood.  However, some of the ``reionization-driven" models are consistent with current observations.  We return to this in the next subsection.

\begin{figure*}
		\includegraphics[width=0.5\textwidth]{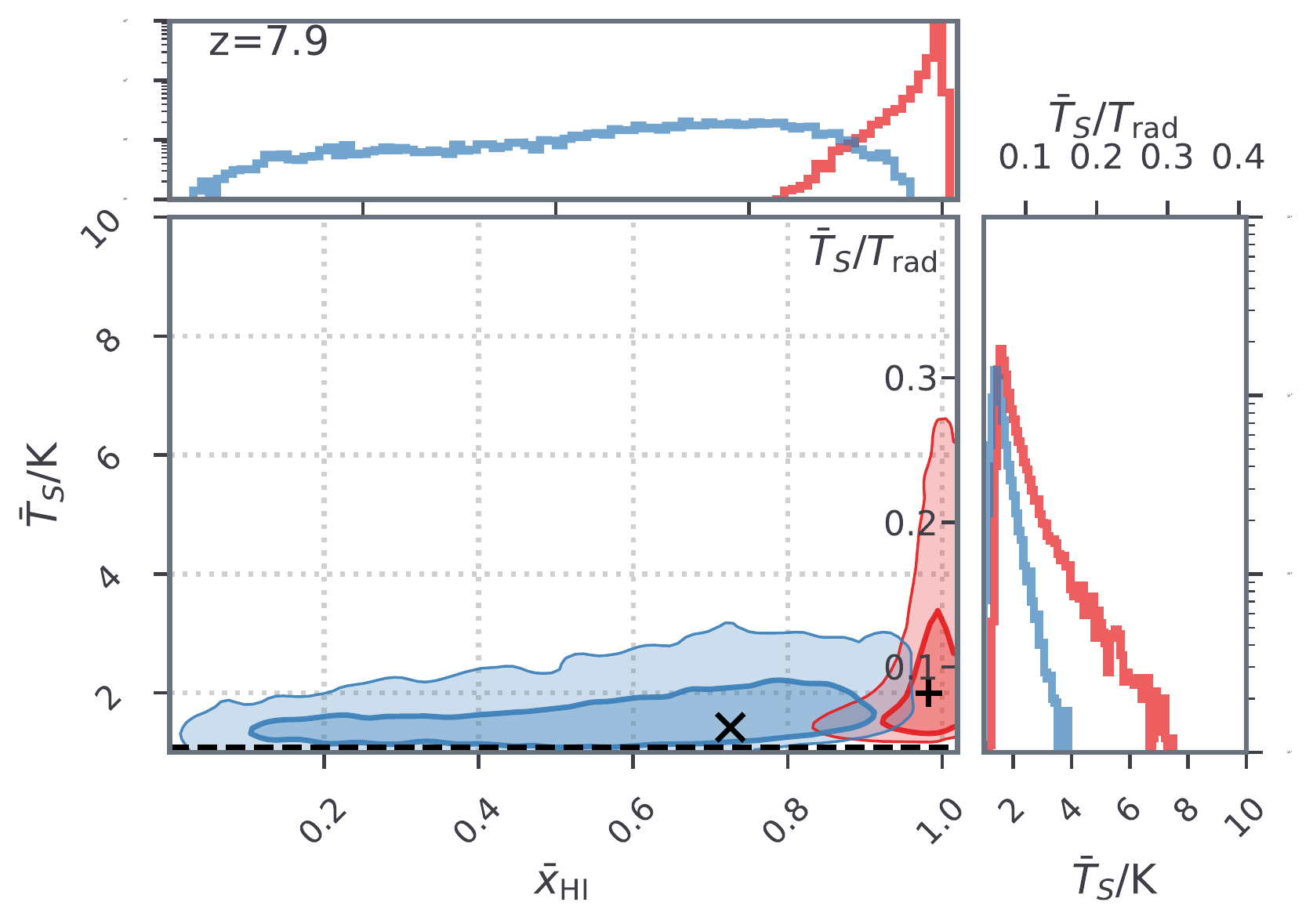}
		\includegraphics[width=0.5\textwidth]{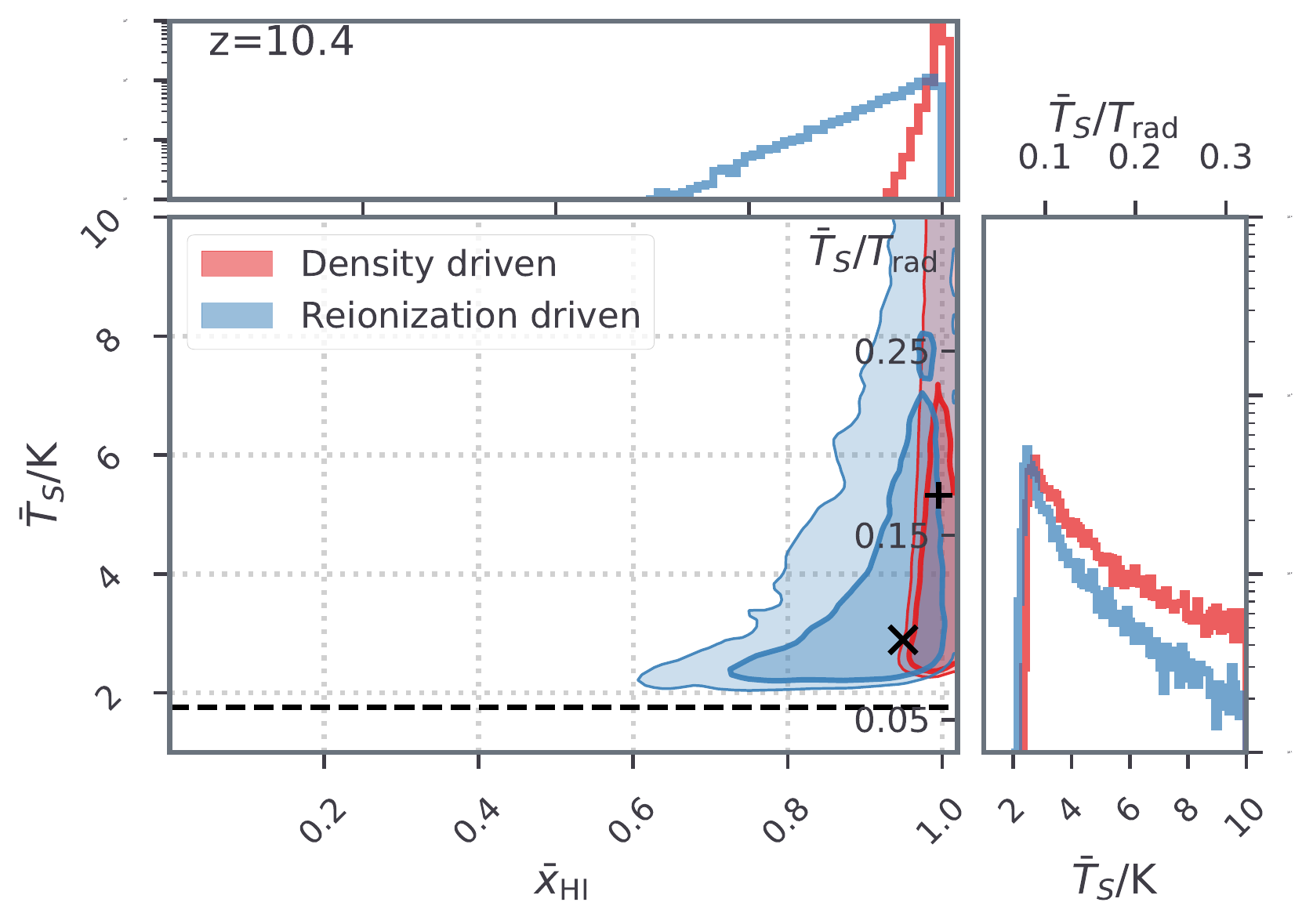}
		\vspace*{-6pt}
		\caption{\label{fig:reverse_xT}
Distributions of $\avenf$ and $\aveTs$ corresponding to the two classes of HERA-disfavored models from Figure \ref{fig:reverse_posterior}, obtained using the inverse likelihood.  The left (right) panel corresponds to $z=7.9$ (10.3). 
The adiabatic cooling limit is shown with the dashed horizontal line, and the two example models from figure 1 are denoted with 'x' and '$+$'.  The left axis corresponds to the usual assumption of the radio background being dominated by the CMB, while the right axis translates these values into the more generic ratio, $\aveTs / T_{\rm rad}$, valid for any homogeneous radio background. For $\aveTs$ we perform averaging only over the neutral IGM (specifically those cells with $x_{\rm HI}>0.95$).}
\end{figure*}

In Figure~\ref{fig:reverse_xT}, we show where these HERA-disfavored models sit in the marginalized 2D space of $\avenf$ vs $\aveTs$.\footnote{Note that the values of $\aveTs$ we quote throughout section \ref{sec:21cmmc} are averaged only over the neutral IGM component ($T_S$ is undefined for ionized gas).  Because in the standard picture reionization is approximately ``inside out" on large scales, averaging over the neutral IGM means that the $\aveTs$ limits are slightly biased towards underdense volumes.} The left and right panels correspond to $z=7.9$ and 10.4.  The two modes discussed above are clearly seen to emerge by $z=8$; at present, the lower-redshift data provide most of the constraining power.

At $z\approx8$ we see that HERA-disfavored models have low spin temperatures: $\aveTs \lsim 3$ K (or more generically for any uniform radio background $\aveTs/\bar{T}_{\rm radio} \lsim 0.1$ for $0.1 \lsim \avenf \lsim 0.9$.
These constraints are somewhat tighter than analogous ones based on recent LOFAR \citep{Mertens20} and MWA \citep{Trott20} upper limits: $\aveTs \lsim 2$--2.5 K, over narrower ranges in $\avenf$ (c.f. Fig. 4 in \citealt{Greig21LOFAR} and Fig. 6 in \citealt{Greig21MWA}).  Thus, as expected from the stronger PS upper limits, the \citetalias{HERA2021} limits rule out more models than previous power spectrum limits. Furthermore, the density-driven modes were not ruled out by the previous LOFAR limits, which had a larger amplitude and were performed at a higher redshift ($z=9.1$; at which the adiabatic-cooling temperature is larger by a factor of $(1+9.1)^2/(1+7.9)^2$).

At $\avenf\gsim 0.9$, the range of temperatures for the disfavored models broadens.  This is due to the negative contribution of the ionization-density cross power term, that dominates the large-scale 21-cm power in this regime \citep{Lidz08, Zahn11}.  The first galaxies drive HII regions that are very biased in the early stages of the EoR.  These quickly cover up the largest matter overdensities, which had earlier dominated the 21-cm power spectrum.  Thus for models with negligible temperature fluctuations, the large-scale power drops in the early stages of the EoR before rising again as it transitions from being sourced by the matter fluctuations to ionization fluctuations.

\subsection{How do the HERA limits improve upon previous complementary data?}
\label{sec:normal}

\begin{figure*}
		\includegraphics[width=\textwidth]{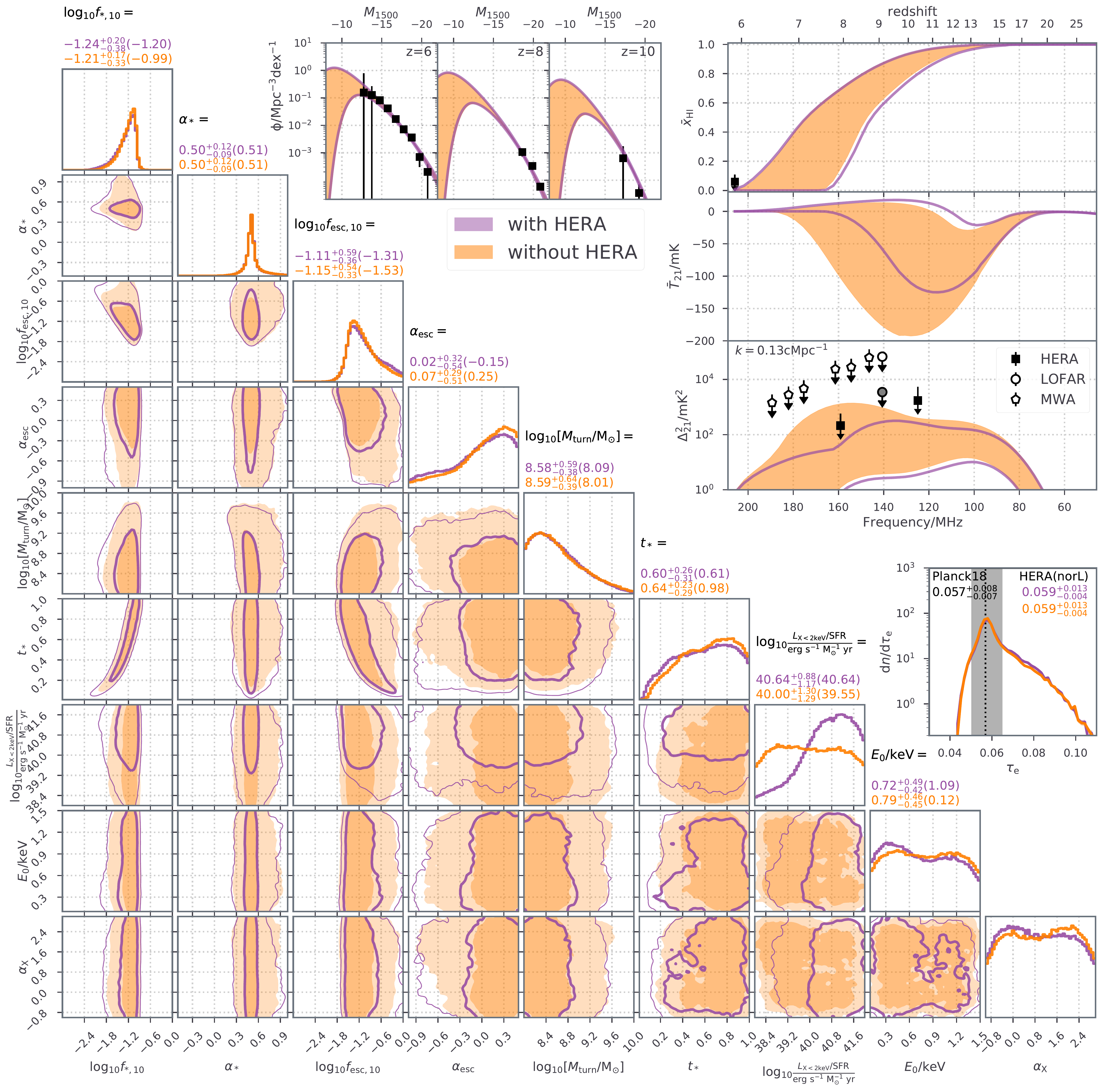}\\
		\vspace*{-6pt}
		\caption{
		\label{fig:posterior} Posteriors with and without the \citetalias{HERA2021} limits.  The 1D and 2D marginalized posteriors are shown in the bottom left, while the corresponding UV LFs, EoR histories, global 21-cm signal, evolution of the power spectrum at $k=0.13$ cMpc$^{-1}$, and the CMB optical depth are shown in the top right ({\it clockwise from the top middle}).  The {\it without HERA} posterior ({\it tan}) is computed using previous observations: (i) galaxy UV LFs from $z=6-10$ ({\it filled squares in LF panels}); (ii) upper limit on $\bar{x}_{\hone}$ from the QSO dark fraction ({\it filled square at $z=5.9$ in the EoR history panel}); and (iii) CMB optical depth from {\it Planck} ({\it shaded region in the $\tau_e$ panel}).  The {\it with HERA} posterior ({\it purple}) uses the HERA limits from Fig. 1 in addition to (i) -- (iii).  Although we use all data points in the HERA likelihood, we show the two deepest limits from Band 1 ($k=0.17$ cMpc$^{-1}$) and Band 2 ($k=0.13$ cMpc$^{-1}$) in the PS evolution inset panel.  Here for comparison we also show the recent 1$\sigma$ limits at $k=0.1$ cMpc$^{-1}$ from MWA (pentagons; \citealt{Trott20}) and  $k\approx0.1$ (0.05) cMpc$^{-1}$ from LOFAR (upper/lower circle; \citealt{Mertens20}); the MWA and LOFAR limits are not included in the likelihood.  We assume flat priors over the astrophysical parameter ranges shown in the subpanel axes.  This figure illustrates two important points: (i) current observations already exclude a large majority of our prior volume; (ii) HERA limits constrain the X-ray luminosities of the first galaxies.
		}
\end{figure*}

As already mentioned, many of the models that are disfavored by the current HERA limits are already inconsistent with existing observations of the $z>6$ Universe.  Here we put the HERA constraints in context with these other observations by computing the Bayesian posterior over our parameter space with and without the new HERA limits.  In particular, we run two inferences (see also section \ref{sec:inference}):
\begin{itemize}

\item {\it without HERA}: This run corresponds roughly to our current state of knowledge, without including 21-cm observations.\footnote{Here we restrict ourselves to arguably the most model-independent EoR constraints.  In the future, as the 21-cm data improves, we will fold in additional constraints from Lyman-$\alpha$ emitting galaxies, QSO damping wing analysis, opacity fluctuations in the Lyman forests, the patchy kinetic Sunyaev-Zeldovich effect (e.g. \citealt{Stark10, Schenker12, Pentericci14, Mason17, becker15, bosman18, Banados18, Wang21, Reichardt21}).  These require more subtle modeling of associated systematics, but could have a non-negligible impact on the recovered EoR history (e.g. \citealt{GM17, Dai19, Qin21, CMP21}).
We also do not include previous 21-cm upper limits from MWA and LOFAR since these weaker PS limits would not change our {\it with HERA} posterior (see the PS evolution inset in Fig. \ref{fig:posterior}).  Thus by comparing {\it without HERA} and {\it with HERA}, we highlight the impact of 21-cm measurements.
}  As detailed in section~\ref{sec:inference}, the likelihood incorporates observations of (i) the galaxy UV luminosity functions at $z{=}6{-}10$; (ii) the upper limit on $\avenf$ at $z\sim5.9$, inferred from the Lyman forest dark fraction; and (iii) the CMB optical depth  $\tau_e$.

\item {\it with HERA}: Here the likelihood is computed using both the complementary observations in {\it without HERA} above, as well as the HERA limits from Bands 1 and 2.  Specifically, we use the (regular) HERA likelihood as defined in equation~(\ref{eq:likelihood_func}).

\end{itemize}

\begin{figure}
		\includegraphics[width=0.5\textwidth]{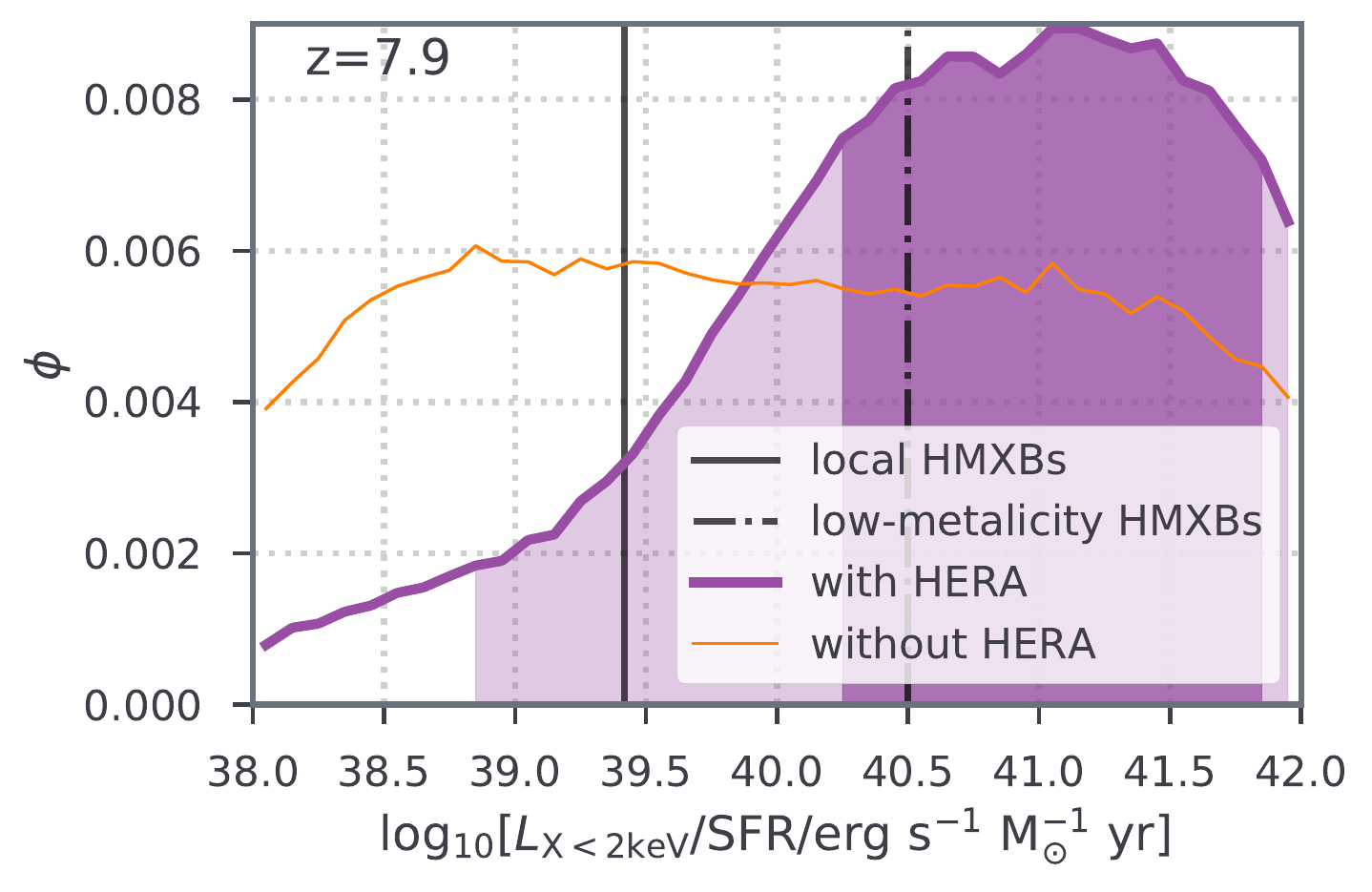}
		\vspace*{-6pt}
		\caption{\label{fig:Lx}
		The marginalized 1D PDFs of the soft band X-ray luminosity to SFR, $L_{\rm X < 2keV}$/SFR, from the {\it with HERA} and {\it without HERA} posteriors.  The highest posterior density (HPD) 68\% (95\%) confidence intervals are denoted under the {\it with HERA} posterior with dark (light) shading.
		The left vertical line denotes the average value of this quantity observed from HMXBs in the sample of local, star-forming galaxies from \citet{MGS12_HMXB}.  The right vertical line corresponds to the theoretical result from \citet{Fragos12} for a metal-free HMXB population, expected to be more representative of the first galaxies.
	    HERA is the first observation to constrain the X-ray luminosities of Cosmic Dawn galaxies over this range, disfavoring the values seen in local, metal-enriched galaxies at $>1 \sigma$.
}
\end{figure}

\subsubsection{Galaxy properties: disfavoring X-ray faint galaxies with HERA}

The corner plot of these two posteriors is shown in Figure~\ref{fig:posterior}, with tan (purple) denoting {\it without HERA} ({\it with HERA}).  As discussed in detail in \citet{Park19}, we see that current observations ({\it without HERA}) already rule out a significant fraction of our prior volume, which highlights the power of our \cmmc\ approach's inclusion of complementary galaxy observations.  Observations of high-$z$ UV luminosity functions shown in the top-middle sub-panels of Figure~\ref{fig:posterior} constrain the stellar-to-halo mass relation and its scaling with halo mass ($f_{\ast, 10}$ and $\alpha_\ast$), as well as place an upper limit on the characteristic turnover scale ($M_{\rm turn}$).  On the other hand, observations of the EoR timing through the CMB optical depth (c.f. bottom-right sub-panel) and the Lyman forest dark fraction (c.f. upper limit in the EoR history sub-panel) constrain the ionizing escape fraction normalization ($f_{\rm esc, 10}$) to within 1 dex and place very weak constraints on its evolution with halo mass ($\alpha_{\rm esc}$).  Using such complementary observations in the likelihood is especially important when sampling from a high dimensional parameter space with flat priors, for which most of the prior volume is sourced by ``extreme" corners of parameter space that are already ruled out by existing observations (as is immediately evident from Fig. \ref{fig:posterior}).

Comparing the {\it without HERA} and {\it with HERA} posteriors, we see that the \citetalias{HERA2021} limits do not have a notable impact over most of the astrophysical parameter space.  The new models that HERA rules out, discussed in the previous section, occupy a modest prior volume.\footnote{We use a narrower prior range on $L_{\rm X < 2keV}$/SFR and $M_{\rm turn}$ in Fig. \ref{fig:posterior} compared to Fig. \ref{fig:reverse_posterior}.  This is because Fig. \ref{fig:posterior} is a true posterior requiring physically reasonable prior ranges, which we discuss further below when presenting galaxy inference.  In contrast, Fig. \ref{fig:reverse_posterior} is only meant to illustrate where HERA-disfavored models are expected to reside in our parameter space.}

However, note that the three X-ray parameters ($L_{\rm X <2keV}$/SFR, $E_0$, $\alpha_{\rm X}$) are largely unconstrained by the complementary observations over our prior ranges, because none of the \emph{without HERA} observations are sensitive to the IGM temperature, the observable most strongly affected by the X-ray emissivity. In this part of parameter space, HERA does have a notable impact by ruling out models with weak X-ray heating,
which in our parametrization is predominately determined by the integrated soft-band X-ray luminosity to SFR, $L_{\rm X <2keV}$/SFR.
The exclusion of these models is also evident in the 21-cm panels at the upper right, where the recovered signal ranges decrease significantly when including HERA data.

We show a zoom-in of the marginalized 1D PDFs of $L_{\rm X <2keV}$/SFR in Figure \ref{fig:Lx}.
The marginalized {\it without HERA} posterior is consistent with the flat prior over the range shown.  Current observations do not constrain this quantity aside from disfavoring extreme values of $L_{\rm X < 2keV}$/SFR $\gsim 10^{42}$ erg s$^{-1}$ $M_\odot^{-1}$~yr,  which is so large that X-rays can significantly contribute to reionization (e.g. \citealt{MFS13}), making it too early in many models.  However, the {\it with HERA} posterior is able to rule out the lower end of this range, resulting in a 68\% highest posterior density (HPD) confidence interval of $L_{\rm X <2keV}$/SFR = $\{ 10^{40.2}, 10^{41.9} \}$ erg s$^{-1}$ $M_\odot^{-1}$ yr.  \citetalias{HERA2021} is the first observation to place constraints over this range; the analogous analysis of MWA and LOFAR observations (c.f. Fig.~1 in \citealt{Greig21LOFAR} and Fig.~2 in \citealt{Greig21MWA}) disfavored models with lower luminosities.\footnote{This comparison is only approximate, because the earlier analyses were based on the inverse likelihood rather than the proper  marginalized posterior shown here.}

\begin{figure*}
		\includegraphics[width=0.5\textwidth]{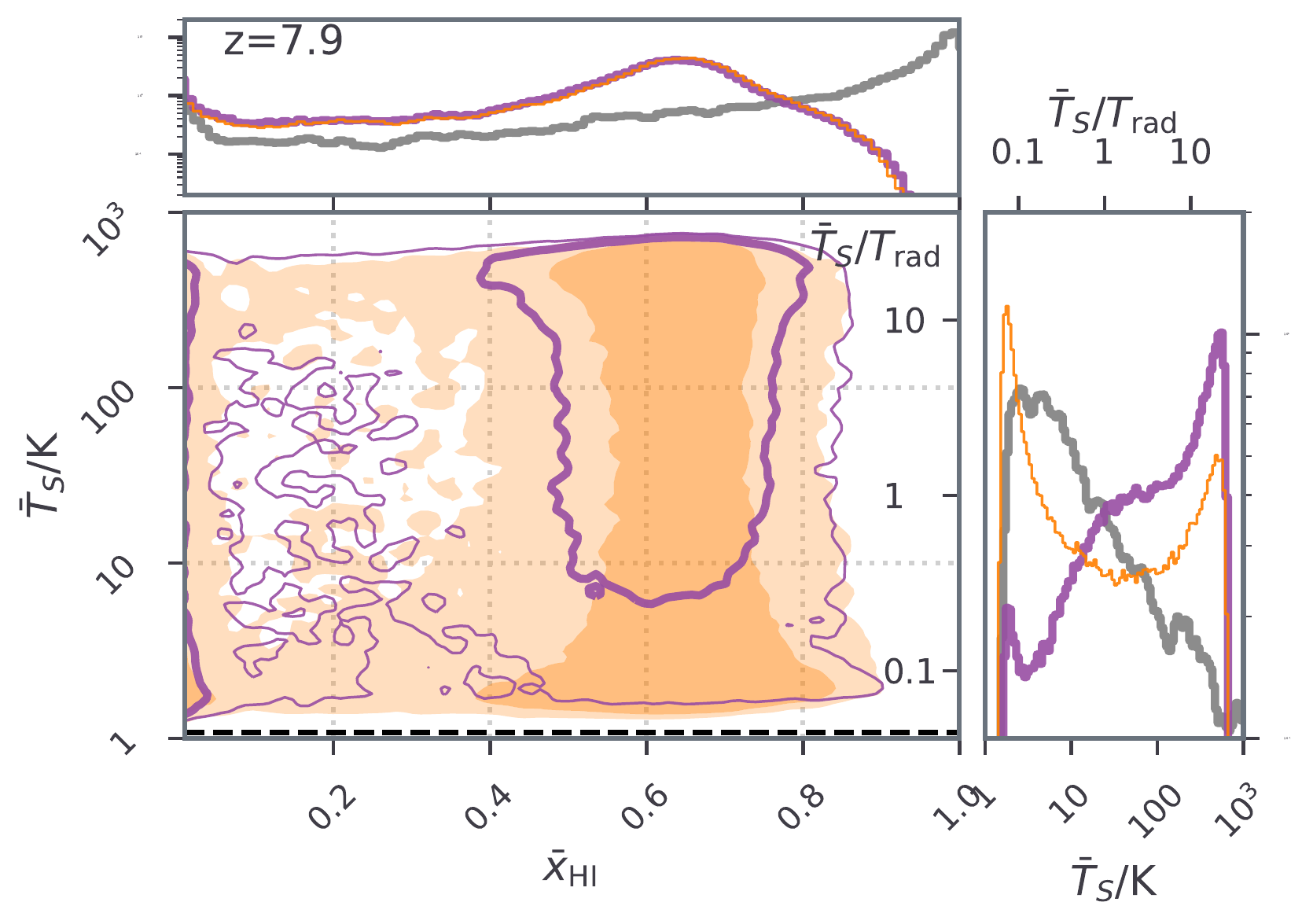}
		\includegraphics[width=0.5\textwidth]{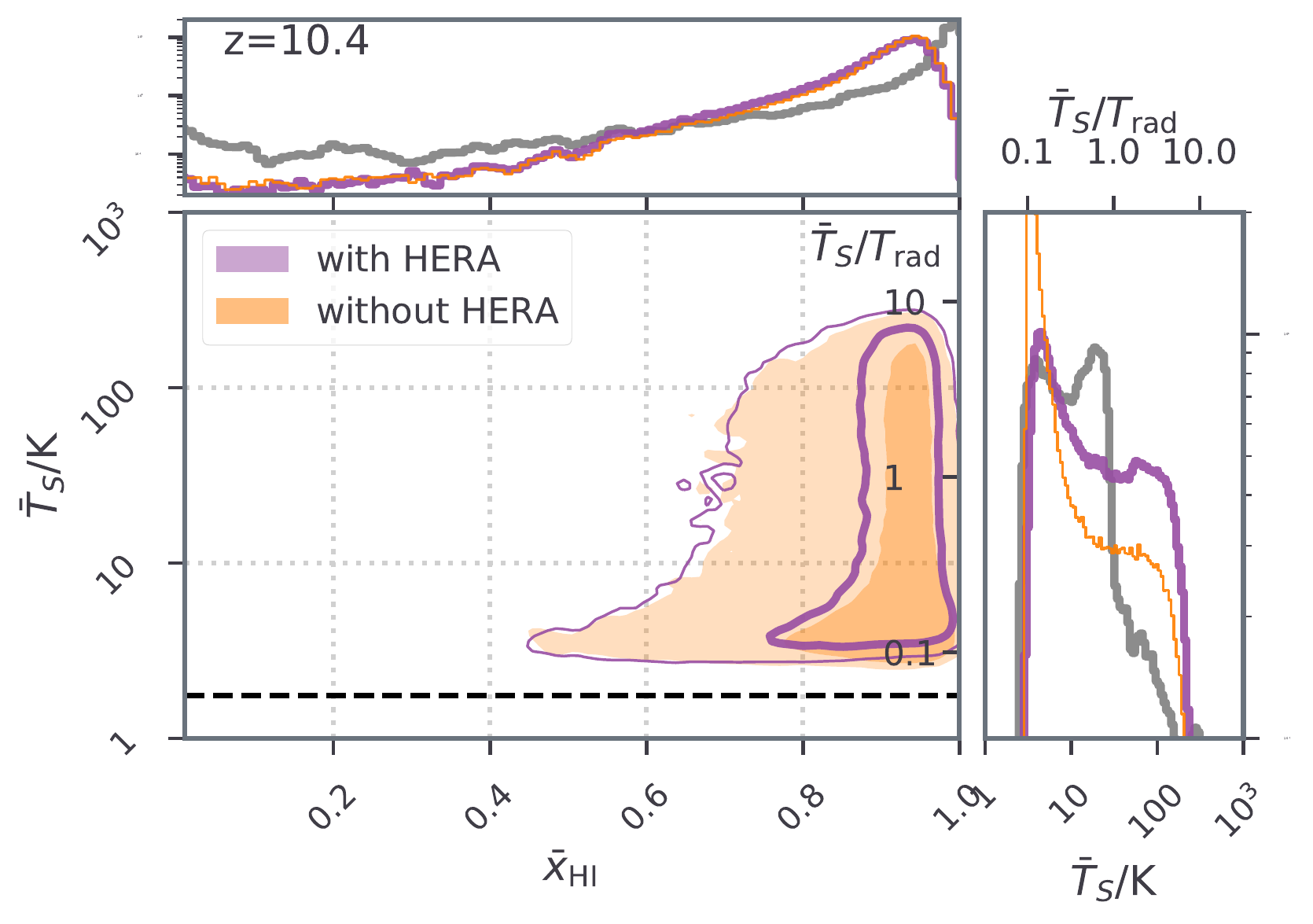}
		\vspace*{-6pt}
		\caption{\label{fig:xT}
Marginalized IGM properties corresponding to the posteriors from Figure \ref{fig:posterior}.  As throughout, $\aveTs$ is computed by averaging only over the neutral IGM.  Note that unlike Fig. \ref{fig:reverse_xT}, these are true Bayesian posteriors, as they were generated using a regular likelihood and marginalizing over physical priors.  The left (right) panel corresponds to $z=7.9$ (10.4).  \citetalias{HERA2021} limits increase the preference for hotter temperatures of the neutral IGM component.  The gray curves shown in the 1D marginalized panels show our prior distribution.  Our galaxy priors do not result in flat priors over $\avenf$ and $\aveTs$.}
\end{figure*}

In Figure \ref{fig:Lx} we also compare the {\it with HERA} limits with estimates based on high-mass X-ray binaries (HMXBs), thought to be the dominant X-ray sources in high-$z$ galaxies (e.g. \citealt{Fragos12}). 	The left vertical line denotes the average value observed from HMXBs in local, metal-enriched, star-forming galaxies (\citealt{MGS12_HMXB}; see also e.g. \citealt{Lehmer10}).  Because the HMXB luminosity increases with decreasing metallicity (e.g. \citealt{Basu-Zych13, Douna15, Brorby16}), we do not expect the first, metal-poor galaxies to sit on the left side of this line.  And indeed, this local scaling relation is outside of the {\it with HERA} 68\% confidence interval; thus HERA data already suggests that the first galaxies were more X-ray luminous than their local counterparts.  In contrast, the right vertical line in Figure \ref{fig:Lx} corresponds to the theoretical result from \citet{Fragos13} for a metal-free HMXB population, expected to be more representative of the first galaxies.
Our recovered 1D posterior of $L_{\rm X, <2keV}$/SFR supports theoretical predictions (e.g. \citealt{Fragos13}) and the observed evolution with metalicity and redshift (\citealt{Basu-Zych13, Douna15, Brorby16, Lehmer16}), that this quantity increases towards high redshifts.

Finally, we caution that our limits on $L_{\rm X <2keV}$/SFR could weaken if alternate heating mechanisms play a significant role.  Although we include adiabatic, ionization, X-ray and Compton heating/cooling, in some extreme models alternate heating sources could dominate.  These could include shock heating (e.g. \citealt{Furlanetto06, MO12}), dark-matter annihilation heating (e.g. \citealt{EMF14}; though see \citealt{Honorez16}), CMB heating (e.g. \citealt{Venumadhav18}; though see \citealt{Meiksin21}), and Lyman-$\alpha$ heating (e.g. \citealt{Chuzhoy:2007}).  However, the amount of heating required by the HERA limits at $z\approx8$ is generally beyond what most of these alternate sources can achieve without violating constraints from other high-$z$ observations in our model likelihood.  For example, Lyman-$\alpha$ heating 
only dominates for a relatively large, slowly evolving star-formation density coupled with a low X-ray efficiency. This region of parameter space  is ruled out by the combination of complementary observations and HERA limits (e.g. compare the narrower range of our {\it with HERA} posterior in the top right panels of Fig. \ref{fig:posterior} to the range of blue curves in Fig. 10 of \citealt{Reis:2021}).  Thus it is unimportant for the data-constrained {\it with HERA} posterior in this section, though it can be important in ruling out extreme models when not considering complementary observational data (c.f. \citealt{Reis:2021} and \S \ref{sec:radio}).

\subsubsection{IGM properties: disfavoring a cold IGM with HERA}

In Figure~\ref{fig:xT} we show the marginalized {\it without HERA} and {\it with HERA} posteriors in the space of $(\avenf, \aveTs$) (tan and purple regions, respectively). 
In gray we also show the prior distribution over this space.
Comparing the tan to the gray regions, we see that previous observations disfavor a notable prior volume also in the space of IGM properties.\footnote{We note that  our priors over galaxy parameters do not translate into flat priors over $(\avenf, \aveTs$).  It is easier to theoretically and empirically motivate priors on (fundamental) galaxy properties than on (derived) IGM properties. Thus choosing flat priors directly over mean IGM properties could result in biased posteriors when using weakly constraining data (e.g. \citealt{Ghara2020, Ghara21}).}  Most notably, current observations shift the posterior so that the midpoint of the EoR is occurring around $z\sim8$ to match EoR constraints from Planck and QSO spectra.

Now introducing the \citetalias{HERA2021} limits with the purple curves, we see that the \referee{HERA disfavors this region of low temperatures for $0.4 \lsim \avenf \lsim 0.8$ at $z=8$.  These are the previously-mentioned "reionization-driven" models: having large fluctuations in the ionization field combined with a cold IGM.  The impact of HERA is most strongly seen in the marginalized temperature PDFs in the right side panel:}
 {\it with HERA} and {\it without HERA} exhibit qualitatively different distributions, with the HERA limits strongly disfavoring the low $\aveTs$ peak seen in the posterior without HERA.  This demonstrates that the HERA limits are ruling out otherwise viable models.

\begin{figure}
		\includegraphics[width=0.5\textwidth]{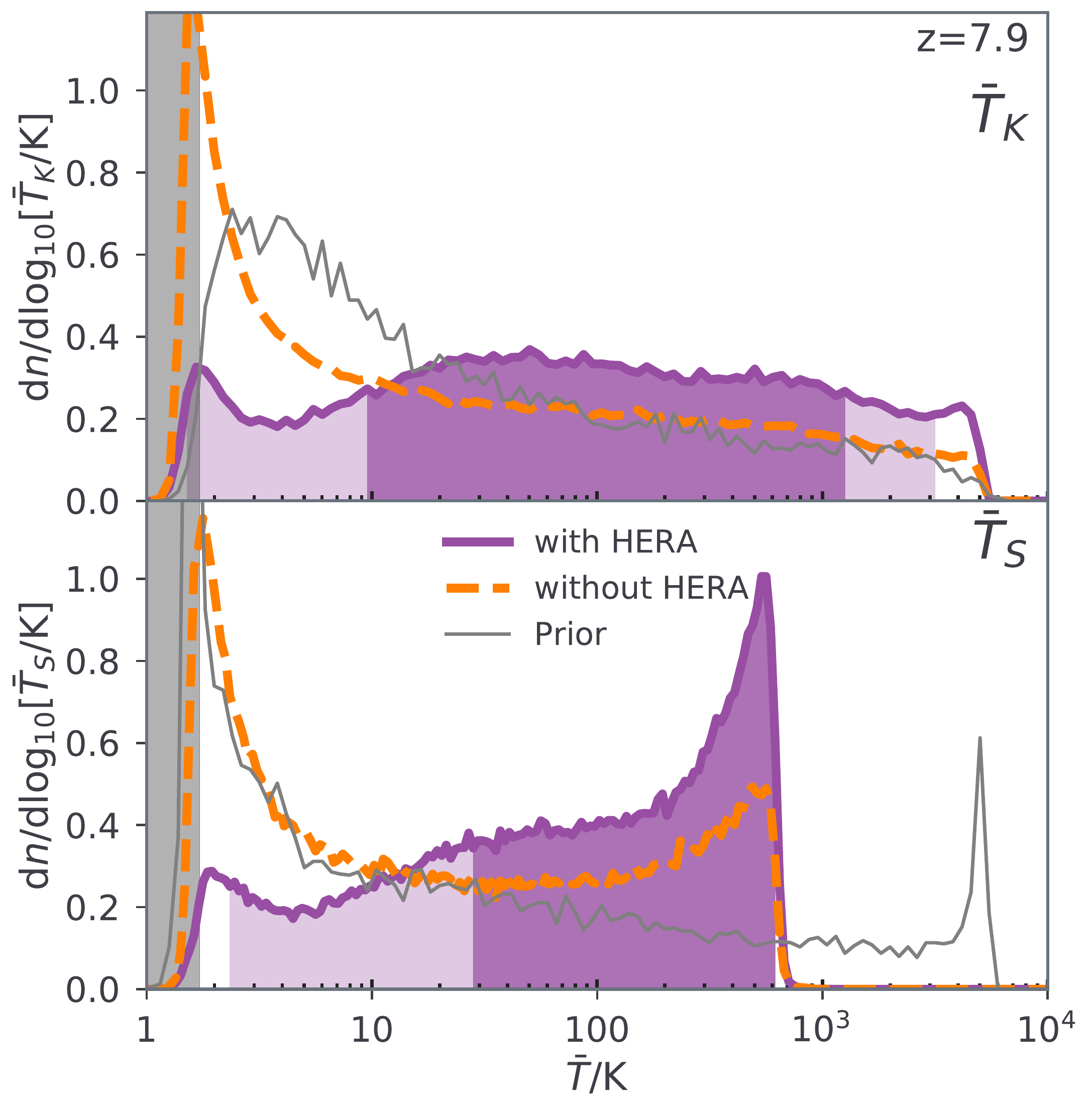}
		\vspace*{-6pt}
		\caption{\label{fig:temp}
Marginalized 1D PDFs of the spin temperature (bottom) and the kinetic temperature (top) of the neutral IGM at $z=7.9$. The colors are the same as in the previous figure.   The HPD 68\% (95\%) C.I. are denoted under the {\it with HERA} posterior  with  dark  (light)  shading. Since by definition, the averaging is performed only over the neutral cells (with $x_{\rm HI}>0.95$), there are no kinetic temperatures above $\bar{T}_K \sim 10^4$ K.  For models without any neutral cells at $z=7.9$ ($13$\% of the {\it with HERA} posterior), we take the mean temperatures from the last snapshot that contains neutral cells. The gray region on the left denotes values below the adiabatic cooling limit for the IGM at mean density.
}
\end{figure}

In Figure \ref{fig:temp}, we further investigate the physical origins of the temperature PDFs, plotting the spin temperature distributions in the bottom panel and the corresponding kinetic temperature distributions in the top panel.  Both are averaged only over the neutral IGM, specifically those cells with $x_{\hone} > 0.95$.  We see that the kinetic temperature of the neutral IGM smoothly extends to $T_K \sim 10^4$ K, without the bimodality seen in the spin temperature distributions for {\it without HERA}.  This is because the spin temperature is inversely weighted between the kinetic ($T_K$) and radio background ($T_{\rm rad}$) temperatures (c.f. eq. \ref{eq:Tspin}).  As $T_K \rightarrow \infty$, the spin temperature asymptotes to $T_S \rightarrow (1+x_\alpha) T_{\rm rad} \sim (1+x_\alpha) \times 24$ K for the standard assumption of a CMB-dominated radio background at $z=8$.  Although $x_\alpha$ scales with the Lyman-$\alpha$ background, it cannot exceed values of $x_\alpha \sim 300$ in our data-constrained models of $f_{\rm esc}$ without the gas in the simulation cell becoming ionized.  This results in the sharp upper limit of $\aveTs\lsim$ 600--10$^3$ K for the neutral IGM seen in Figure~\ref{fig:temp}.\footnote{Indeed the marginalized prior on $\aveTs$ (shown with the gray curve in the bottom panel of Figure~\ref{fig:temp}) extends out to $\aveTs\sim$ 10$^4$ K as the prior volume includes low values of $f_{\rm esc}$ that do not reionize the Universe.  Observations exclude these models from the posterior.}

Comparing the purple and the tan curves in Fig. \ref{fig:temp}, we reach the main conclusions of this subsection.    \citetalias{HERA2021} observations substantially improve our understanding of the $z=8$ neutral IGM temperatures,\footnote{We want to re-emphasise that our temperatures are averaged over the {\it neutral} IGM component, for which the spin temperature is a defined quantity.  The ionized IGM component, likely comprising tens of percent of the IGM volume at $z\sim8$ (c.f. the EoR history panel in the top right of Fig. \ref{fig:posterior}) would have $T_K \sim$10$^4$ K (e.g. \citealt{DAloisio19}).  Thus the kinetic temperature averaged {\it over all volume} would be roughly $\langle T_K \rangle_V \approx (1 - Q_{\rm HII}) \bar{T}_K + Q_{\rm HII} \times 10^4$ K, where $Q_{\rm HII} \approx (1 - \avenf)$ is the volume filling factor of the ionized IGM component, and $\bar{T}_K$ is the average IGM temperature of the neutral IGM component (plotted in the top panel of Fig. \ref{fig:temp}).}
allowing us to place 68\% (95\%) high posterior density confidence intervals on the spin temperature of $27$ K $ < \aveTs < 630$ K (2.3 K$<\aveTs < $ 640 K) and the kinetic temperature of 8.9 K $< \bar{T}_K < 1.3\times10^3$ K (1.5 K $< \bar{T}_K < 3.3\times10^3$ K).  Other observations of the early Universe and high-$z$ galaxies are unable to constrain these temperatures on the low end.

Indeed because these temperature constraints of the neutral IGM come almost exclusively from the 21-cm signal (where they depend only on the ratio $T_S / T_{\rm rad}$; c.f. eq. \ref{eq:21cm_signal}), we can generalize our temperature limits for any {\it homogeneous} radio background even if the standard assumption of  $T_{\rm rad} = T_{\rm CMB}$ is incorrect.  In the regime of $T_{\rm rad} > T_{\rm CMB}$ our {\it with HERA} limits can thus be generalized as 1.1 (0.095) $< \aveTs/T_{\rm rad} <$ 26 (26) and 0.37 (0.062) $< \overline{T}_K/T_{\rm rad} <$ 54 (140) at 68\% (95\%) confidence. 

\section{Constraints on IGM properties using a reionization-driven phenomenological model} 
\label{sec:phenom_models}

Here we introduce simple, phenomenological models for reionization-driven 21-cm power spectra and compare the resulting constraints on IGM properties to those obtained with \cmfast\ in the previous section. Although very simple, these phenomenological models help build physical intuition for the most important effects to consider when interpreting upper limits on the 21-cm power spectrum. We summarize the functionality of this model briefly below, and we defer a more complete description to Mirocha et al., in preparation.

Our principal goal is to examine a model built directly from IGM structures rather than galaxy models, so that we do not make any \emph{explicit} assumptions about the heating and ionizing sources during reionization. To that end, we parameterize the process not with galaxy properties but with the IGM temperature and with the ionized bubble size distribution (BSD) directly. Note that this approach does make \emph{implicit} assumptions about the sources of reionization, e.g., through the assumed BSD parameterization; it is just non-trivial to determine what these assumptions are. However, they are certainly different from physical models like \cmfast, and as a result, help to determine how robust IGM constraints are to modeling assumptions.

For an idealized two-phase IGM in which the BSD is known, the two-point statistics of the ionization field can be worked out analytically following \citet{FZH04}. In \cmfast\ and similar models, the excursion set approach is used to forward model the BSD, but we parameterize it more flexibly here with a log-normal distribution and allow the characteristic bubble size, $R_b$, and the width of the distribution, $\sigma_b$, to vary as free parameters. Note that BSDs derived from excursion set or semi-numerical models generally have broader tails to low $R_b$ than even a log-normal \citep{Furlanetto:2004,Paranjape2014,Ghara2020}, but for fits to a single $k$ and a wide prior on $\sigma_b$, as we perform here, we do not expect the detailed shape of the BSD to be important. We further assume that the ``bulk IGM'' outside of bubbles is fully neutral and is of uniform spin temperature, $\overline{T}_S$. The fourth and final free parameter is the volume-filling fraction of ionized gas, $Q_{\rm HII} \equiv 1 - \avenf$, which normalizes the BSD. 

To model the 21-cm power spectrum within this simplified framework, one must model the ionization field \textit{and} its correlation (or anti-correlation) with the density field. Because we abstract away assumptions about astrophysical sources completely, and instead work in terms of the BSD and mean IGM properties $Q_{\rm HII}$ and $\overline{T}_S$, it is not immediately obvious how to do this. While it is possible to estimate the behaviour of cross-terms using the halo model \citep{FZH04} or perturbation theory \citep{Lidz2007b}, here, we take a simpler approach that avoids explicit assumptions about astrophysical sources. 
If we assume for simplicity that the structure of the density field mirrors that of the ionization field, i.e., it is a binary field, cross-terms involving ionization and density can be re-written in terms of the ionization power spectrum given the typical density of ionized regions, $\langle \delta \rangle_i$. To estimate $\langle \delta \rangle_i$, we assume that reionization is ``inside-out," or in other words that the ionized volume fraction $Q_{\rm HII}$ is made up of the densest fraction $Q_{\rm HII}$ of the volume. Then, to complete this ``volume matching'' procedure, we assume the density PDF is log-normal \citep{Coles91} with a variance given by the density field smoothed on the scale at which the BSD peaks. This naturally leads to a model in which the typical bubble density declines with time, so the importance of cross-terms is greatest in the early stages of reionization. Finally, as in \S\ref{sec:bias}, we assume $\mu=0.6$ to match the spherical averaging done in {\tt 21cmMC} simulations.

We perform two MCMC fits using \textsc{emcee} \citep{Foreman-Mackey2013} -- one using the inverse likelihood (eq.~\ref{eq:comp_likelihood}) and one with the regular likelihood (eq.~\ref{eq:final_marg_likelihood}) -- to the $k=0.134 \ \mathrm{cMpc}^{-1}$ limit from Band 2 at $z=7.9$ using 192 walkers for a total of $\sim$500,000 steps. We adopt flat priors on each model parameter: $0 \leq \overline{T}_S / \mathrm{K} \leq 10^3$, $0 \leq Q_{\rm HII} \leq 1$, $0 \leq R_b/\mathrm{cMpc} \leq 30$, and $0.5 \leq \sigma \leq 2$. Note that while the 21-cm signal is insensitive to $\overline{T}_S$ once $\overline{T}_S \gg \Tcmb$, our lower limits on $\overline{T}_S$ \textit{are} sensitive to the prior range. Our choice of $\overline{T}_S \leq 10^3$ K is motivated by the maximum allowed spin temperature in standard scenarios (see section~\ref{sec:normal} and Fig. \ref{fig:temp}), though we broaden the lower bound from the expected adiabatic cooling limit of $\overline{T}_S \simeq 1.7$ K to zero so that more exotic scenarios may be considered.

\begin{figure}
	    \includegraphics[width=0.48\textwidth]{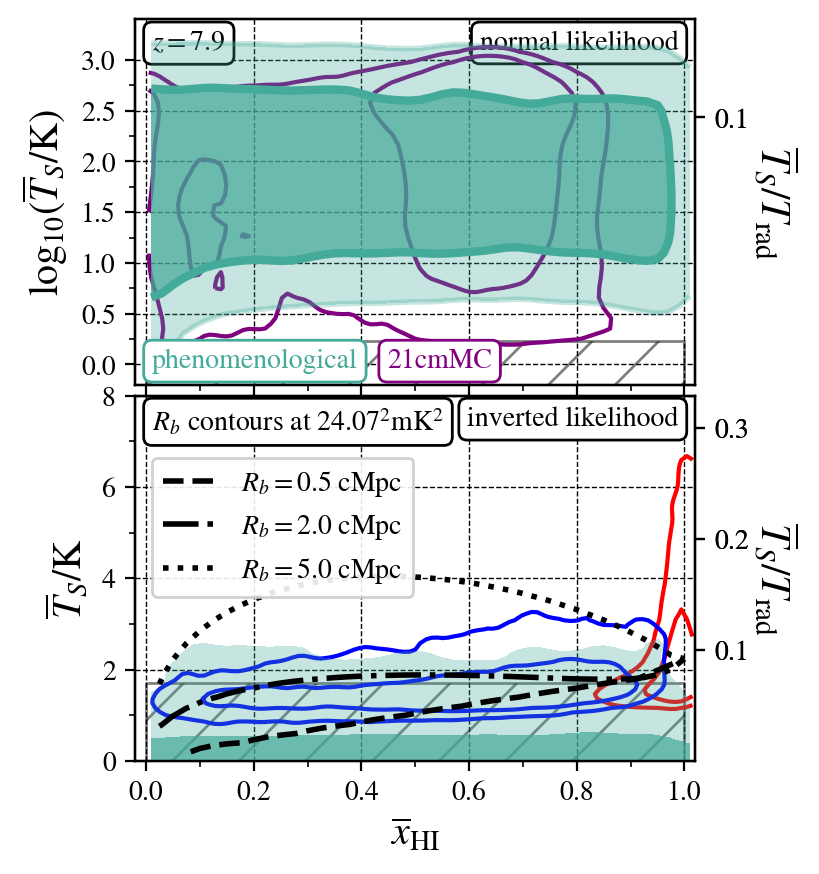}
		\vspace*{-6pt}
		\caption{\label{fig:posterior_phenom}
Constraints on the mean properties of the $z \sim 8$ IGM using phenomenological models (see \S\ref{sec:phenom_models}) compared to the \textsc{21cmFAST} results. \textit{Top:} Filled cyan contours are 68\% (dark) and 95\% (light) confidence levels obtained with the phenomenological model, while purple contours are those from \textsc{21cmMC} (as in Fig.~\ref{fig:xT}). \textit{Bottom}: As in Fig. \ref{fig:posterior}, the blue and red open contours in each panel correspond to reionization- and density-driven scenarios, respectively, while the filled contours show the disfavoured region determined by the phenomenological model. Additional black contours in this panel trace the phenomenological model's predictions at fixed $k=0.134$ power, $\Delta_{21}^2 = 24.07^2 \ \mathrm{mK}^2$, corresponding to the HERA measurement $+1\sigma$, for three different bubble sizes assuming a fixed $\sigma_b=0.6$. The cross-hatching along the bottom of each panel indicates regions with temperature below that of a homogeneous and unheated high-$z$ IGM.
}
\end{figure}

In the top panel of Figure~\ref{fig:posterior_phenom}, we show constraints on the mean spin temperature and ionized fraction of the IGM obtained from this model after marginalizing over the parameters of the bubble size distribution ($R_b$ and $\sigma_b$). We obtain 95\% (68\%) lower limits on the spin temperature of the $z=7.9$ IGM of $\overline{T}_S \geq 5.3$ K ($\overline{T}_S \geq 25$ K).  Qualitatively, these results are in good agreement with those derived using \cmmc\ (the {\it with HERA} posterior is shown with purple contours; see also Fig.~\ref{fig:xT}). As discussed in the previous section, the data-constrained \cmmc\ posterior is dominated by ``reionization driven" fluctuations, since the ``density driven" models have a neutral fraction at $z\sim8$ that is too high and are disfavored by EoR observations.  It is therefore encouraging that our ``reionization driven" phenomenological model is broadly consistent with the {\it with HERA} posterior from \cmmc.  This implies that our claims of  HERA's upper limits disfavouring models in which the IGM has not been heated at $z\sim 8$ are not sensitive to the nature of the EoR fluctuations.

In the bottom panel, we show the results obtained via the inverse likelihood.  Note again that we require only that the mean temperature of the IGM be positive, which is why the disfavoured region in this panel extends all the way to $\overline{T}_S = 0$. This is one of the advantages of the phenomenological approach: it can constrain more exotic scenarios without invoking a particular physical model (c.f. section~\ref{sec:exotic_DM}, where we introduce some such physically motivated models).  Here again, our results are broadly consistent with the analogous ones from \cmmc\ (the ``reionization driven" modes are shown with the blue curves; note the red contours are ``density driven" modes that are not considered by our phenomenological BSD model).

To further explore this agreement,
in the bottom panel of Figure~\ref{fig:posterior_phenom} we show iso-power contours for several different bubble sizes, holding fixed the width of the BSD at $\sigma_b=0.8$. The rationale here is simple: iso-likelihood contours should trace iso-power contours for inference based on a single $k$ mode. From this plot we see that if bubbles are generally small, $R_b \lesssim 2$ cMpc, the phenomenological model predicts that warmer temperatures are needed to preserve the large-scale power as $\avenf \rightarrow 1$. However, if bubbles are generally larger, with $R_b \gtrsim 2$~cMpc, this trend is reversed. These results suggest that physical models like \cmfast\ effectively have a low prior probability assigned to models with large bubbles at early times. Indeed, excursion set calculations suggest that typical bubbles sizes $R_b \gtrsim 2$~cMpc generally do not emerge until reionization is underway at the $\sim 20-30$\% level \citep{FZH04}. However, because the phenomenological model can have arbitrarily large bubbles at any time, the density-driven mode is washed out when marginalizing over $R_b$ and $\sigma_b$. Though the ``density-driven" models are ultimately disfavoured given that they do not complete reionization by $z \sim 6$ (see Fig. \ref{fig:reverse_posterior}), they serve as interesting test case nonetheless (see section~\ref{sec:bias}).


\section{Constraints on dark matter \\ and adiabatic cooling using \\ density-driven models}
\label{sec:exotic_DM}

\begin{figure}
		\includegraphics[width=0.48\textwidth]{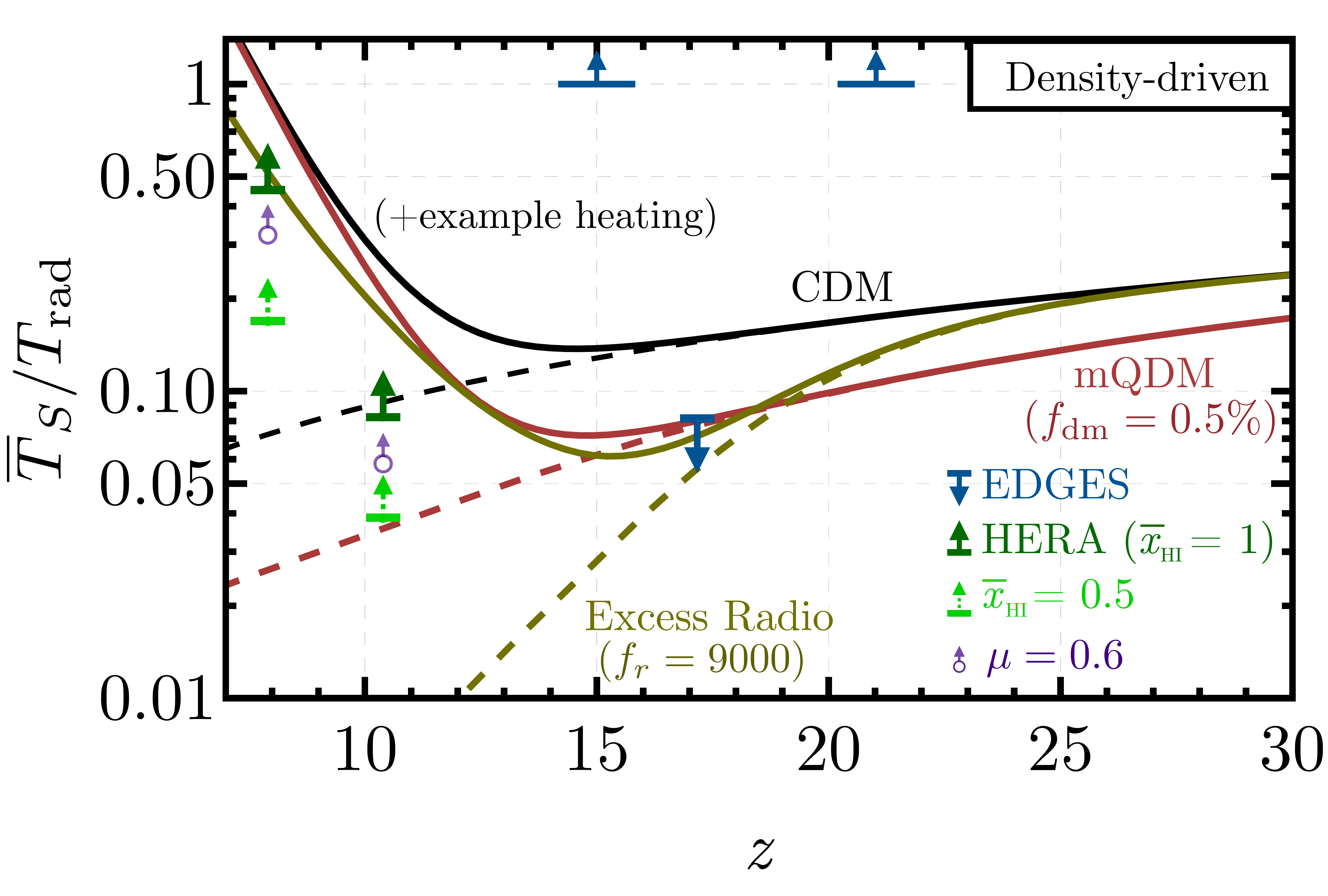}
    \caption{\label{fig:TgasLimits}	Analytically derived lower
	limits on $\aveTs/T_{\rm rad}$ from HERA data (95\% confidence, green arrows) compared to the upper limit from EDGES (blue arrow, which also implies the two lower limits at $z=15$ and 21 given their profile shape).
	The \citetalias{HERA2021} limits have been obtained by assuming density-driven fluctuations (see section~\ref{sec:bias} for details) and two different constant values of the neutral-hydrogen fraction $\avenf=1$ (dark green) and 0.5 (light green).
	The black line shows the standard CDM prediction assuming full coupling ($T_S=T_K$), with the dashed line corresponding to zero heating and the solid line to a \referee{ toy model of } X-ray heating. 
	The HERA Band 2 data
	can rule out adiabatic cooling under our assumptions, requiring some X-ray heating to take place before $z=7.9$.
	The red line includes a fraction $f_{\rm dm}=0.5\%$ of millicharged DM (mQDM), so as to explain the EDGES depth.
	Without any heating (dashed), this curve is ruled out by HERA Band 1, showing that there must be heating between $z=17$ and $z=10.4$ if EDGES is explained by mQDM.
	The conclusion is similar for a model with excess radio emission (with a radio fraction $f_{\rm r}=9000$, see section~\ref{sec:radio}), shown as the tan line.
	The empty symbols represent spherically averaged RSDs (purple, which were shown in Fig.~\ref{fig:FirstLimits}).
	}
\end{figure}

In previous sections we have obtained limits on the IGM spin temperature $\overline{T}_{S}$ using different approaches.
Here we study how these limits compare to predictions in the standard CDM cosmological model, as well as models of millicharged DM (mQDM).
We will also briefly explore how our assumptions about RSDs affect the limits imposed on the IGM.
Throughout this section we will use our density-driven phenomenological model, equation~\eqref{eq:biasm} (in all cases assuming $T_S=T_K$). 
While this approach has limited validity, it provides a useful test bed of our assumptions, as it allows us to obtain analytic limits under different RSD assumptions, as well as extend the temperature range studied below the adiabatic cooling threshold to probe mQDM models, neither of which are currently included in the usual {\tt 21cmFAST} simulation-based approach of section~\ref{sec:21cmmc}.

\subsection{The Impact of RSDs on the $\overline{T}_S$ Limits}
\label{sec:adiabatic}

First we study how our analytic limits change under different RSD assumptions.
Within our bias approach this can be readily implemented by varying $\mu$ in equation~\eqref{eq:biasm}.
For simulations, on the other hand, it is challenging to study the $\mu\to 1$ limit, given the geometry of the Fourier modes populating a square box. The analytic limits obtained in section~\ref{sec:bias} ($\aveTs \geq \{7.8, 1.9\}\,$ K   at $z=\{7.9,10.4\}$), assumed $\mu=0.6$ to match the spherical averaging done in {\tt 21cmMC} simulations.
Under the assumption that modes lie predominantly along the line of sight ($\mu\approx1$), as actually observed by HERA (see section~\ref{sec:ObsCampaign}), these limits strengthen to $\aveTs \geq \{11, 2.6\}\,$ K   (for $z=\{7.9,10.4\}$) at 95\% confidence, which are $\sim 50\%$ stronger, as shown in Figure~\ref{fig:TgasLimits}.
If we had ignored RSDs ($\mu=0$), but kept the same assumptions otherwise, the 95\% CL limits would weaken to
$\aveTs > \{3.1, 0.74\}$ K at $z=\{7.9,10.4\}$, a factor of $\sim 3$ smaller.
The difference between these three assumptions highlights the importance of properly modelling RSDs in 21-cm power-spectrum analyses.  
We note, however, that these results assume the density field drives the 21-cm fluctuations, in which case RSDs always increase the 21-cm power spectrum.
This trend can be reversed if radiation fields are the main source of anisotropy (e.g.~in ``reionization driven" scenarios as in Fig.~\ref{fig:example_slices}), though it is not expected to change our conclusions, see Sec.~\ref{sec:inference}.

We also show the impact of varying the neutral-hydrogen fraction $\avenf$ on our analytic results. 
Unlike the galaxy-driven models of previous sections -- in which patchy reionization {\it enhances} the 21-cm power spectrum because of the bubble structure -- here we assume uniform reionization (which could result from exotic processes; e.g. \citealt{EMF14, Honorez16}),
in which case $\avenf<1$ \emph{suppresses} the 21-cm power spectrum, as clear from equation~\eqref{eq:biasm}. 
Had we assumed $\avenf =0.5$ (instead of fixing $\avenf =1$), 
we would arrive at the 95\% confidence limits $\aveTs > \{4.1, 1.2\}$ K at $z=\{7.9,10.4\}$ (both with $\mu\approx1$).
While it is unlikely that $\avenf$ deviates significantly from unity at $z=10.4$, a global value of $\avenf=0.5$ is in line with our expectations for $z=7.9$.

These limits have interesting implications for the thermal state of the IGM at high redshifts, as well as for the first EDGES claimed detection~\citep{Bowman18}.
We compare all the $\aveTs$ limits (divided by $T_{\rm CMB}$)  in Figure~\ref{fig:TgasLimits}
against the $\aveTs/T_{\rm rad}$ prediction for the standard CDM model, both in the absence of heating and with a fiducial X-ray heating model, akin to the ones implemented within {\tt 21cmFAST} in previous sections.
The HERA Band 2 95\% confidence limit is above the adiabatic cooling prediction at $z=7.9$, both for $\avenf=0.5$ and 1 (and in fact for any $\avenf\geq 0.3$ in this bias approach).
Thus, HERA requires some heating by $z=7.9$ given our assumptions.
Moreover, the HERA limits for Band 1 ($z=10.4$), while below the adiabatic limit at that $z$, can be used to set constraints on dark-matter induced cooling of the gas, which we now explore.

\begin{figure}
		\includegraphics[width=0.45\textwidth]{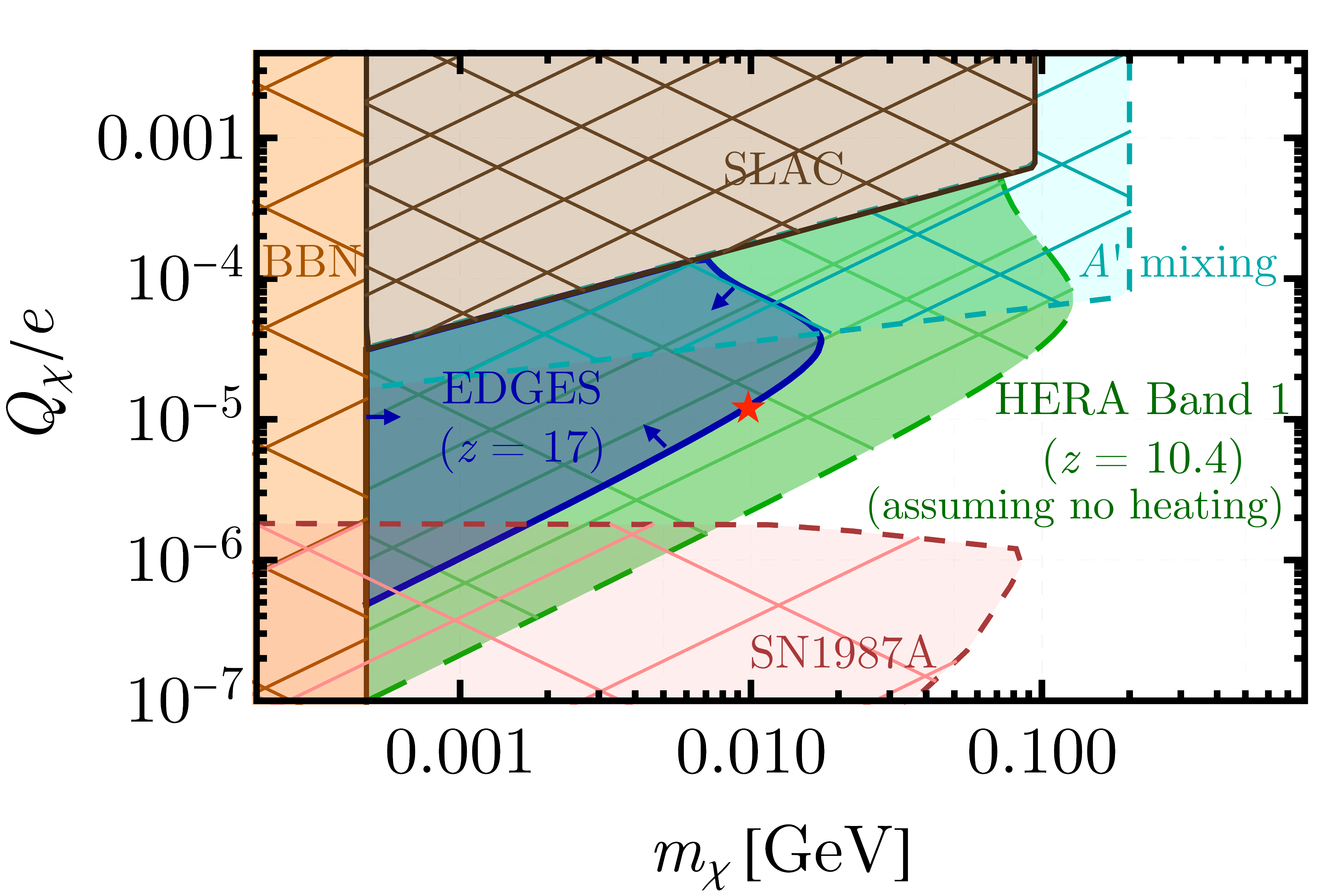}
		\caption{\label{fig:MQparams}
		HERA constraints on the millicharged-DM (mQDM) parameter space. 
		Charge $Q_\chi$ of the particles (divided by the electron charge $e$) versus their mass $m_\chi$ in GeV.
		Hatched regions are ruled out by different experiments, in brown we show limits from the SLAC experiment~\citep{Prinz:1998ua}, in orange the most conservative BBN constraints~\citep{Jaeckel:2010ni}, in red the constraint from cooling of SN1987A from~\cite{Chang:2018rso} (see, however, \citealt{Bar:2019ifz} for criticisms),
		and in cyan the region disfavored if there is a dark photon $A'$ mediating the millicharge~\citep{Munoz:2018pzp} (see also \citealt{Vogel:2013raa}). 
		Blue shows the EDGES-preferred region ($z\approx17$, following \citealt{Munoz:2018pzp}), and the green region is ruled out by HERA band 1 ($z=10.4$) at 95\% confidence, assuming density-driven fluctuations.
		An EDGES detection of millicharged DM is only compatible with HERA if heating takes place between $z\approx17$ and $z=10.4$.
		We have taken a fraction $f_{\rm dm}=0.5\%$ of DM to be millicharged, but our conclusions extend to all relevant fractions.
		The red star is the point that gives rise to the red line in Fig.~\ref{fig:TgasLimits}.
		We remind the reader that the HERA Band 2 data ($z=7.9$) already rules out adiabatic cooling at 95\% confidence, so by construction it also rules out any DM model that produces additional cooling.
		}
\end{figure}

\subsection{Dark matter-baryon interactions}

The first claimed 21-cm detection from the Cosmic Dawn in~\cite{Bowman18} shows a surprisingly deep absorption feature at $z\sim 17$.
The depth of this absorption, if interpreted to be cosmological (see, however, e.g.~\citealt{Hills18, SP19, Tauscher20}), can be translated into a requirement that $\aveTs/T_{\rm rad} \leq 0.08$ at $z=17$, a factor of two smaller than allowed by the standard cosmological model.
Reducing the Wouthuysen-Field coupling in this case only exacerbates the tension, as it would bring the spin temperature closer to that of the CMB, producing shallower absorption.

A possible explanation for this anomalous depth consists of lowering the temperature of baryons in the IGM by allowing them to interact with the cosmologically abundant---and kinetically cold---DM.
Elastic scattering between these two fluids would bring them closer to thermal equilibrium, cooling down the baryons and heating up the DM.
These interactions could take the form of a new fundamental force~\citep{Tashiro:2014tsa,Munoz:2015bca,Barkana18,Fialkov2018}, which however would be in conflict with fifth-force constraints and stellar-cooling bounds.
Alternatively, part of the DM can be electrically charged, for instance through a dark-photon portal~\citep{Holdom:1985ag}, a scenario dubbed millicharged DM (mQDM).
In this case there are no new charges for baryons, and therefore fifth-force and stellar-cooling bounds are naturally evaded~\citep{Munoz:2018pzp,Barkana:2018qrx,MDL18,Slatyer:2018aqg,Berlin:2018sjs,Kovetz:2018zan,Liu:2019knx}.
Here we briefly study how well DM-baryon interactions, in the form of mQDM\footnote{
We note that current HERA data does not allow us to place limits on DM annihilation or decay \citep{EMF14,Lopez-Honorez:2016sur,Liu:2018uzy}, as we only have lower bounds on the gas kinetic temperature.
A 21-cm detection is required for those analyses.}, can be constrained by the \citetalias{HERA2021} limits.

To illustrate the effect of mQDM, we show in Figure~\ref{fig:TgasLimits} the prediction for an example model using the software developed in~\cite{Munoz:2018pzp}.
\referee{ We solve the coupled differential equations for the mQDM and hydrogen-gas temperatures starting at recombination. The interactions due to millicharges produce thermalization of the (initially cold) DM and the hydrogen gas, therefore cooling the latter. } For this figure we have chosen mQDM with a charge $Q_\chi=1.3\times 10^{-5} \, e$, where $e$ is the electron charge, and mass $m_\chi=10$ MeV, composing a fraction 0.5\% of the total DM.
These parameters are chosen to (barely) explain the EDGES depth 
and, as is clear from the figure, the cooling induced at later times lowers $\overline{T_S}$ below the HERA limit both at $z=10.4$ and $z=7.9$, ruling out this model in the absence of heating.

We generalize this result by performing a 2D scan of mQDM charges $Q_\chi$ and masses $m_\chi$, 
assuming that mQDM particles compose a 0.5\% fraction of the DM, which is at the edge of \referee{the 95\% confidence interval region allowed by }CMB constraints~\citep{Boddy:2018wzy}, with the remainder being neutral and non-interacting CDM.
We show the results in Figure~\ref{fig:MQparams}, where we also show the region that produces enough cooling to explain the EDGES depth (defined to be $\aveTs <4$ K as in~\citealt{Munoz:2018pzp}). 
This region is entirely contained by the HERA Band 1 constraint ($\aveTs > 2.6$ K at 95\% confidence), which shows that all the mQDM models that explain the EDGES depth also require heating before $z=10.4$ in order to avoid conflict with HERA.
Our conclusions hold for all other mQDM fractions in the relevant range $f_{\rm dm}=0.05$--$5\%$.

\section{Astrophysical Constraints in Models with an Extra Radio Background}
\label{sec:radio}

\subsection{The 21-cm Signal in the Presence of Radio Sources}
\label{sec:radiontro}

In this section we use HERA data to constrain models in which either astrophysical or exotic high-redshift radio sources contribute to the total radio background, in addition to the  CMB. Such an excess radio background above the CMB level has been observed at $z=0$, with the data consistent with a synchrotron radio background of a spectral index $-2.58$ and a brightness temperature  $\sim 603$ K at the rest-frame 21-cm frequency  \citep{fixsen11, seiffert11, dowell18}.  The nature of this excess is still undetermined \citep[e.g.][]{Subrahmanyan:2013}, and it could partially be accounted for by a population of unresolved high-redshift sources of either astrophysical or exotic origin \citep{ewall18, jana19, Fraser:2018, Pospelov:2018, Brandenberger:2019, Superstrings2021}.

An excess high-$z$ radio background would have important implications for the 21-cm signal, because the stronger background amplifies the absorption  \citep[via the temperature term in eq.~\ref{eq:21cm_signal} including an effect on coupling coefficients in eq.~\ref{eq:Tspin}; see complete discussion in ][]{FB19,Reis:2020}. Such models have been presented as potential explanations of the anomalously strong EDGES Low Band detection \citep[][]{Bowman18}; for example, \citet{FB19} found that the EDGES signal can be explained if the cosmological (high redshift) contribution of such a background is between 0.1\% and 22\% of the CMB
at 1.42 GHz (see also \citealt{mirocha19, jana19, ewall18, ewall20, Mebane2020, Reis:2020, Superstrings2021}). These explanations are challenging, however, requiring either unconfirmed exotic sources or astrophysical sources that are far stronger than expected based on local observations \citep{ewall18, mirocha19, Mebane2020, ewall20} and necessitating rapid X-ray heating to match the steep recovery in the EDGES signal \citep{Reis:2020}.

More interestingly for our purposes, the presence of a radio background can also enhance fluctuations in the 21-cm signal \citep{FB19,Reis:2020}, so that \citetalias{HERA2021} can place limits on such a background (whether generated by discrete sources or more exotic processes) at $z \sim 8$ and 10. In this section we will consider both such scenarios, including the resulting limits in the context of other observations of the low-frequency radio background \citep{fixsen11, dowell18} and X-ray background \citep{Lehmer:2012}.

\subsection{Modelling}
\label{subsec:modelling_radio}

We generate spherically averaged 21-cm power spectra as a function of astrophysical parameters using the semi-numerical simulation method described in  \citet{Visbal12, Fialkov:2014, FB19, Cohen:2020, Reis:2020, Reis:2021}. Our simulations are 384 cMpc on a side and have a resolution of 3 cMpc. Initial large-scale perturbations in density and relative velocity between dark matter and gas \citep{TH10}   are linearly evolved from the Dark Ages ($z\sim60$) to $z = 5$. Using the modified Press-Schechter
 mass function which takes into account the effect of large-scale overdensity and velocity fields \citep{Barkana:2004, Fialkov:2012}, we calculate the halo abundance in each voxel of the simulation. Each halo is then populated by stars, and emissivities in different bands are calculated \citep[see, e.g.][for details]{Cohen:2020}. 
 RSDs are computed by multiplying the real space isotropic 21-cm signal by $(dv_{\rm r}/d{\rm r})^{-1}$ which is the radial component of the velocity gradient created by structure formation  \citep{Fialkov:2020}. Using coeval simulation cubes we calculate the spherically-averaged power spectrum at every redshift.
 
The key radiation backgrounds are all driven by the cosmic star formation rate, which in the simulations depends on two parameters. First, we choose a minimum circular velocity of star forming halos $V_c$, which determines the halo population that can form stars. We then choose a  star formation efficiency $f_{\rm *}$, which measures the amount of collapsed gas that turns into stars for halos above the atomic cooling limit, imposing an extra suppression in smaller halos \citep[e.g.][]{Fialkov13, Cohen:2020}. The code includes the suppression of star formation by Lyman-Werner feedback \citep{Fialkov13}, relative velocities between dark matter and gas  \citep{Fialkov:2012} and photoheating feedback \citep{Cohen:2016}. To calculate the Ly-$\alpha$ background, we  assume Pop~II star formation following \citet{BL05_WF}. For completeness we note that here we include multiple scattering of Ly-$\alpha$ photons and Poisson fluctuations in the number of first star forming halos  \citep[however, these effects are not significant at the redshift range observed by HERA,][]{Reis:2021}.

X-ray heating of the IGM is powered by a population of X-ray binaries with the ratio of bolometric luminosity to SFR of
\begin{equation}
    \frac{L_{{\rm X}, ~0.2-95~ {\rm keV}}}{{\rm SFR}} = 3\times 10^{40} f_{\rm X}~\frac{{\rm erg}~{\rm s}^{-1}}{ M_{\odot} {\rm yr}^{-1}}
\end{equation}
calculated between 0.2 keV and 95 keV assuming a hard X-ray SED of X-ray binaries \citep{Fragos13}. The free parameter $f_{\rm X}$  normalizes the X-ray efficiency relative to a population of present day binaries \citep[but  including an order-of-magnitude increase in this ratio at the low metallicity expected for high-redshift galaxies,][]{Fragos13}. The assumed SED is relatively hard, peaking at $\sim 1$ keV \citep{Fragos13, Fialkov:2014}. Note that  assuming a different SED could affect our final results \citep{Fialkov:2014, Pacucci14, Das17, Reis:2020}. The unresolved X-ray background observed by {\it Chandra} imposes an upper limit on this contribution \citep{Fialkov:2017}, as we will explore later, but we allow a broad range of $f_X$ between $10^{-4}$ and $10^3$ in the estimation framework.

This model also includes heating by scattering of Ly-$\alpha$ photons \citep{Chen2004, Chuzhoy:2007, ciardi10, Mittal:2021, Reis:2021} and the CMB (\citealt{Hirata:2007, Venumadhav:2018, FB19, Reis:2021}, though see \citealt{Meiksin21}). With the onset of the first stellar population, the extra heating processes raise the IGM temperature above the adiabatic limit even in the absence of X-ray heating, reducing the 21-cm background at the relevant redshifts in some scenarios by a factor of a few \citep[see][for more details]{Reis:2021}.

The process of reionization is implemented using the excursion set formalism \citep{Furlanetto:2004} and is described by two parameters: the ionizing efficiency of sources $\zeta$, which is normalized via the total CMB optical depth $\tau$, and the horizon of ionizing photons, $R_{\rm mfp}$. Although the latter parameter does affect the intensity of the 21-cm fluctuations at the end of the EoR, we fix it at 40 Mpc here as it plays  a secondary role in our constraints. 

Finally, we explore two types of radio backgrounds (beyond the CMB): 
\begin{enumerate}
    \item A fluctuating, time-variable radio background generated by galaxies, parameterized by $f_{\rm r}$. We assume that the galaxy radio luminosity per unit frequency in units of W Hz$^{-1}$ is proportional to the SFR   \citep[following][]{mirocha19}
\begin{equation}
    \frac{L_{r,\nu}}{\rm SFR} = 10^{22}\left(\frac{\nu}{150{\rm MHz}}\right)^{-\alpha_r} f_{\rm r}  ~\frac{{\rm W}~{\rm Hz}^{-1}}{M_{\odot} {\rm yr}^{-1}}
\end{equation}
where  $\alpha_r$ is the spectral index in the radio band,
\referee{which we set to the typical value of 0.7 \citep{mirocha19}, which is compatible with observations \citep{Hardcastle2016,gurkan18}.}
We calculate $T_{\rm rad}$ at redshift $z$ by summing up over the past light-cone contribution of all the radio galaxies  \citep[see ][for more details]{Reis:2020}.
\item A smooth synchrotron background that decays with time, for which we replace $T_{\rm CMB}$ by 
\begin{equation}
    T_{\rm rad} = T_{\rm CMB}(1+z)\left[1+A_{\rm r} \left(\frac{\nu_{\rm obs}}{78{\rm MHz} }\right)^{\beta}\right]
\label{eq:Arad}
\end{equation}
where $\nu_{\rm obs}$ is the observed frequency, $A_{\rm r}$ is defined relative to the CMB temperature and $\beta = -2.6$ is the spectral
index in agreement with the ARCADE2 \citep{fixsen11} and LWA1 \citep{dowell18} observations. Here we treat this background as phenomenological, but it could have been produced by exotic radio sources, e.g. radiative decay of relic neutrinos into sterile neutrinos \citep{Chianese:2018},  light dark matter
decays \citep{Fraser:2018, Pospelov:2018} and superconducting cosmic strings \citep{Brandenberger:2019}.
\end{enumerate}
In this work, we allow a broad range of $f_{\rm r}$ and $A_{\rm r}$ parameters. However, as we discuss later \citep[and as was shown by][]{FB19, Reis:2020} models with strong radio backgrounds, e.g. $ f_{\rm r} \times f_* >10^3-10^4$ for the radio from galaxies,
are constrained by ARCADE2/LWA1 data.

To summarize, the models considered here include an extra radio background in addition to the CMB either produced by radio galaxies or emitted by exotic sources. Our models  build on the following parameters: $V_c$ varied between 4.2 and 100 km s$^{-1}$, $f_{\rm *}$ between 0.001 and 0.5, $f_{\rm X}$ in the range between $10^{-4}$ and $10^3$, $\tau$ between 0.035 and 0.088,  $f_{\rm r}$ from 1 to $10^5$, and  $A_{\rm r}$ between $10^{-2}$ and $10^5$. Owing to the large dynamic ranges, we assume uniform priors on the parameters $\log_{10} f_{\rm *}$, $\log_{10}V_c$, $\log_{10} f_{X}$, $\tau$, and $\log_{10} f_{\rm r}$ or $\log_{10} A_{\rm r}$. For completeness we also include our constraints on the standard models (i.e. with no extra radio background above the CMB). 

Although similar in spirit to the \cmfast\ simulation suite described in section \ref{sec:21cmmc}, there are differences between the two sets of simulations. We refer the reader to the relevant papers for details on the physics and implementation differences. Broadly, the simulations described in this section include some additional heating processes, such as Lyman-$\alpha$ heating (which can affect the IGM temperature when it is very cold in some models) and (of course) radio emission, but they have a more prescriptive source model with fewer free parameters (and they are not constrained by ancillary observations such as the galaxy luminosity functions). A detailed code comparison is beyond the scope of this paper; instead, we focus on how these distinct codes can address the issues to which they are each best suited.

\subsection{Parameter estimation}
\label{sec:parameter_estimation}
We explore the parameter space of models compatible with the HERA upper limits
based on the likelihood $\mathcal{L}_m$ defined in equation 
\eqref{eq:final_marg_likelihood}. We also decimate the
power spectra as described in Section \ref{sec:datapoint_decimation},
using the even wave numbers ($k=0.086, 0.17$, ... cMpc$^{-1}$) of Band 1 and odd wave numbers ($k=0.13, 0.21$, ... cMpc$^{-1}$) of Band 2.
Note that we only compute the power spectrum up to $k=1.1\,\mathrm{cMpc}^{-1}$, limited by the simulation resolution. Larger wavenumbers (smaller scales), however, are irrelevant as the HERA limits rise much steeper at small scales than realistic models so those scales do not contribute towards the constraints.

Because individual simulations take a few hours to complete, we instead use an emulator to interpolate the power
spectra from $\sim 10^4$ existing simulation runs
(for each of the two types of radio background that we investigate here).
We implement the emulator using neural networks: taking  the astrophysical parameters as an input, a network predicts the logarithm of the power spectra for the HERA redshifts  ($z=7.93$ and  $10.37$) and wave numbers (from $k=0.086$  to $1.1$ cMpc$^{-1}$). The architecture of the emulator includes a multi-layer perceptron with 4 hidden layers of 100 nodes each, implemented using scikit-learn
\citep{Pedregosa2011}.
The power spectra are predicted
with a relative error of 20\%; we take this
uncertainty into account by adding it in quadrature to the observational 
error $\sigma_i$ in equation~\eqref{eq:final_marg_likelihood}. Although this is an approximation, the associated error is negligible in the context of current analysis. A detailed discussion of the emulator and its accuracy can be found in Appendix \ref{appendix:emulator_error}.

We explore the parameter space using the MCMC Ensemble sampler \texttt{emcee}
\citep[][]{Foreman-Mackey2013}, and we visualize and analyze the results
using \texttt{anesthetic} \citep{Handley2019} and \texttt{GetDist} \citep{Lewis2019}.

\subsection{Results: A radio background generated by galaxies}
\label{sec:radio_galaxies}

\begin{figure}
    \centering
    \includegraphics[width=0.5\textwidth]{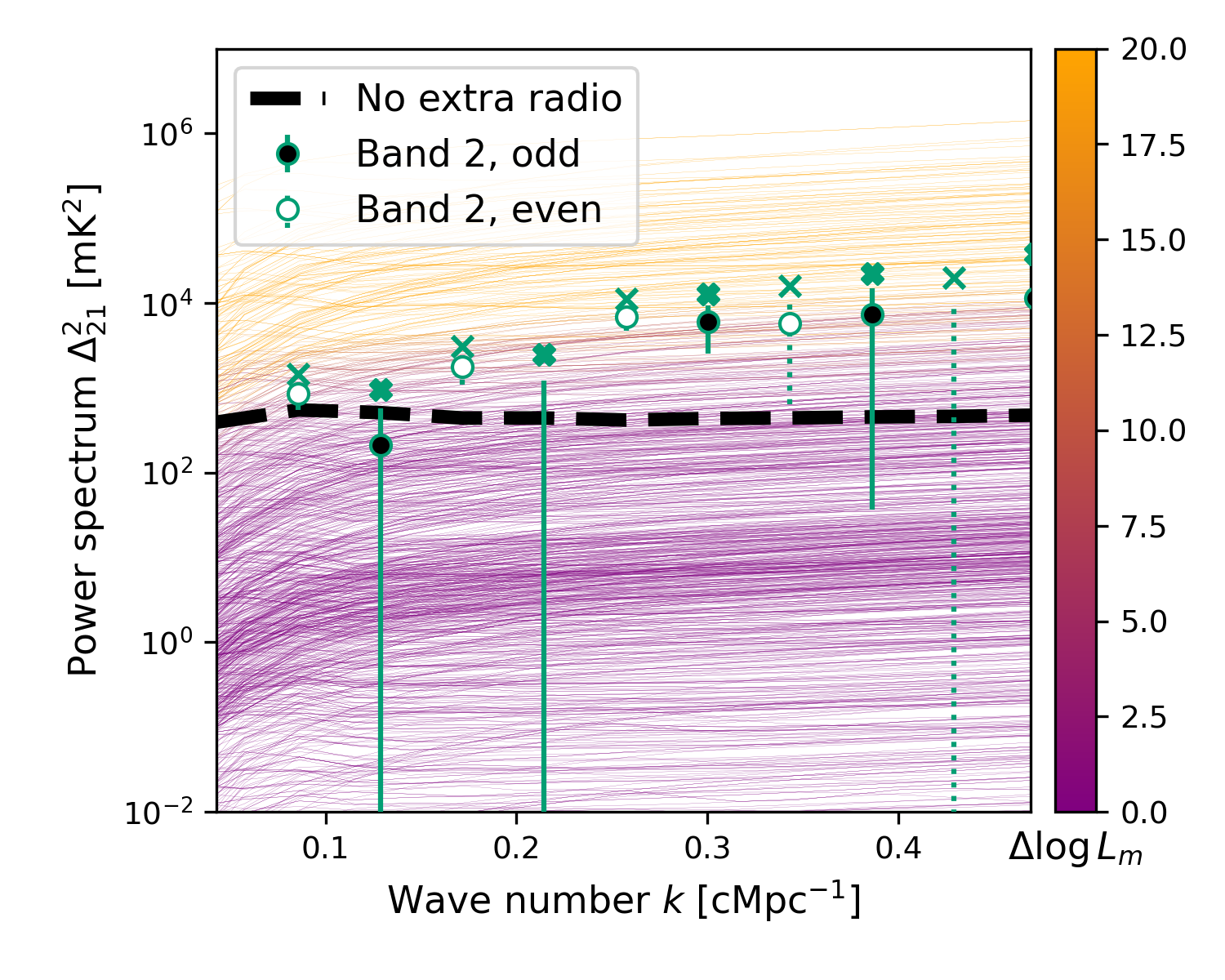}
    \caption{Power spectra of 1000 randomly selected models with an  extra radio background created by galaxies, at $z=7.93$ (HERA Band 2).
    A number of these models can be ruled out by the HERA data
    as shown by the color of the lines
    indicating the likelihood of each model.
    We use the
    decimated data points as described in section \ref{sec:datapoint_decimation},
    taking only the
     ``even'' points of Band 1 and ``odd'' points of Band 2.
    We show the Band 2 points with black circles for the points we take into account
    and white circles for the unused data.
    The error bars show $1\sigma$ errors and the crosses indicate the $2\sigma$ upper limits.
    For comparison, the thick dashed line shows the maximal possible amplitude of the ensemble of standard models at each wave number $k$ (i.e. the envelope).
    }
    \label{fig:extraradiobackground_radiopowerspectra}
\end{figure}

As was alluded to above, models with an additional radio background can easily, unlike most standard scenarios, exceed the HERA upper limits. To illustrate this point, we show a random subset of the simulated power spectra for the case of a radio background from galaxies in Figure \ref{fig:extraradiobackground_radiopowerspectra}. The power spectra are shown at $z=7.93$ and colored with respect to their compatibility with HERA constraints, indicating the
difference in log-likelihood $\Delta \log \mathcal{L}_m$ compared to  the best fit
 (which is $\Delta^2_{21}\approx0$~mK$^2$).
For comparison, we also plot 
the current HERA limits marked by data points with error bars. Clearly, a substantial fraction of the models (shades of orange)  are excluded by the current HERA limits with high significance.  In comparison, corresponding standard models (no additional radio background) have much lower amplitudes. We show the envelope of  these models (i.e. the maximal possible amplitude of the ensemble of standard models at each $k$)  with the thick dashed line in Figure~\ref{fig:extraradiobackground_radiopowerspectra} and discuss astrophysical implications of HERA for these cases in section~\ref{sec:8-standard}.

\begin{figure*}
    \centering
    \includegraphics[width=\textwidth]{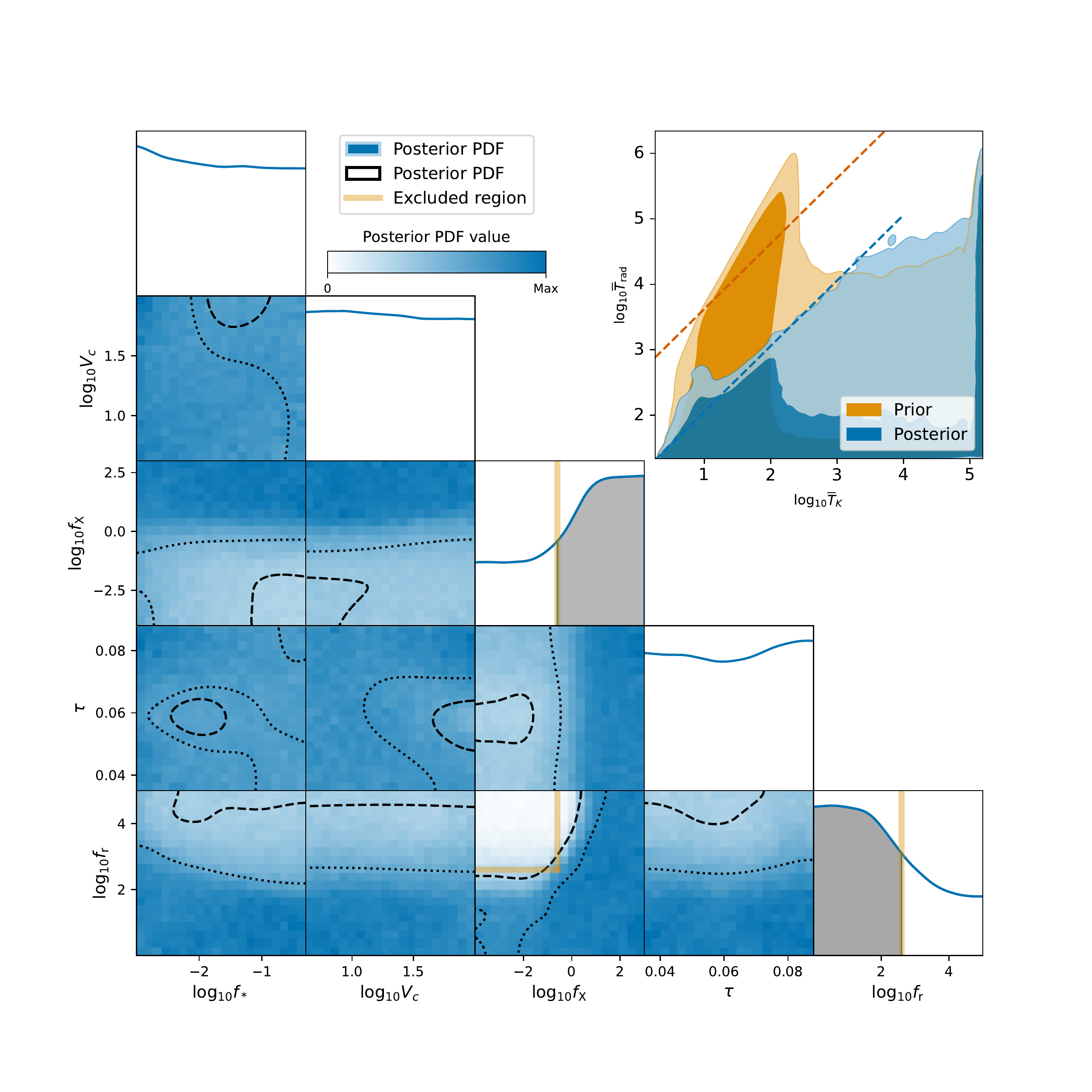}
    \caption{HERA constraints on models with extra radio background from galaxies. The triangle plot shows the marginalized 1D and 2D posteriors for all model parameters, with 68\% and 95\% confidence contours indicated by black dotted and dashed lines, respectively.
    The backgrounds show the 2D histograms of posterior values to illustrate how strongly a region is excluded, white background
    indicates a posterior close to zero while dark blue corresponds to the maximum posterior value of a panel.
    The grey areas in diagonal panels, and the solid orange lines show the individual 68\% confidence limits on $f_{\rm X}$ and $f_{\rm r}$ (numbers quoted in text). Combined, these approximately correspond to our main result, the 95\% confidence excluded region in the $f_{\rm X}$-$f_{\rm r}$ space, as shown in the center bottom panel.
    The inset in the top right corner shows the corresponding constraints on the derived parameters, $\overline{T}_{\rm rad}$ and  $\overline{T}_{K}$, at $z=8$ for both prior  and posterior. The prior is uniform in the model parameters ($\log_{10} f_{\rm *}$, $\log_{10} V_c$, $\log_{10} f_{\rm X}$, $\tau$, $\log_{10} f_{\rm r}$) but therefore non-uniform in the derived parameters, $\overline{T}_{\rm rad}$ and  $\overline{T}_{K}$. The blue and orange dashed lines show upper bounds (at 95\% confidence) on $\log_{10}(\overline{T}_{\rm rad}/\overline{T}_{K})$ from prior and posterior, respectively. 
    }
    \label{fig:extraradiobackground_posteriors}
\end{figure*}

Using the HERA likelihood alone ($\mathcal{L}_m$) we show the marginalized constraints  on the parameters  in Figure \ref{fig:extraradiobackground_posteriors}. The diagonal panels show the 1D marginalized posterior PDFs while the others show the 2D marginalized PDFs with the dotted and dashed lines indicating the 68 and 95\% confidence contours (containing 68 and 95\% of the 2D posterior probability, respectively).
The 2D marginalized posteriors involving $f_*$, $V_c$, and $\tau$  are relatively flat with the ratio between minimum and maximum posterior probability\footnote{These are measured from the bins in Figure \ref{fig:extraradiobackground_posteriors} using the minimum/maximum bin sample count ($\propto$ posterior value). A value close to one shows that the PDF is largely flat and therefore does not provide a strong constraint.}
being between 0.3 and 0.6. This results in confidence contours which could be easily affected by fluctuations due to the random sampling and are  strongly dependent on the prior. On the contrary, we find the 2D posterior in the $f_{\rm X}$--$f_{\rm r}$ plane to show a strong contrast between minimum and maximum regions  (with minimum/maximum posterior ratio of 0.02, i.e. dropping by more than three e-folds). There is a vanishing probability  for models with both a strong radio background (large $f_{\rm r}$) and weak X-ray heating (low $f_{\rm X}$). The large contrast in the probability across this sub-space indicates that the constraints on the combination of $f_{\rm X}$ and $ f_{\rm r}$ are expected to be robust,
i.e. even for different priors
(which would cause a small shift in the contour lines)
most of the high-$f_{\rm r}$ low-$f_{\rm X}$ region will still be excluded.

Marginalizing over the rest of the model parameters,  we constrain $f_{\rm X} $ to be greater than $ 0.25$ and $f_{\rm r}$ less than $397$ at 68\% confidence level individually, which maps to the excluded (at 68\%) $L_{{\rm X}, ~0.2-95~ {\rm keV}}/{\rm SFR} < 7.6 \times 10^{39}$ erg s$^{-1}$ M$_\odot^{-1}$ yr and $L_{r,\nu}/{\rm SFR} > 4 \times 10^{24}$ W Hz$^{-1}$ M$_\odot^{-1}$ yr (calculated at reference frequency $\nu=150\,\mathrm{MHz}$). The 2D region where both  $f_{\rm X} < 0.25$ and $f_{\rm r}>397$ (marked by orange solid lines on the corresponding 2D PDF in Figure  \ref{fig:extraradiobackground_posteriors})
approximately corresponds to the region
excluded at 95\% confidence level (exact contours shown as dashed black line).

To relate the constraints on model parameters to the physical state of the IGM at the observed redshifts, we calculate limits on the 
radiation temperature $\overline{T}_{\rm rad}$ and gas temperature $\overline{T}_K$ at redshift $z=8$ (note that these are derived, i.e. not sampled, parameters in the context of this model). The temperatures are averaged over all the regions that are not fully ionized (i.e. over the volume outside of HII regions).\footnote{For completeness, here the model suite includes scenarios that have extremely strong heating corresponding to $f_*\times f_X\gg1$ which are ruled out by {\it Chandra} observations. In such scenarios gas outside of the HII regions could be partially ionized up to $\sim 50$\% by X-ray photons  \citep[e.g.][]{Fialkov:2017} which leads to a suppressed 21-cm signal. The maximal possible temperature of such partially ionized gas is set to be the ionizing temperature of hydrogen, $1.57\times 10^5$ K, equivalent to the energy of 13.6 eV.} 
These temperatures are stored for every simulation 
and emulated similarly to the power spectra (see Appendix \ref{appendix:emulator_error} for details).
We obtain 2D posterior PDFs for these temperatures, marginalized over all other parameters, and we show the results in the top right panel of Figure \ref{fig:extraradiobackground_posteriors} (blue contours). 
Because the sampling is  uniform in terms of the simulation parameters (or their log for $f_{\rm *}$, $V_c$, $f_{\rm X}$, $\tau$,  $f_{\rm r}$), 
the priors on the derived parameters (temperatures) are not uniform
and shown as orange contours in the background.

The  \citetalias{HERA2021} limits significantly affect the temperature distribution, excluding high values of $\overline{T}_{\rm rad}$ when $\overline{T}_K\lesssim 1000$ K. This is expected, as, typically, a combination of high $\overline{T}_{\rm rad}$ and low $\overline{T}_K$  leads to a strong  21-cm signal exceeding the observed HERA limits. Exceptions are scenarios in which gas is either  significantly ionized already by $z=8$  or in which the radio background has a non-negligible effect on the
spin temperature $\aveTs$, driving it towards the background radiation temperature $\overline{T}_{\rm rad}$,
and, as a result, suppressing the total power spectrum.  In such models the 21-cm power spectrum is below the HERA limits even when gas is predominantly neutral and is colder than the radio background. These models populate the lower-left corner of the temperature plot and are not ruled out by HERA even when $\overline{T}_K \ll \overline{T}_{\rm rad}$.
Overall we derive a constraint on the ratio $\log_{10}(\overline{T}_{\rm rad}/\overline{T}_K)<1.1$ at $z=8$ (at 95\% confidence), compared to the prior constraint of $\log_{10}(\overline{T}_{\rm rad}/\overline{T}_K)<2.6$ (95\% confidence),
noting that these limits can be exceeded at low $\overline{T}_K$ due to the effect described above.
The bounds are shown as dashed lines in the temperature plot in Figure~\ref{fig:extraradiobackground_posteriors}.

\begin{figure}
    \centering
    \includegraphics[width=0.5\textwidth]{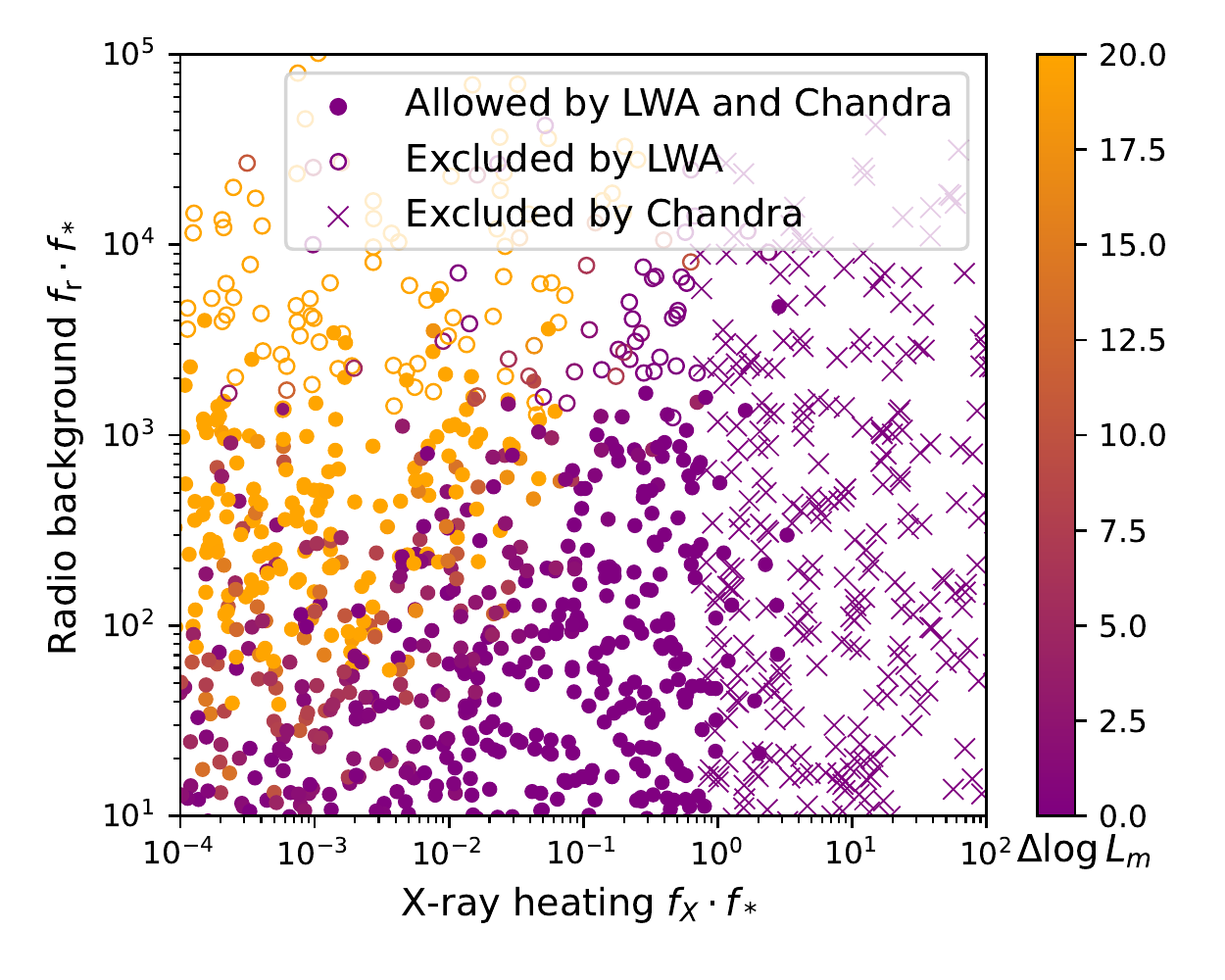}
    \caption{HERA constraints compared to the radio background constraints from LWA1
    and the X-ray background constraints from {\it Chandra}. The plot color-codes the models
    by their HERA likelihood $\mathcal{L}_m$ (orange corresponds to the models excluded at $\Delta \log \mathcal{L}_m\geq 20$ relative to the best fit and purple are the allowed models), and models excluded by LWA and {\it Chandra} are shown as empty circles and crosses, respectively. The latter constraints are computed from the level of X-ray and radio background produced population of sources at $z>8$ and approximately correspond to $f_X\cdot f_{\rm *}\lesssim 1$ and $f_{\rm r}\cdot f_{\rm *}\lesssim10^3$, respectively.     We see that HERA can exclude a substantial fraction of the otherwise unconstrained parameter
    space.}
    \label{fig:extraradiobackground_LWA}
\end{figure}

\citet{Reis:2020} showed that models with a strong radio background, corresponding to large values of the product  $f_*\times f_{\rm r}$, violate the ARCADE2/LWA1 measurement with only a weak dependence on the values of other astrophysical parameters. Additionally, models with large values of  $f_*\times f_{X}$ are ruled out by the {\it Chandra} X-ray background \citep{Fialkov:2017}, though in detail the constraints are sensitive to the assumed SED at high energies. To compare HERA constraints to the ARCADE2/LWA1 and {\it Chandra} limits we calculate the contributions to the observed present-day radio and X-ray backgrounds created by a population of sources at $z>8$ and flag as rejected models in which these high-redshift contributions exceed the observations. In Figure \ref{fig:extraradiobackground_LWA} we show that the \citetalias{HERA2021} limits are complementary to both the ARCADE2/LWA1 and {\it Chandra} X-ray background limits in that they rule out a different section of the parameter space, which is characterized by a high radio background and low X-ray heating (orange filled circles).

\subsection{Synchrotron radio background} \label{sec:synchrotron}

\begin{figure*}
    \centering
    \includegraphics[width=\textwidth]{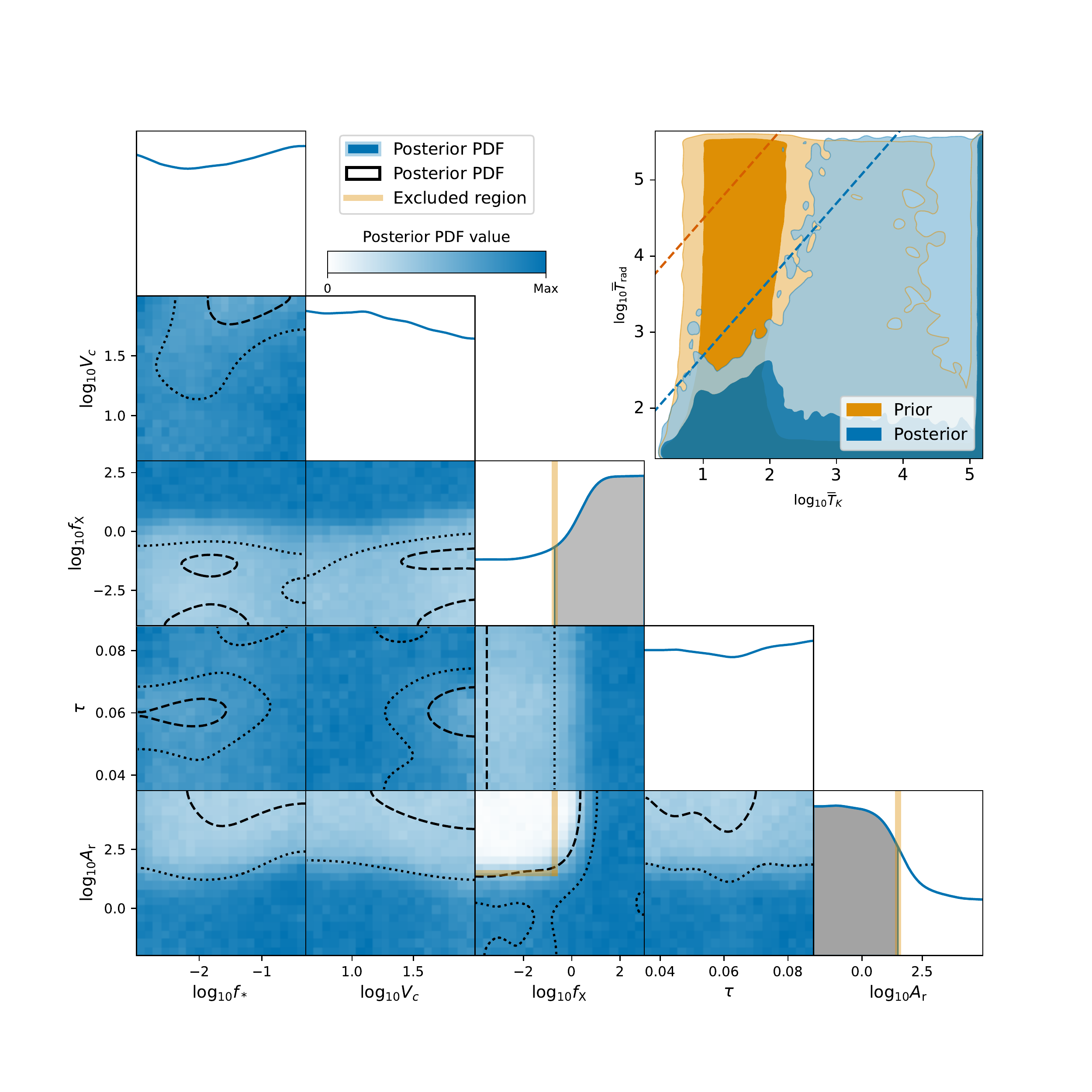}
    \caption{HERA constraints on models with a smooth synchrotron extra radio background, with analogous notation to Figure \ref{fig:extraradiobackground_posteriors}.
    Dotted and dashed lines indicate 68\% and 95\% confidence contours, the background shows the histogram of the 2D posterior.
    The grey areas and orange solid lines show the 1D 68\% percent limits on $f_{\rm X}$ and $A_{\rm r}$ individually, the combination of the latter approximately corresponds to the 95\% combined exclusion region, our main result.
    The upper right inset shows  the prior (orange) and posterior (blue) distributions of the derived model parameters $\overline{T}_{K}$ and $\overline{T}_{\rm rad}$ at $z=8$, with the dashed lines indicating the respective 95\% confidence
    limits on $\log_{10}(\overline{T}_{\rm rad}/\overline{T}_{K})$.
    }
    \label{fig:extraradiobackground_Ar_post}
\end{figure*}

Next, we calculate limits on the model parameters for the case of a  smooth phenomenological synchrotron radio background (given by eq.~\ref{eq:Arad} with $\nu_{\rm obs} = 1420/(1+z)$ MHz) that is stronger at higher redshifts. Here we build on the formalism developed by \citet{FB19} of the excess radio background, taking into account its effect on the background intensity, spin temperature and coupling coefficients, as well as heating by radio/CMB photons. We expand over the model used in \citet{FB19} by adding  Ly-$\alpha$ heating, RSDs, multiple scatterings of Ly-$\alpha$ photons, and Poisson fluctuations as described in \citet{Reis:2021}.  We use the HERA likelihood alone and obtain qualitatively similar results (Figure \ref{fig:extraradiobackground_Ar_post}) to the ones found in models where the radio background is generated by galaxies. Differences arise because the synchrotron background decays with time, while the radio background from galaxies grows. In addition, the former background is smooth, while the latter traces galaxies and, thus, includes spatial fluctuations. 

We find that models with a high excess radio background of $A_{\rm r}>31$ (calculated relative to the CMB at the reference frequency of 78 MHz, corresponding to 1.6\% of the CMB at 1.42 GHz) are excluded at 68\% confidence;
and so are models with a low X-ray efficiency of $f_{\rm X}<0.20$ (corresponding to the X-ray  luminosity per SFR of $L_{{\rm X}, ~0.2-95~ {\rm keV}}/{\rm SFR} < 5.9\times 10^{39}$ erg s$^{-1}$ M$_\odot^{-1}$ yr).
Models fulfilling both criteria are clearly excluded by the data as seen in the $f_{\rm X}$-$A_{\rm r}$ panel in Figure \ref{fig:extraradiobackground_Ar_post},
they mostly lie beyond the 95\% 2D confidence contour (dashed line).

We also constrain the ratio $\log_{10}(\overline{T}_{\rm rad}/\overline{T}_{K})<1.7$ at $z=8$ at 95\% confidence, compared to the prior constraint of $\log_{10}(\overline{T}_{\rm rad}/\overline{T}_{K})<3.5$,
again noting that these limits are less strict due to the shape of the posterior distribution as shown in the temperature plot in Figure \ref{fig:extraradiobackground_Ar_post}.

A similar analysis with these models was done by \citet{Mondal:2020} using the upper limit at  $z=9.1$  derived from  141 hours of LOFAR HB data  \citep{Mertens20}. These data rule out (at 68\% confidence) a strong contribution above 0.83\% of the CMB at 1.42 GHz (corresponding to  $A_{\rm r}>15.9$) of the high-redshift Universe to the ARCADE2 and LWA1 measurements  as well as scenarios with weak X-ray heating $f_{\rm X}<0.01$. Because the assumed synchrotron radio background decays with time, tighter constraints at higher redshifts are expected. However, the theoretical framework used here has been upgraded since the analysis of the LOFAR data (specifically,  RSDs and  extra heating by Ly-$\alpha$ and radio were added). This difference impedes a quantitative comparison between the two works. Moreover, compared to \citet{Mondal:2020}, here we  used a broader prior range on $A_{\rm r}$, which could partially explain our higher limit. 

\subsection{Implications for standard models} \label{sec:8-standard}

For completeness we note that only a small fraction of the prior volume used for this simulation suite is ruled out if the only source of the background radio photons is the CMB. We find that every excluded model  (at the level of $\Delta\log \mathcal{L}_m > 1$) belongs to the region of parameter space with $f_X\lesssim1$ (corresponding to X-ray luminosity per unit SFR of $3\times 10^{40}$ erg s$^{-1}$ M$_\odot^{-1}$ yr), $V_c\gtrsim50$ km s$^{-1}$, $\tau\lesssim0.064$ and  $f_*\lesssim 0.07$.
Note that varying $R_{\rm mfp}$, which is fixed here to 40 Mpc, could slightly modify the standard model's constraints \citep[see][]{Mondal:2020}.
The excluded models are characterized by a low X-ray heating efficiency and star formation rate, resulting in a partially reionized IGM (close to 50\%) at $z=8$ and creating the strongest reionization-driven power spectra at the frequencies observed by HERA. We do not rule out density-driven models with HERA observations, likely due to Ly-$\alpha$ heating which lowers the power spectrum amplitude in those models.

\begin{figure}
    \centering
     \includegraphics[width=0.5\textwidth]{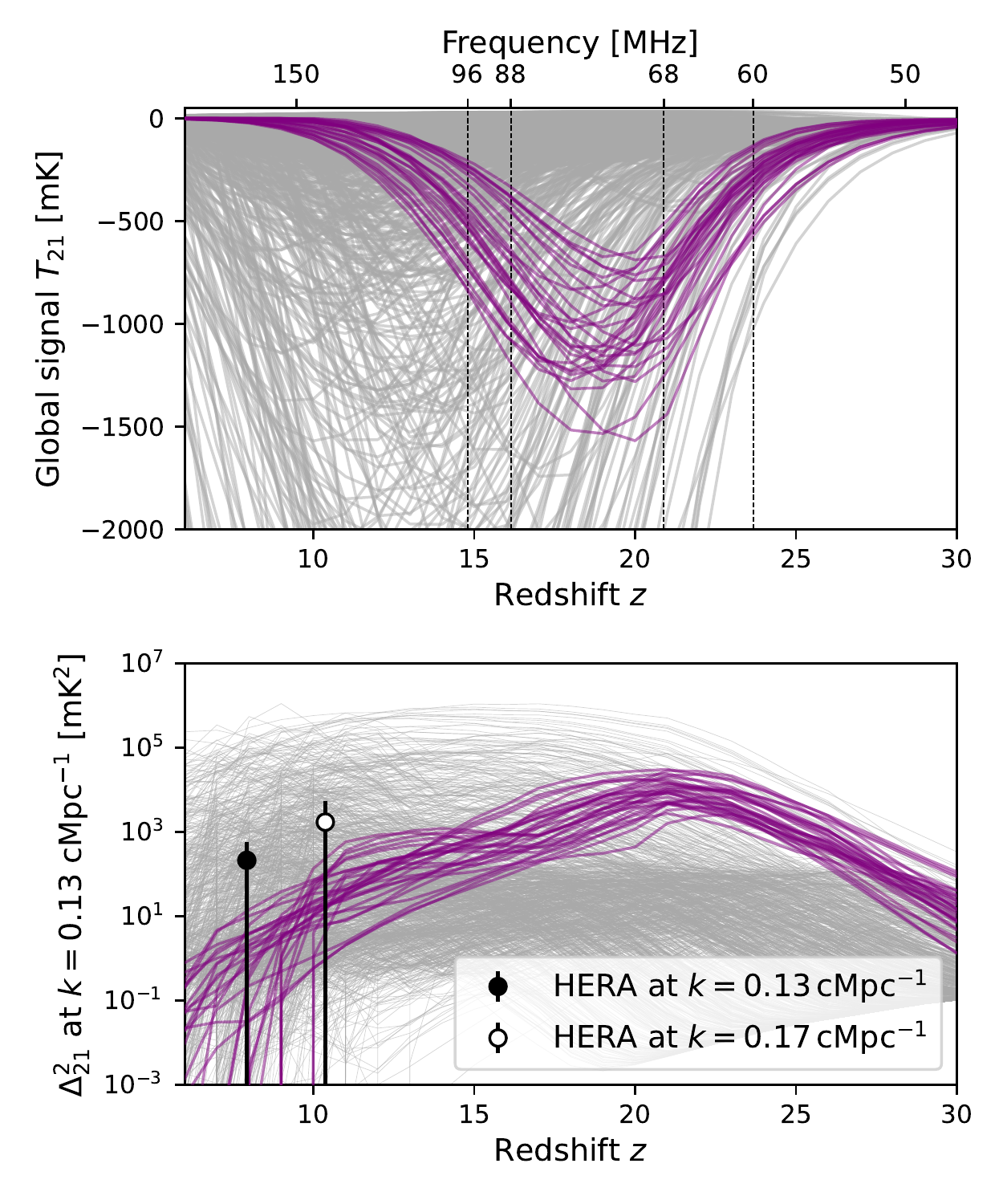}
    \caption{Models compatible with a cosmological interpretation of the EDGES detection in the context of HERA limits. \emph{Top:} Global signals of randomly selected models (grey) and those that are compatible with EDGES (purple), selected by their absorption feature close to 78\,MHz and the narrowness of the profile. The models here include an extra radio background from radio galaxies.
    \emph{Bottom:} Corresponding  power spectra at $k=0.13\,\mathrm{cMpc}^{-1}$. HERA measurements at $k=0.13$ and $0.17\,\mathrm{cMpc}^{-1}$ with $1\sigma$ error bars are shown
    for comparison. These wavenumbers are the most constrained by HERA
    but still cannot constrain the EDGES compatible models as their power spectra peak at a much higher redshift.
    }
    \label{fig:extraradiobackground_edges}
\end{figure}

\subsection{EDGES-motivated scenarios}
As mentioned in section \ref{sec:radiontro},  models with a strong extra  radio background were proposed as  a potential explanation of the anomalously deep absorption trough detected by EDGES. In such models, an accompanying enhancement of the power spectra (compared to the corresponding cases with the CMB as the background radiation) is expected \citep[][ see also Figure   \ref{fig:extraradiobackground_edges}]{FB19, Reis:2020}, and, thus, the EDGES detection could be verified by interferometers. However, in the case of the EDGES-compatible scenarios, the expected  enhancement of the power spectrum mostly falls around the central frequency of the detected global signal (i.e. 78 MHz, corresponding to  $z\sim17$), and, thus, far outside the HERA band. This is because, in order  to create a deep and narrow signal at  $z\sim17$ to explain the EDGES detection, a combination of a strong radio background and strong X-ray heating is required \citep[][]{FB19, Reis:2020}, which is a different corner of the astrophysical parameter space than that excluded by HERA. Models with strong X-ray heating tend to be suppressed at low redshifts, and, therefore, the EDGES-compatible models are also compatible with the limits set by HERA.

To illustrate this behavior, we check which of our simulated global signals are broadly compatible with EDGES\footnote{Following \citet{Fialkov2018} we adopt the criterion that EDGES-compatible global signals should have a drop of 300 to 1000\,mK at frequencies between 68 and 88 MHz, compared to the adjacent regions (60-68 and 88-96\,MHz, indicated by the dotted lines).} and mark them as purple lines in Figure \ref{fig:extraradiobackground_edges}. The rest of the models are painted in grey. In the bottom panel we plot corresponding power spectra. As we can see from the figure, indeed, the power spectra of the EDGES-compatible models (purple) peak at much higher redshifts than those probed by HERA and have low power in the HERA band. 
None of the EDGES compatible models (neither with the radio background from galaxies as shown here, nor with the phenomenological synchrotron radio background) are excluded by the HERA data.


\begin{figure}
    \centering
     \includegraphics[width=0.48\textwidth]{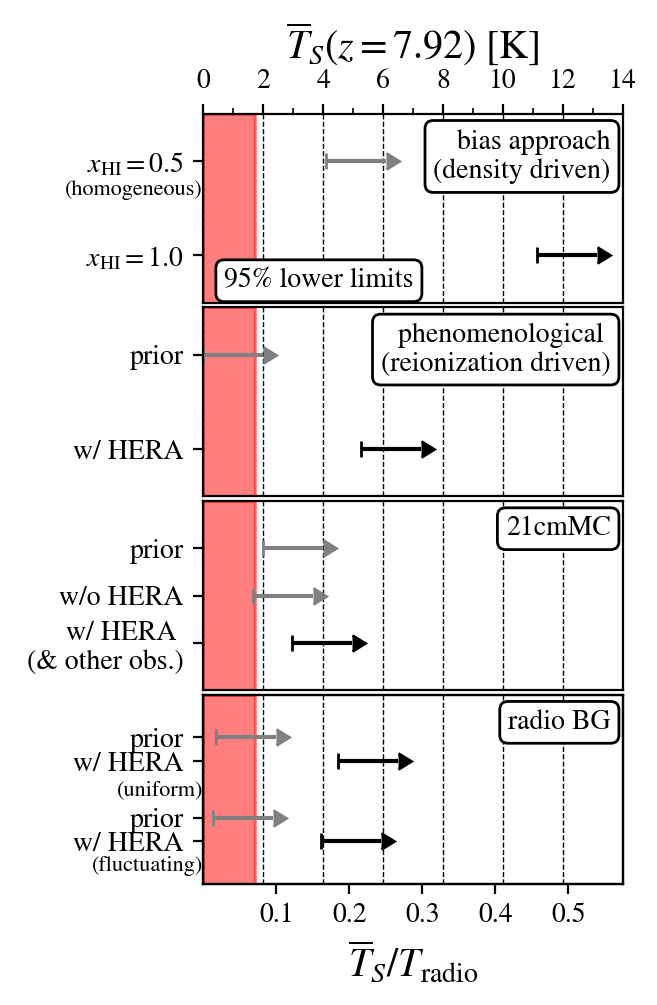}
    \caption{Summary of lower limits on the IGM spin temperature at $z=7.9$ enabled by HERA \citetalias{HERA2021} limits. From top to bottom, we show the 95\% lower limits derived via density-driven models (\S\ref{sec:bias}), phenomenological reionization-driven models (\S\ref{sec:phenom_models}), {\tt 21cmMC} (\S\ref{sec:21cmmc}), and radio background models (\S\ref{sec:radio}). The red region shows temperatures below the adiabatic cooling limit. Black points show our final constraints, while gray points indicate priors, or constraints derived with simplifying assumptions or without HERA, e.g., the top panel adopts two contrasting scenarios, a fully neutral case and a case with \emph{uniform} 50\% ionization. The bottom three panels all indicate the prior range, as well as the limits obtained via independent measurements alone in the {\tt 21cmMC} case. The bottom panel includes both homogeneous (\S\ref{sec:synchrotron}) and inhomogeneous (\S\ref{sec:radio_galaxies}) radio backgrounds -- in the latter case, the correspondence between top and bottom $x$ axes assumes a uniform $T_{\rm rad} = \Tcmb(z=7.9)$, but we caution that  $T_S/T_{\rm rad}$ constraints cannot simply be converted to $\overline{T}_S$ constraints.
    }
    \label{fig:punchline}
\end{figure}

\section{Discussion}
\label{sec:disc}

The HERA Collaboration has recently presented our first power spectrum limits at $z=7.9$ and $z=10.4$ in \citetalias{HERA2021}. In this paper, we have used a suite of theoretical models to quantify the implications of these upper limits for the landscape of IGM and galaxy formation models of the Cosmic Dawn. The most fundamental of our conclusions, about the IGM temperature at $z \sim 8$, are illustrated in Figure~\ref{fig:punchline}.  We have shown that:

\begin{itemize}

    \item The HERA limits at $z=7.9$ strongly disfavor otherwise viable models with weak or no IGM heating by astrophysical sources.  In the fiducial scenario where the CMB dominates the high-$z$ radio background, the HERA limits combined with galaxy and EoR observations constrain the spin temperature of the neutral IGM to 27 (2.3) K $< \aveTs <$ 630 (640) K at 68\% (95\%) highest posterior density confidence interval.  This corresponds to kinetic temperatures of the neutral gas of 8.9 K $< \overline{T}_K < 1.3 \times10^3$~K at 68\% confidence.  We have demonstrated this with flexible galaxy evolution models (using $\cmfast$; Fig.~\ref{fig:temp}), a phenomenological model that directly parameterizes the IGM properties  (Fig.~\ref{fig:posterior_phenom}),
    and a simple bias model (Fig.~\ref{fig:TgasLimits}).

    \item Using \cmfast\ and \cmmc\, in which astrophysical heating pre-reionization is sourced by galactic X-ray emission, we found that the combination of other high-$z$ observations and the recent limits from HERA allows us to constrain the X-ray luminosity per unit star formation rate of Cosmic Dawn galaxies, $L_{\rm X, <2keV}$/SFR.  Our resulting 68\% confidence interval of $L_{\rm X <2keV}$/SFR = $\{ 10^{40.2}, 10^{41.9} \}$ erg s$^{-1}$ $M_\odot^{-1}$ yr supports theoretical predictions (e.g. \citealt{Fragos13}) and the observed evolution with metallicity and redshift (\citealt{Basu-Zych13, Douna15, Brorby16, Lehmer16}), that this quantity increases towards high redshifts. 
    
    \item Using 
    a different
    set of simulations (section \ref{sec:radio}),
 we showed that HERA's limit (on its own) places a  constraint on an early radio background
 for models in which galaxies have substantial radio emission but weak X-ray heating. We find that the models with 
 $L_{r,\nu}/{\rm SFR} > 4 \times 10^{24}$ W Hz$^{-1}$ M$_\odot^{-1}$ yr (calculated at reference frequency $\nu=150\,\mathrm{MHz}$) \emph{and}  $L_{{\rm X}, ~0.2-95~ {\rm keV}}/{\rm SFR} < 7.6 \times 10^{39}$ erg s$^{-1}$ M$_\odot^{-1}$ yr are excluded at 95\% confidence.
The ratio of radio to gas temperatures must satisfy $\log_{10}(\overline{T}_{\rm rad}/\overline{T}_K)<1.1$ at $z = 8$ to be consistent with HERA at 95\% confidence.

 Considering
 a phenomenological model where the radio background (in addition to the CMB) has a synchrotron spectrum, we find that
 an extra radio background of 1.6\% of the CMB at 1.42 GHz or stronger \emph{and} weak heating by X-ray binaries with luminosity 
 $L_{{\rm X}, ~0.2-95~ {\rm keV}}{\rm SFR} <  5.9\times 10^{39}$ erg s$^{-1}$ M$_\odot^{-1}$ yr, are robustly excluded by the data (at 95 \% confidence). For this model we find that the ratio $\log_{10}(\overline{T}_{\rm rad}/\overline{T}_K)<1.7$ at $z=8$ is required for consistency with HERA at 95\% confidence.

 The HERA constraints on X-ray and radio efficiencies are complementary to the constraints from the unresolved X-ray background  (measured by {\it Chandra}) and the low-frequency radio background (detected by ARCADE2/LWA), ruling out otherwise unconstrained parameter space.

    \item There has been considerable recent interest in exotic cosmological models that might explain the EDGES measurement at $z \sim 18$ \citep{Bowman18}, either through cooling the IGM below the adiabatic limit or imposing an additional radio background. The \citetalias{HERA2021} limits cannot confirm or disprove the EDGES signal, simply because it observes at a lower redshifts. However, the limits do independently rule out
    a very cold IGM at $z=10.4$ (as well as an IGM at or below the adiabatic limit at $z=7.9$), so that heating must occur before $z \sim 10$ if millicharged dark matter is behind the anomalous EDGES depth. 
    The EDGES signal itself, if taken at face value, already requires such heating at $z \ga 15$, so cosmological explanations of that signal are compatible with the \citetalias{HERA2021} limits. Future measurements from HERA at lower frequencies will be important for further testing such scenarios.

\end{itemize}

These conclusions, validated by multiple independent approaches, demonstrate that the \citetalias{HERA2021} upper limits provide important astrophysical constraints on early galaxies. This is the first time IGM heating has been required by observations, independent of any assumptions about galaxy formation during this era. Nevertheless, the upper limits are sufficient to show how state-of-the-art galaxy models can offer new insights, as demonstrated by the difference in temperature constraints in ``density-driven" models. 

Moreover, our results emphasize how 21-cm measurements are highly complementary to other probes of the reionization era. No other existing probe can constrain the X-ray or radio emissivity of early-galaxy populations as effectively. The requirement of at least minimal heating (or substantially more if radio emission is strong or if dark-matter interactions are at play) translates to a lower limit on radiative heating by star-forming galaxies (conventionally interpreted as X-ray heating, though at very low temperatures Lyman-$\alpha$ heating also plays a role). While the current upper limits do not strongly constrain the expected parameter space of early galaxies, they show the promise of 21-cm interferometers. We also note that the combination of HERA data with complementary data is most powerful when they probe similar redshifts, so that assumptions about redshift evolution can be avoided.

Of course, there are several caveats to this interpretation and ways in which our methods can be improved: \emph{(i)} they are driven largely by a very small number of measurements (single bandpowers at $z=7.9$ and $z=10.4$) so are more subject to systematic concerns; \emph{(ii)} our constraints do not fully capture the correlations between bandpowers; and \emph{(iii)}  our simulation-based analyses also do not treat the angular dependence of the signal properly, as current observations are sensitive only to modes nearly along the line-of-sight, where the power spectrum is most strongly affected by redshift-space distortions (see Fig.~\ref{fig:TgasLimits}). Our assumptions are strictly conservative for density-driven scenarios, but the effect of redshift-space distortions is more complicated during reionization. We expect to improve the match between simulations and the data for future observing campaigns.

The \citetalias{HERA2021} constraints are based on a single observing season with a small number of antennae and the first iteration of HERA's infrastructure. HERA has continued to expand and improve, and later campaigns will offer much higher sensitivity. This analysis provides a foundation for the interpretation of these forthcoming datasets. 

\acknowledgments

This material is based upon work supported by the National Science Foundation under Grant Nos. 1636646 and 1836019 and institutional support from the HERA collaboration partners.
This research is funded by the Gordon and Betty Moore Foundation
through grant GBMF5215 to the Massachusetts Institute of Technology.
HERA is hosted by the South African Radio Astronomy Observatory, which is a facility of the National Research Foundation, an agency of the Department of Science and Innovation.
A.M. acknowledges funding from the European Research
Council (ERC) under the European Union’s Horizon 2020
research and innovation programme (grant agreement No
638809 – AIDA). The results presented here
reflect the authors’ views; the ERC is not responsible for
their use.

J.S.D. gratefully acknowledges the support of the NSF AAPF award \#1701536.
A.~Liu acknowledges support from the New Frontiers in Research Fund Exploration grant program, the Canadian Institute for Advanced Research (CIFAR) Azrieli Global Scholars program, a Natural Sciences and Engineering Research Council of Canada (NSERC) Discovery Grant and a Discovery Launch Supplement, the Sloan Research Fellowship, and the William Dawson Scholarship at McGill.
N.K. acknowledges support from the MIT Pappalardo fellowship.
G. B. acknowledges funding from the INAF PRIN-SKA 2017 project 1.05.01.88.04 (FORECaST), support from the Ministero degli Affari Esteri della Cooperazione Internazionale - Direzione Generale per la Promozione del Sistema Paese Progetto di Grande Rilevanza ZA18GR02 and the National Research Foundation of South Africa (Grant Number 113121) as part of the ISARP RADIOSKY2020 Joint Research Scheme, from the Royal Society and the Newton Fund under grant NA150184 and from the National Research Foundation of South Africa (grant No. 103424).
A.F. is supported by the Royal Society University Research Fellowship. 
J.B.M.~is supported by a Clay Fellowship at the Smithsonian Astrophysical Observatory.
Y.Q. would like to thank Cathryn M. Trott for providing MWA data and acknowledge the support from the High Performance Computing centers of the Scuola Normale Superiore (Italy), the Council for Scientific and Industrial Research (CHPC South Africa), and the OzSTAR national facility (Australia) for computational resources. Parts of this research were supported by the Australian Research Council Centre of Excellence for All Sky Astrophysics in 3 Dimensions (ASTRO 3D), through project number CE170100013. MGS acknowledges support from the South African Radio Astronomy Observatory(SARAO) and the South Africa National Research Foundation (Grant No. 84156).

\begin{center}
 CONTRIBUTIONS OF THE AUTHORS
\end{center}

S.R.F. leads the HERA Theory team and organized this effort. Y.Q. led the \cmfast\ and \cmmc\ simulation and analyses, with particular support from B.G. and A.M. (\S \ref{sec:21cmmc}). J.M. led the analysis using phenomenological models (\S \ref{sec:phenom_models}), with particular support from J.B.M. and S.R.F. J.B.M. led the bias analysis (\S \ref{sec:bias}) and the dark-matter interpretation (\S \ref{sec:exotic_DM}). S.H. led the analysis described in \S \ref{sec:radio} using models created by A.F.,
and developed the emulators based on the architecture by S.S.
S.G.M. and N.K. developed the derivation of the likelihood (\S\ref{sec:likelihood}). R.B. is in the Builder's list of the 21-cm code used in section \ref{sec:radio},
I.R. led the development of the radio-background from galaxies model  used in Section \ref{sec:radio}.
The HERA Builder's list is included  as authors because of the dependence of this work on the combined efforts of collaborators on the various software repositories as well as the necessity of using HERA data to constrain the models used in this paper.

\newpage
\appendix{}

\section{Fluctuations Tracing the Velocity Field}
\label{sec:vao}

In section~\ref{sec:bias}, we assumed that the 21-cm signal traced the matter power spectrum for simplicity. 
However, in some regimes the 21-cm power spectrum may more closely trace the velocity-induced acoustic oscillations (VAOs) from the dark matter-baryon relative velocities $v_{\rm cb}$.
This regime can be triggered by a strong dependence of galaxy formation on these streaming velocities at small scales~\citep{Tseliakhovich:2010bj,Dalal:2010yt,Naoz:2011if}, which however are not expected to play a key role at low redshifts~\citep{Fialkov2012,McQuinn:2012rt,Munoz:2019rhi} (see however~\citealt{Park:2020ydt,Cain:2020npm} for the effect during the epoch of reionization).
Exotic interactions between dark matter and baryons can modulate the gas temperature with the same velocity feature~\citep{Munoz:2015bca,MDL18,Fialkov:2018xre}.
Regardless of whether its origin is astrophysical or exotic, a detection of VAOs would provide us with a standard ruler~\citep{Munoz:2019fkt}.

To model such a scenario, we replace the matter power spectrum in equation~(\ref{eq:d21-analytic}) with $\Delta^2_{v_{cb}}(k)$ and constrain the corresponding $b_v$.
The quantity $\Delta^2_{v_{cb}}(k)$ is the power spectrum of $\delta_{v_{\rm cb}} = \sqrt{3/2}\, (v_{\rm cb}^2/v_{\rm rms}^2-1)$, which is dimensionless, $z$-independent, and has unit variance~\citep{Dalal:2010yt,Ali-Haimoud:2013hpa}.
Using the \citetalias{HERA2021} data we can set the 95\% confidence limits $b_{v_{\rm cb}} < \{49, 175\}$ mK at $z=\{7.9,10.4\}$, respectively.
For comparison, a typical VAO contribution to the 21-cm power spectrum found in the simulations of~\cite{Munoz:2019rhi} and \cite{Fialkov:2012su} is roughly $\Delta^2_{21}\approx10$ mK$^2$ (which peaks at $z\approx15$ during their X-ray heating epoch). This would convert into a bias $b_{v_{\rm cb}}\approx 5$ mK, much below the limits above.

\section{Emulator for extra radio background simulations}
\label{appendix:emulator_error}

Emulators are widely used to rapidly evaluate 21-cm power spectra across the broad astrophysical parameter space \citep{Kern17, Schmit2019, Jennings2019, Ghara2020, Mondal:2020}. This technique allows us to interpolate existing simulations instead of running a new simulation at every point of the  parameter space.
Here we choose to emulate the power spectrum using a neural network (multi layer perceptron) regression, implemented in {\tt scikit-learn} \citep{Pedregosa2011}.

We train emulators for
the two cases of radio backgrounds separately: one data set corresponding to the models with radio background created by galaxies (10700 samples), the other corresponding to the smooth synchrotron radio background (10300 samples). These sets were created by randomly sampling the parameters $V_c$, $f_{\rm *}$, $f_{\rm X}$, $\tau$, and $f_{\rm r}$ or $A_{\rm r}$ within the bounds described in section \ref{subsec:modelling_radio}.
To calculate the likelihood of each parameter set we go through a two-step process:  First, we emulate the power spectrum
assuming no RSDs (emulators based on the corresponding $\sim 10000$ models mentioned above), then we use another emulator to account for the effect of RSDs (using another $\sim 2000$ simulations calculated separately for each type of radio background), boosting or suppressing the power spectra by a factor between 0.8 to 4 depending on the values of the astrophysical parameters. The final power spectra are used in the likelihood calculation.  
We also emulate the values of spin, gas, and radiation temperatures (training data $\sim 10000$ cases for each type of radio background) to infer constraints on the  physical properties of the IGM (e.g. Figure \ref{fig:TgasLimits} and \ref{fig:extraradiobackground_posteriors}, insert).
For completeness we also create a data set ($\sim 2000$ models) with no radio background to test the  emulator in this limit
and to calculate the envelope of these models in
Figure~\ref{fig:extraradiobackground_radiopowerspectra} (black dashed line).

The input parameters for all our emulators are taken to be natural logarithms
of model parameters ($\ln f_{\rm star}$, $\ln V_c$, $\ln f_{\rm X}$,
$\ln \tau$, and $\ln f_{\rm r}$ or $\ln A_{\rm r}$) to cover the dynamic range
(using $\ln \tau$ for simplicity; we use linear $\tau$ in the analysis).
As part of the emulator, the inputs are also shifted and scaled such that
the training input values have zero mean and unit variance.
The network itself consists of 4 hidden layers for power spectra and RSDs and  2 hidden layers for
temperatures. Each layer has  100 nodes, employs  a \texttt{ReLU} activation function \citep{nair2010} and is trained using the \texttt{adam} optimizer \citep{Kingma2014}. The  power spectra emulators return 78 outputs corresponding to the two log-power spectra
(or RSD boost factors) at $z=7.93$ and 10.37 and at the 39 wave numbers measured  by HERA; the temperature emulators have a single output that is the  logarithm of temperature. The simulation power spectra are only computed at the wave numbers $\le 1.1\,\mathrm{cMpc}^{-1}$ due to the limited resolution of the simulation. However, values at smaller scales are not relevant as the HERA limits there exceed the expected power spectra. To take into account the large dynamic range of possible power spectra we use the logarithm
$\log (\Delta^2_{21}+1\,\mathrm{mK}^2)$ as the target for the emulator. Adding the baseline  $1\,\mathrm{mK}^2$ term helps to improve the performance of the emulator for the values of power spectra that can be constrained by HERA.

\begin{figure*}
    \centering
    \includegraphics[width=0.8\textwidth]{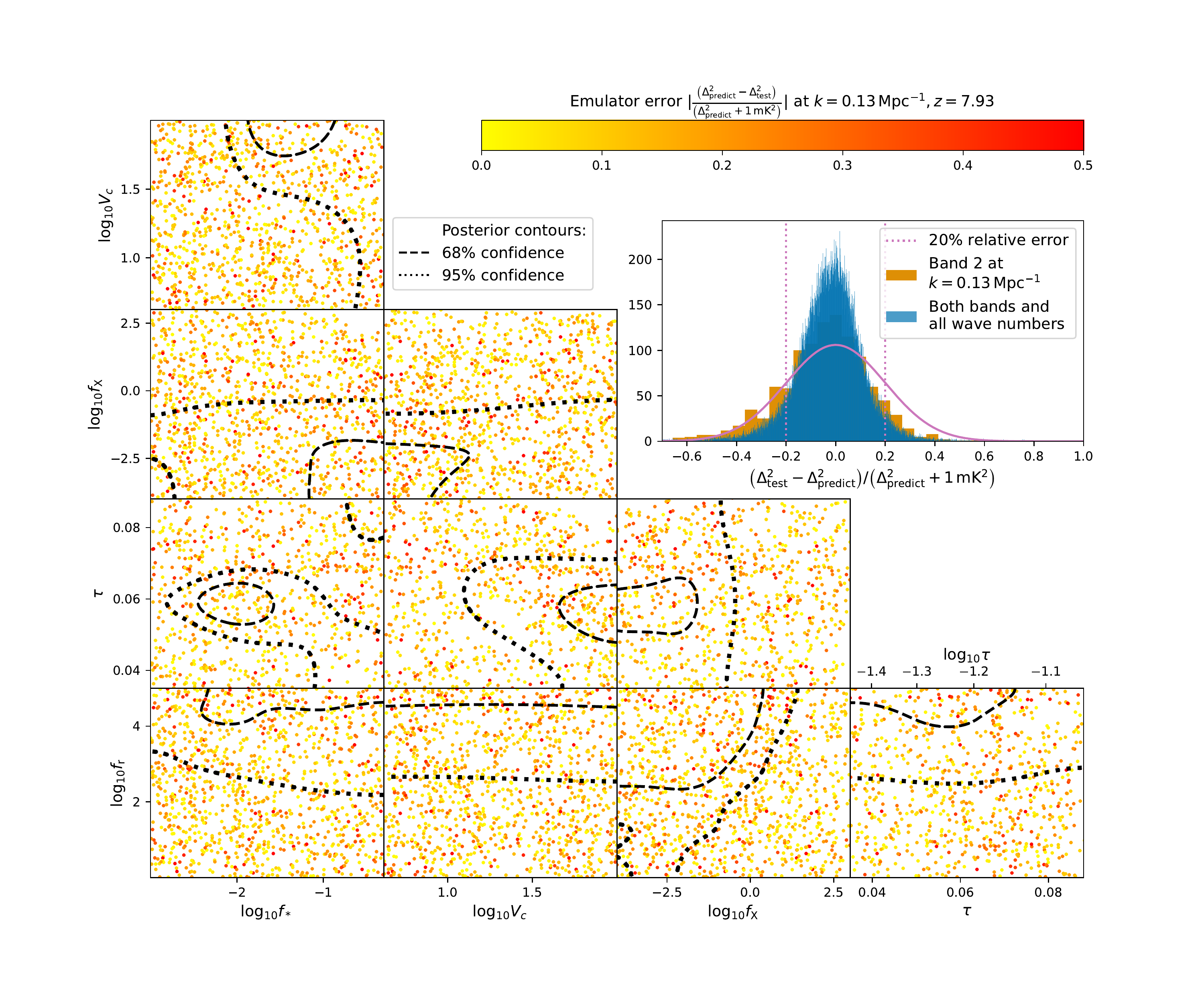}
    \includegraphics[width=0.32\textwidth,trim={0 1cm 0 0},clip]{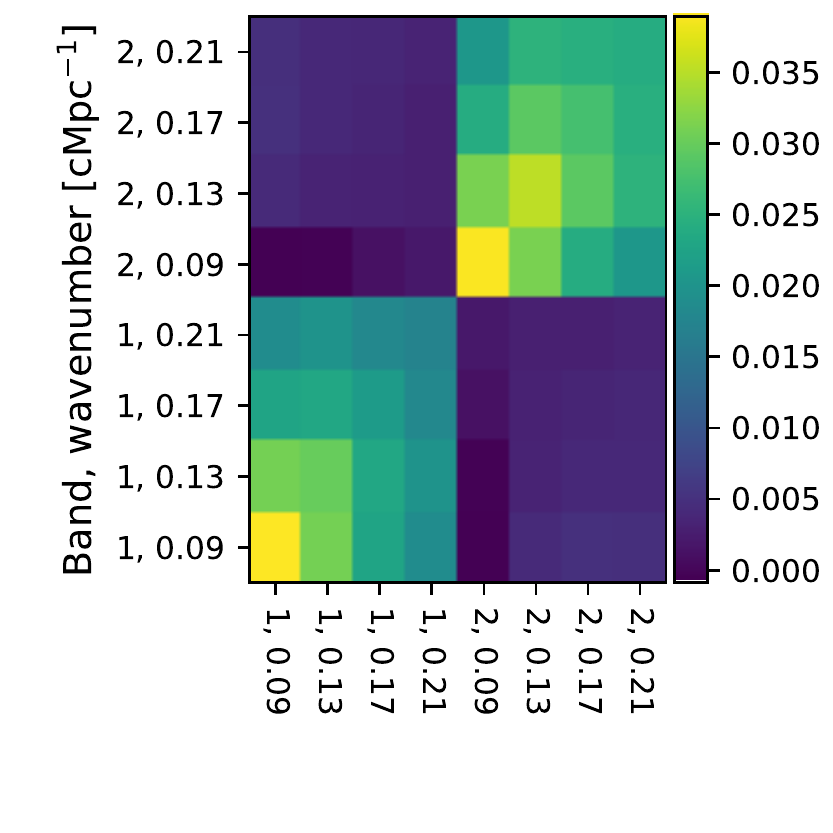}
    \includegraphics[width=0.32\textwidth]{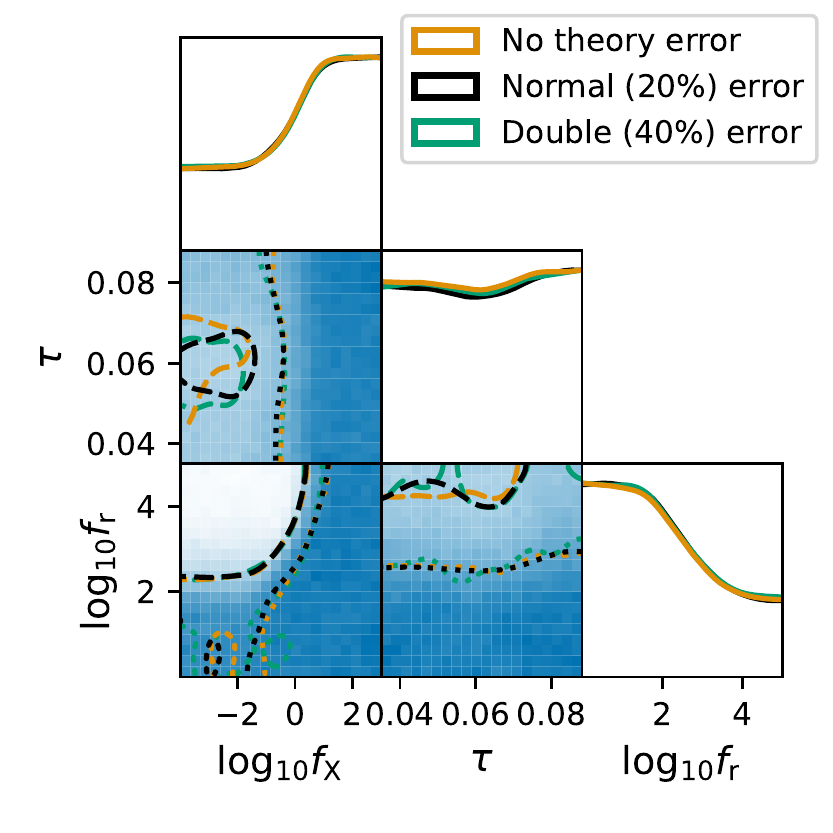}
    \includegraphics[width=0.32\textwidth]{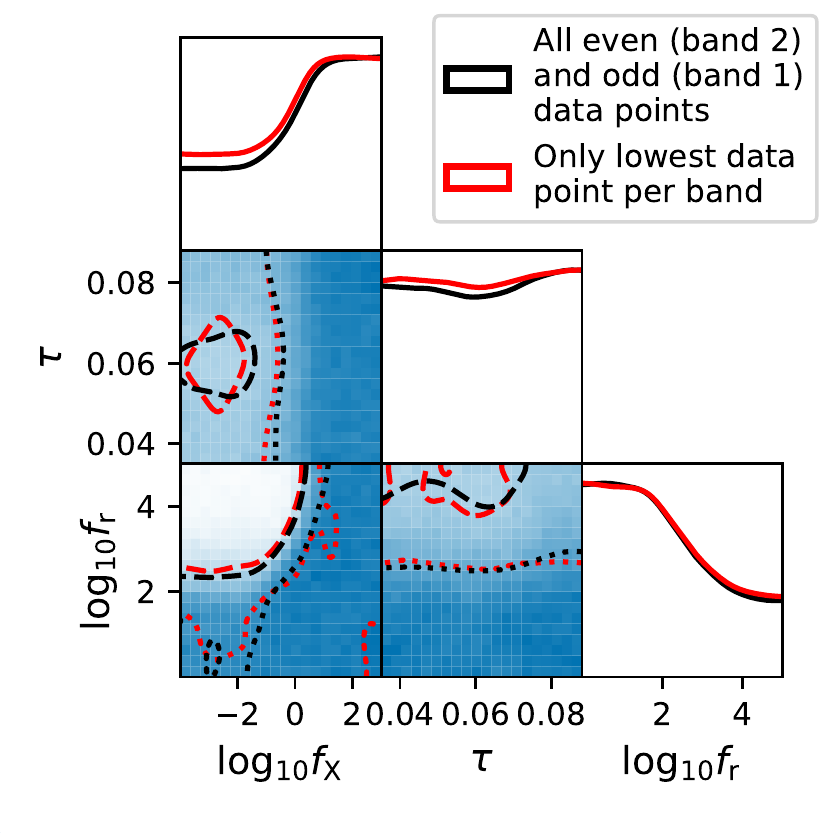}
    \caption{\emph{Top:} Accuracy of the neural network predictions. The triangle plots show the relative accuracy
    of predicting the power spectrum at $z=7.93$, $k=0.13\,\mathrm{cMpc}^{-1}$, for all parameters in the test set.
    The relative emulator error is shown by the color of the points, from yellow (no error) to red (50\% error).
    For comparison we add the black contour lines from Figure \ref{fig:extraradiobackground_posteriors}
    corresponding to  68\% and 95\% parameter confidence intervals, respectively.
    The histogram insert shows the overall distribution of the error $\sigma_{\rm rel} = {\left(\Delta^2_{\rm predict}-\Delta^2_{\rm test}\right)}/{\left(\Delta^2_{\rm predict}+1\,\mathrm{mK}^2\right)}$, for $z=7.93$, $k=0.13\,\mathrm{cMpc}^{-1}$ (orange histogram) and for all points combined (blue). The orange histogram can be approximated by a normal distribution with scale $\sigma=0.2$ (pink curve).
    \emph{Botton:} The left panel shows the covariance matrix of the relative emulator
    error ($\sigma_{\rm rel}$ as used above, variance of 0.03 corresponds to 17\% error) over both bands and the first few wavenumbers.
    The emulator error correlation between the different bands is negligible, but the errors within a band
    can be strongly correlated with a high correlation coefficient $r \approx 0.9$.
    The middle panel shows the overall impact of taking the emulator uncertainty into account
    by looking at the marginalized $f_{\rm r}$, $f_X$ and $\tau$ constraints.
    We see that assuming no emulator error (orange), 20\% (black, as used in results) or 40\% (green) does not
    noticeably affect the parameter constraints as the observational uncertainty dominates the error budget in the
    relevant region.
    The right panel shows parameter constraints using a restricted likelihood, based on just a single data point
    per band (red contours, using $k=0.17$ and $0.13\,\mathrm{cMpc}^{-1}$ for band 1 and 2, respectively)
    compared to the full likelihood (black contours,
    as used for our scientific results). The constraints with the single data points are just slightly weaker than
    using the full likelihood which shows that the parameter constraints are mainly driven by the lowest point
    in each band.
    }
    \label{fig:extraradiobackground_appendix_emulator_error}
\end{figure*}

The accuracy in  predicting the power spectra is evaluated using a test data set (2000 test samples for power spectrum and temperatures, 1000 for RSD boost factor and final power spectra). Since we train the emulator on logarithms of the power spectra, the emulator errors on these logarithmic values are approximately the same over the parameter space.
This implies that the error on the power spectrum $\Delta^2_{21}$ is
proportional to the power spectrum value itself.
We find that the deviations
are well within the 20\% relative error in the theoretical modelling that  we assume here, as seen from the histogram insert
in  Figure \ref{fig:extraradiobackground_appendix_emulator_error} which  shows the
distribution of the relative error (in $\Delta^2_{21}+1\,\mathrm{mK}^2$) for all the predicted 
power spectra points (blue) at the most relevant $k=0.13\,\mathrm{cMpc}^{-1}, z=7.93$
band (orange), along with the assumed Gaussian uncertainty (pink curve, $\sigma=0.2$).

The triangle plot in Figure \ref{fig:extraradiobackground_appendix_emulator_error} shows the relative emulator error
(including RSD, and again as $\Delta^2_{21}+1\,\mathrm{mK}^2$) for every parameter
point in the test set, together with the contour lines from Figure \ref{fig:extraradiobackground_posteriors}
in the background. 
This shows that no parameter region has a particularly high emulator error, and that the test (and also training) data are evenly distributed over the whole parameter space.

We assume this relative emulator uncertainty to be approximately Gaussian and independent, and add it in
quadrature to the observational error. We note that the emulator error is strongly correlated within each band
as shown in the lower left panel of Figure \ref{fig:extraradiobackground_appendix_emulator_error},
however we neglect this correlation to derive the marginal likelihood in Section \ref{sec:marginal_likelihood}. This is  justified as
(i) the emulator uncertainty is  subdominant compared to the large observational errors and barely affects the parameter constraints (lower central panel of Figure \ref{fig:extraradiobackground_appendix_emulator_error}),
and (ii) the constraints are mainly driven by the single lowest data points in each band 
(the lower right panel of Figure
\ref{fig:extraradiobackground_appendix_emulator_error}
shows that using just these 2 points only slightly weakens the constraints to
$f_X>0.15$ and $f_{\rm r}<504$ compared to $f_X>0.33$ and $f_{\rm r}<391$  when the complete data set is used) implying that the correlations have only a small, secondary effect.

\bibliographystyle{aasjournal}
\bibliography{ms, additional}

\end{document}